\documentclass[CIT,biber]{nowfnt} 

\usepackage{tikz}
\usetikzlibrary{arrows.meta, positioning}

\usepackage[utf8]{inputenc}

\usepackage{empheq}
\usepackage{psfrag}
\usepackage[process=all]{pstool}
\usepackage{subfigure}
\usepackage{graphicx}
\usepackage{algorithm,algorithmic,tcolorbox}

\title{Directed Information}
\subtitle{Estimation, Optimization and Applications in Communications and Causality}

\maintitleauthorlist{
Dor Tsur \\
Ben-Gurion University of the Negev \\
dortz@post.bgu.ac.il
\and
Oron Sabag \\
The Hebrew University of Jerusalem \\
oron.sabag@mail.huji.ac.il
\and
Navin Kashyap \\
Indian Institute of Science \\
nkashyap@iisc.ac.in
\and
Haim Permuter \\
Ben-Gurion University of the Negev \\
haimp@bgu.ac.il
\and
Gerhard Kramer \\
Technical University of Munich \\
gerhard.kramer@tum.de
}

\usepackage{mwe}
\usepackage{Fnt_macros}

\author[1]{Dor Tsur}
\author[2]{Oron Sabag}
\author[3]{Navin Kashyap}
\author[1]{Haim Permuter}
\author[4]{Gerhard Kramer}

\affil[1]{Ben-Gurion University of the Negev}
\affil[2]{The Hebrew University of Jerusalem}
\affil[3]{Indian Institute of Science}
\affil[4]{Technical University of Munich}

\articledatabox{\nowfntstandardcitation}
\allowdisplaybreaks

\makeatletter
\providecommand\HyPL@Entry[1]{}
\AddToHook{env/document/begin}{%
\@ifpackageloaded{preview}{
\ifPreview
 \let\Hy@FirstPageHook\relax
 \let\Hy@EveryPageAnchor\relax
\fi}{}}
\makeatother

\begin{document}

\makeabstracttitle

\begin{abstract}
Directed information is an information measure that attempts to capture directionality in the flow of information from one random process to another. The definition of directed information, in its present form, is due to Massey (1990), but the origins can be traced back to the work of Marko in the 1960s. It is closely related to other causal influence measures, such as transfer entropy and Granger causality. This monograph provides an overview of directed information and its main application in information theory, namely, characterizing the capacity of channels with feedback and memory. We begin by reviewing the definitions and basic properties of directed information, particularly its relation to mutual information, the classic information measure introduced by Shannon. We then explain how directed information is related to transfer entropy, Granger causality, and Pearl's causality. 
 
Directed information is often used to identify causal relationships in natural processes, such as neural spike trains. Several methods have been developed in the literature to estimate the directed information between two processes from time series data. We provide a survey, ranging from classic plug-in estimators to modern neural-network-based estimators. Considering the information-theoretic application of channel capacity estimation, we describe how such estimators numerically optimize directed information rate over a class of joint distributions on input and output processes.

A significant part of the monograph is devoted to techniques to compute the feedback capacity of finite-state channels (FSCs). The feedback capacity of a strongly connected FSC is given by a multi-letter expression involving the maximization of the directed information rate from the channel input process to the output process, the maximization being performed over the class of causal conditioned probability distributions on the input process. When the FSC is also unifilar, i.e., the next state is given by a time-invariant function of the current state and the new input-output symbol pair, the feedback capacity is the optimal average reward of an appropriately formulated Markov decision process (MDP). This MDP formulation has been exploited to develop several methods to compute exactly, or at least estimate closely, the feedback capacity of a unifilar FSC. This monograph describes these methods, starting from the classic value iteration algorithm, moving on to $Q$-graph methods, and ending with reinforcement learning algorithms that can handle channels with large input and output alphabets.
\end{abstract}

\chapter{Introduction}
\label{chap:intro}
\section{Information Theory and Causality}
Information theory began as a mathematical theory of communication as developed by Shannon \cite{shannon1948mathematical}. The theory deals fundamentally with manipulating finite sequences $x^n=(x_1,\dots,x_n)$ of symbols and how to efficiently store and communicate them. Shannon used a statistical approach where each sequence $x^n$ is a realization of a random sequence $X^n=(X_1,\dots,X_n)$ drawn from some probability distribution. 

Shannon's theory deals primarily with treating long blocks of symbols. For instance, in source coding, an encoder maps a block $x^n$ of symbols to a block $b^k$ of bits and wishes to determine the smallest rate $R=k/n$ so that a decoder can recover $x^n$ from $b^k$ with vanishing error probability. However, there are many problems where one observes multiple sequences 
and wishes to determine how the symbols relate at different positions.

For example, if the symbol index $i$ refers to time, one may wish to determine which effect is larger; the change that $X^i$ \emph{causes} in $Y_i$ given the past symbols $Y^{i-1}$ or the one $Y^i$ \emph{causes} in $X_i$ given the past symbols $X^{i-1}$. However, the concept of \emph{causality} is fraught with pitfalls since it is easy to mistake causality for \emph{correlation}.\footnote{This is a central argument in Pearl's framework of causality \cite{pearl2009causality}, which we connect with directed information in Section \ref{sec:causality}.} For example, in source coding it is natural to state that $X^n$ causes $B^k$, rather than the reverse, because $B^k$ is a \emph{function} of $X^n$ but not necessarily the reverse. However, many statistical measures, such as mutual information, do not elucidate information flow's \emph{direction}.

The idea, then, is to have a standard framework to measure the \emph{direction} of information flow, be it temporal, spatial, or otherwise. This monograph reviews a measure known as \emph{directed information} (DI).

\section{Marko's Bidirectional Communication}
Hans Marko is considered to be the pioneer of DI through his work \cite{marko1966bidirectional} (the English translation is \cite{marko1973bidirectional}). In this monograph, he defines the information flow in a bidirectional setting that comprises two interacting systems. Such systems represent `conscious processors', and the work aims to mathematically characterize human interaction, noting that each person is represented with their `own' stochastic process and entropy; the bidirectional communication is referred to as a `dialogue'.
According to Marko, bidirectional communication is based on work about the relation between information theory and cybernetics \cite{marko1967information} and Shannon's two-way channel \cite{shannon1961two}.

Marko defines three quantities for directional information flows. Consider symbol pairs $(X^n,Y^n)$ from a joint distribution $P_{X^n,Y^n}$, $n\ge 1$, and suppose the symbols $(X_i,Y_i)$ are available at transmitters $M_x,M_y$, respectively. Define the following quantities\footnote{The corresponding limits exist whenever the underlying stochastic processes are stationary \cite{CovThom06}.} for $M_x$ and $M_y$: 
\begin{enumerate} \itemsep 0pt
    \item \textit{Total} information at $M_x$, $H_x = \lim_{n\to\infty}\EE[-\log P(X_{n+1}|X^n)]$, which is closely related to the entropy rate of $\XX\triangleq(X_i)_{i=-\infty}^{\infty}$ \cite{CovThom06}. 
    \item \textit{Free} information at $M_y$, $F_y =\lim_{n\to\infty} \EE[-\log P(Y_{n+1}|X^n,Y^n)]$, which is closely related to the entropy rate of $\YY\triangleq(Y_i)_{i=-\infty}^{\infty}$ causally conditioned on $\XX\triangleq(X_i)_{i=-\infty}^{\infty}$.
    \item Directed transinformation is defined as
    \begin{align}
     T_{x,y} &= \lim_{n\to\infty}\EE\left[-\log\frac{P(Y_{n+1}|Y^n)}{P(Y_{n+1}|X^n,Y^n)}\right] \nn\\
     &= \lim_{n\to\infty} I(X^n;Y_{n+1}|Y^n),
    \end{align}
    which naturally leads to the notion of DI.  
\end{enumerate}
Marko develops relations between these quantities on a graph that describes bidirectional communication between the transmitters $(M_x,M_y)$ and points out the similarity to Kirchhoff's laws.
Interestingly, Marko was the first to discuss the conservation of information (which was later formalized by Massey \cite{massey2005conservation}, as presented in \cite[Eq.~(14)]{marko1973bidirectional}).
Marko also defines the `stochastic degree of synchronization' between the entities, which is given by the ratio
$\sigma_x=H_x/T_{x,y}$, i.e., the ratio between the amount of information generated at $M_x$ and the amount of information that `arrives at' $M_y$.
Finally, Marko applies his paradigm to analyze social interaction between two monkeys. By analyzing the sequence of actions the monkeys convey, Marko establishes two behavioral mechanisms that adhere to distinct sets of information‐flow values, namely \emph{hero} and \emph{dictator} behaviors. In the \emph{dictator} behavior, one agent generates strictly more free information than the other, i.e., $F_{1}>F_{2}$, and simultaneously transmits more directed transinformation, $T_{1,2}>T_{2,1}$, indicating an initial dominance in the bidirectional coupling. In contrast, the \emph{hero} behavior is characterized by comparable free‐information rates, $F_{1}\approx F_{2}$, but with a stronger informational influence, $T_{2,1}>T_{1,2}$.

\section{Overview of Applications of the Directed Information}
DI is a concept in information theory that measures the amount of information flowing from one process to another in a causal manner. It extends the idea of mutual information to capture the directional and time-dependent nature of information transfer between random processes or time series.


\begin{figure}[ht]
    \centering
    \subfigure[Hans Marko]
    {
    \includegraphics[width=0.35\linewidth,scale=0.8]{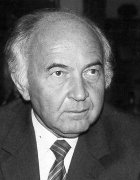}
    }
    \qquad\qquad
    \subfigure[James Massey]
    {
     \includegraphics[width=0.33\linewidth,scale=0.74]{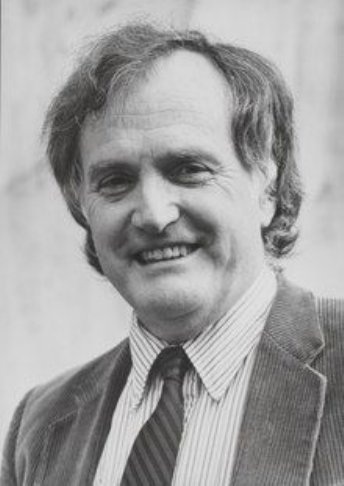}
    }
  \caption{The founders of directed information.}
  \label{fig:di}
\end{figure}


To tackle problems arising in information systems with causally dependent components, James Massey~\cite{massey1990causality}, inspired by
Marko's work \cite{marko1966bidirectional,marko1973bidirectional} on bidirectional communication, coined the
notion of DI from $X^n$ to $Y^n$, defined as
\begin{equation}\label{e_def_directed}
I(X^n\to Y^n) \triangleq \sum_{i=1}^n I(X^i;Y_i|Y^{i-1}),
\end{equation}
and showed that the normalized maximum DI upper bounds the capacity of channels with feedback. We review subsequent work inspired by~\cite{massey1990causality}.

\underline{Capacity and coding of channels with feedback:}
DI, as defined by Massey, characterizes the capacity of channels with feedback~\cite{kramer1998directed,tatikonda_phd, Kramer03, kim2008coding,Tatikonda2009feedbakcapacity, permuter2009finite,PermuterWeissmanChenMAC_IT09, ShraderPermuter09CompoundIT,DaborahGoldsmith10BC_feedback,Kramer-E23}.
In Section \ref{chap:capacity} we describe the role of DI in characterizing the capacity and review several cases of channels with memory that were solved both in terms of capacity and coding with feedback, such as the Trapdoor channel \cite{Permuter2008Trapdoor}, the Ising channel
\cite{ElishcoPermuter2014Ising}, the Binary Erasure Channel (BEC) and the Binary Symmetric Channel (BSC) with no repeated ones \cite{SabagPermuterKashyap2016BEC}, the Dicode Erasure channel \cite{sabag2017single}, the Previous Output is the STate (POST) channel \cite{permuter2014capacity}, and the energy harvesting model \cite{Harvesting_shmuel_sabag_permuter24}.


\begin{figure}[h!]{
\centering
\includegraphics[width=\linewidth]{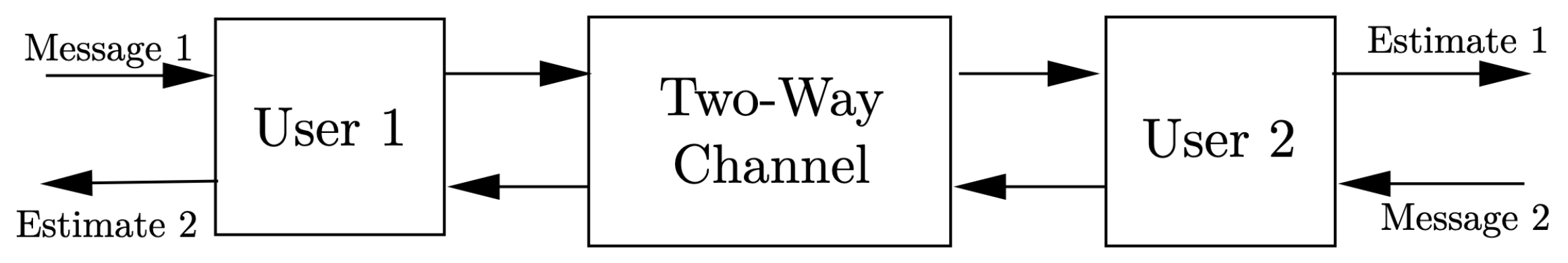}
\caption{Two-way communication. Messages are transmitted in both directions. Feedback can be used
in such communication systems.} \label{f_tow_way_basic} }\end{figure}

\underline{Bidirectional communication:} A channel with feedback is a special case of bidirectional communication in which two parties exchange information over a single channel. A fundamental bidirectional model is the two-way channel introduced by Shannon~\cite{shannon1961two}. The two-way channel (see Figure \ref{f_tow_way_basic}) exhibits the difficulty in searching for a general theory of information flow in networks. There is an inherent tension between the two information flows $(X_1 \to Y_2)$ and $(X_2 \to Y_1)$ over the shared channel $P(y_1, y_2 | x_1, x_2)$. Moreover, within each information flow, the forward stream for one message and the feedback stream for the other message compete for resources. This nested trade-off between competition and cooperation embodies the main complexity of the general problem. It turns out that DI is a fundamental quantity in characterizing the two-way channel capacity region 
 \cite{kramer1998directed}. 

Furthermore, DI characterizes information flow in multi-user networks where bidirectional transmission is allowed in addition to the two-way channel. Examples include:  multiple access channels with memory and feedback~\cite{Kramer03, PermuterWeissmanChenMAC_IT09}, broadcast channels with memory and feedback \cite{DaborahGoldsmith10BC_feedback}, compound channels with feedback \cite{ShraderPermuter09CompoundIT}, general memoryless networks \cite{kramer1998directed,Kramer03}, and networks with in-block memory \cite{Kramer12_InBlockMemory,Kramer-E23}.

\underline{Portfolio theory:} A special case of portfolio theory is horse race gambling, where each day only one stock (horse) increases (wins) and the rest of the stocks (horses) have zero value (lose). 
In \cite{permuter2011interpretations}, the authors show that the normalized DI $\frac{1}{n}I(Y^n\to X^n)$ has a
natural interpretation in gambling as the difference in growth rates due to {\it causal} side information. As a special case, if the outcome of the horse race and the side information are i.i.d., then the (normalized) DI becomes a single-letter mutual information $I(X;Y)$, and coincides with Kelly's result \cite{kelly1956new}. In the stock market, where stocks are sequences or stochastic processes, the DI rate is an upper bound on the difference between growth rates of the optimal investor wealth with and without the side information $Y$.


\underline{Causal estimation:} In \cite{weissman2012directed}, the authors introduce the notion of DI between two continuous-time processes and show that, for the continuous-time additive white Gaussian noise channel where the input $X_t$ may depend on the output with some arbitrary delay feedback, the following relation between causal estimation and DI holds:
\[
      \frac{1}{2}
    \int_{0}^T \EE \left[(X_t - \EE[X_t | Y_0^t])^2 \right] dt = I \left(X_0^T \rightarrow Y_0^T
    \right).
\]
The definition of DI for continuous-time processes, denoted $I( X_0^T \to Y_0^T) $, is obtained by dividing the time interval $(0, T)$ into a finite number of small intervals, then applying discrete-time DI using the random variables representing the processes within these intervals, and finally taking the infimum over all divisions of the interval $(0, T)$ \cite[Def. 1]{weissman2012directed}.


\underline{Compression with causal side information:} In
\cite{simeonePermuter13_source_delay}, the authors show that, in a lossless source coding problem where side information is available at the encoder, the difference between the minimum rate needed when no side information is available at the decoder to the case when causal side information is available at the decoder is the DI from the side information to the source. Furthermore, DI characterizes the rate distortion function with feedforward \cite{venkataramanan2007source}, analogous to channel coding with feedback. 

\begin{figure}[t!]
\centering
\begin{tikzpicture}[>=Latex, node distance=3cm, thick]

  \node[align=center] (forwardlabel) {\small Forward Channel - $P(y^n \| x^n)$};
  \node[draw, rectangle, minimum width=3.5cm, minimum height=1.2cm, below=0.15cm of forwardlabel] (forward) {\small $P(y_i|x^i,y^{i-1})$};

  \node[align=center, below=2.2cm of forwardlabel] (backwardlabel) {\small Backward Channel - $P(x^n \| y^{n-1})$};
  \node[draw, rectangle, minimum width=3.5cm, minimum height=1.2cm, below=0.15cm of backwardlabel] (backward) {\small $P(x_i | x^{i-1}, y^{i-1})$};

  \draw[->] 
    (forward.east) 
    -- ++(1.2,0) 
    -- ++(0,-2.85) 
    -- (backward.east)
    node[midway, right] {\Large };
    \draw[->] 
    (backward.west) 
    -- ++(-1.2,0) 
    -- ++(0,2.85) 
    -- (forward.west)
    node[midway, left] {\Large };
\end{tikzpicture}
\caption{Simple depiction of information flow.} \label{fig:inf_flow}
\end{figure}
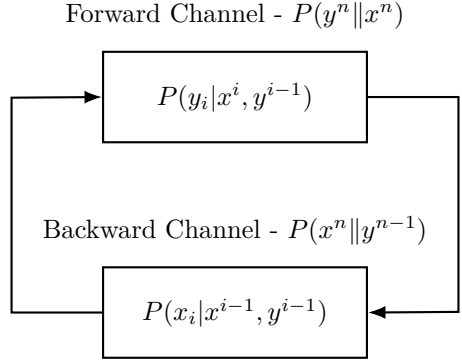

\underline{Statistical physics:} In \cite{VinklerPMerhav_stat_Phys}, the authors show an analogy between the maximal amount of work that can be extracted from a single heat bath using measurement, i.e., using ``information engines'', and Kelly gambling. This analogy led to new results in each field; in particular, the energy extracted due to causal measurements is the DI from the measurement to the placement of the particle.

\underline {Causal inference:} DI is an effective way to detect causality between two sequences. Suppose we have two sequences $\XX$ and $\YY$, and we ask which sequence causes which. That is, does $\XX$ cause $\YY$ or vice versa? DI gives an alternative to Granger causality \cite{granger1969investigating, Kleinberg--Hripcsak2011}, a widely used measure to identify causal inference between two processes.
As an information-theoretic measure, DI is a function of the joint distribution of the processes $(\XX, \YY)$ rather than the specific values they take.
However, Granger causality is a second moment function of $\XX$ and $\YY$ and therefore is influenced by the values that these random sequences take (see Section \ref{sec:gc} for more information).
Furthermore, Granger causality assumes a functional model: a univariate auto-regressive time series of $(\XX, \YY)$. Therefore, the concept is limited to an autoregressive stochastic process. Granger causality coincides with the DI (up to a multiplicative constant)\cite{amblard2012relation} under the assumption of a univariate auto-regressive time series process (see Section \ref{sec:gc} for more details on Granger causality).

Another popular notion for analyzing causal effects is Pearl's causality, which describes the cause-effect relationships between variables through the concepts of interventions and the development of the do-calculus. Section \ref{sec:causality} reviews the literature on the connection between DI and Pearl's causal framework.

\underline{Biology:} DI is a valuable tool when inference about causality is needed in Biology. In \cite{Mathai--Martins--Shapiro2007} and \cite{Rao--Hero--States--Engel2008}, the authors use DI to test influence in gene networks.

\underline{Neuroscience:} DI is useful in neuroscience \cite{wibral2014directed} for the following purposes.
In \cite{liu2009directed,wibral2014directed}, DI was used to investigate how information is stored and transferred around various brain areas. This allows, e.g., to reveal the difference between healthy and diseased areas in the brain, or where information is stored during specific cognitive tasks.
The advantage of DI for neuroscience is that it operates under the assumption of any computational model.
Another utility of DI is to infer causal relationships between neurons \cite{quinn2011estimating,theocharous2024causality}. As DI is a proxy for causal influence, its magnitude can be used to compare the effect of neurons on each other.
Here, traditional cross-correlation and mutual information fail to properly quantify dependence under temporal evolution because of their `memoryless' definition.

\underline{Directed information and control:}
DI has been studied in control theory, where causality and feedback-based decisions are a basis for designing closed-loop control systems. One prominent example is networked control systems (NCS), where the sensing device is not co-located with the control unit. This is often the case in practical control systems, and the main research question is to understand the fundamental trade-off between control performance and communication cost from the sensor to the controller. The DI was shown to characterize fundamental limits of such NCS scenarios \cite{tanaka2017lqg,KostinaHassibiTradeoff,sabag2022reducing,SilvaFramework11,SabagCDC}. The DI also has an interesting relation with Bode's integral sensitivity formula that can be used to design coding schemes \cite{BodeMeetsShannon}. 

In the context of large-language models, the recent work \cite{bai2025forget} proposes to use DI and its variants for several purposes: to quantify the causal influence of context on token generation, guide decoding and alignment strategies, serve as a loss function for semantic embedding in next-token prediction, analyze information flow during prompting and reasoning, and evaluate the model’s utilization of long-range memory.

\section{Challenges}
While DI is significant in various fields, it is not an easy tool to use.
As a \emph{multi-letter expression}, quantifying DI requires taking the expectation of a log-likelihood ratio of a sequence with potentially long memory, which results in complex optimizations. This monograph reviews tools to handle DI, i.e., to quantify, estimate, and optimize it. We categorize these tools into two classes.  


The first class consists of classic methods, which we refer to as `model-based', as they rely on specific model assumptions or require knowledge of a joint distribution.  
For example, we discuss generalizing the Blahut-Arimoto algorithm to DI (Section \ref{sec:BA_DI}).  
We also present reductions of the DI functional to single-letter expressions under certain assumptions on the joint distribution (Section \ref{sec:unifilar_fsc}).  
Such single-letter expressions can be optimized using dynamic programming techniques (Section \ref{sec:MDP}).

The second class encompasses data-driven approaches.  
Beyond various estimation techniques (Section \ref{chap:estimation}), we explore optimization methods that rely on modern paradigms rooted in machine learning.  
Such tools make DI accessible to contemporary settings, where one may lack knowledge of the joint distribution or must work with samples from it.  
Specifically, we examine DI optimization using neural networks, which can be performed in two ways.  
The first involves jointly optimizing two neural networks: one optimized to estimate a DI surrogate, and the other to maximize this estimate.  
The second leverages reinforcement learning.
Specifically, it extends the classic dynamic programming paradigm to large datasets and data-driven contexts by introducing neural network optimization.

\chapter{Directed Information and Causal Conditioning}\label{chap:di}
\section{Notation}
Random variables are denoted with uppercase letters, e.g., $X$, while realizations are denoted with lowercase letters, e.g., $x$. Calligraphic letters, e.g., $\cX$, denote sets, and $\cX^n$ denotes the $n$-fold Cartesian product of $\cX$.
When $\cX$ is finite, we use $|\cX|$ for its cardinality.
For $i,j\in\ZZ$ with $i\leq j$, we use the shorthand $X_i^j \triangleq (X_i,X_{i+1},\ldots,X_j)$ and  $x_i^j\triangleq(x_i,x_{i+1},\ldots,x_j)$; when $i=1$, we drop the subscript and simply write $X^j$ and $x^j$. We denote by $(\Omega,\cF,\PP)$ the underlying probability space on which all random variables are defined, with $\EE$ denoting expectation.


The probability distribution of a random variable $X$ is denoted by $P_X$, the joint distribution of $X$ and $Y$ by $P_{XY}$, and the conditional distribution of $Y$ given $X$ by $P_{Y|X}$. In the case of discrete random variables, $P_X$ and $P_{XY}$ are also used to denote the probability mass functions of $X$ and $(X,Y)$, respectively, while $P_{Y|X}$ is the probability kernel given by $\frac{P_{XY}}{P_X}$. Thus, $P_X(x) = \mathbb{P}(X=x)$, $P_{XY}(x,y) = \mathbb{P}(X=x, Y=y)$, and $P_{Y|X}(y|x) = \mathbb{P}(Y=y \mid X=x) = \frac{P_{XY}(x,y)}{P_X(x)}$. Often, to simplify notation, and in keeping with common practice in information theory (see, e.g., \cite{CovThom06}), we will use $P(x)$, $P(y)$ and $P(x,y)$ to denote the probability mass functions of $X$, $Y$ and $(X,Y)$, respectively, and write $P(y|x)$ instead of $P_{Y|X}$. This practice will also extend to causal conditioned distributions --- see Section~\ref{sec:CCdist}. In a further overload of notation, we will also use $P(x)$, $P(y)$ etc.\ to denote the probabilities $\mathbb{P}(X=x)$, $\mathbb{P}(Y=y)$ etc.; for example, $H(X) = -\sum_{x \in \cX} P(x) \log P(x)$. Finally, capital letters are used for arguments of $P$, such as in $P(X)$ or $P(X|Y)$, only when we take an expectation with respect to the distribution of the random variables involved in the argument. With this as our convention, we write expressions such as $H(X) = -\mathbb{E}[\log P(X)]$ and $H(X|Y) = -\mathbb{E}[\log P(X|Y)] = - \sum_{x,y} P(x,y) \log P(x|y)$. 

Stochastic processes are denoted by blackboard bold letters, e.g., $\XX\triangleq(X_i)_{i\in\mathbb Z}$.
For probability distributions $P,Q$ such that $P \ll Q$, i.e., $P$ is absolutely continuous with respect to $Q$, the Radon-Nikodym derivative of $P$ with respect to $Q$ is denoted by $\frac{dP}{dQ}$. The KL divergence of $P$ and $Q$, with $P\ll Q$, is
$\DKL(P\|Q)\triangleq\EE_P\big[\log\frac{\mathrm{d}P}{\mathrm{d}Q}\big]$. 
The mutual information (MI) of $(X,Y)\sim P_{XY}$ is $I(X;Y) \triangleq \DKL(P_{XY}\|P_X P_Y)$, where $P_X$ and $P_Y$ are the marginals of $P_{XY}$.
The entropy of a discrete random variable $X\sim P$ is $H(X) \triangleq -\EE\left[\log P(X)\right]$, and the conditional entropy of $X$ given $Y$ is $H(X|Y)\triangleq-\EE[\log P(X|Y)]$. 
The conditional MI of $(X,Y)$ given $Z$ is given by $I(X;Y|Z)\triangleq\DKL(P_{XYZ}\|P_{X|Z} P_{Y|Z} P_Z)$.

\section{Definition and Causal Conditioning}
Let $(X^n,Y^n)$ be a pair of sequences of random variables distributed according to $P(x^n,y^n)$.
The DI from $X^n$ to $Y^n$ is defined as 
\begin{equation}\label{eq:di_sum_cmi}
    I(X^n\to Y^n)\triangleq \sum_{i=1}^n I(X^i;Y_i|Y^{i-1}),
\end{equation}
where $I(X;Y|Z)$ is the conditional MI.
DI can be interpreted as the reduction of uncertainty in the elements of $Y^n$ after sequentially observing the past elements of $X^n$ and $Y^n$.
The purpose of this section is to define and characterize DI and to review its properties.

\subsection{Causally Conditioned Distribution} \label{sec:CCdist}
Consider a time-evolving stochastic system, and treat $X^n$ and $Y^n$ as its inputs and outputs, respectively.
We introduce the notion of causal conditioning  \cite{kramer1998directed}, which is a natural form in our setting.
Define the distribution of $Y^n$ \emph{causally conditioned} (CC) on $X^n$ as 
\begin{equation}\label{eq:cc_dist}
    P(y^n\|x^n) \triangleq  \prod_{i=1}^n P(y_i|y^{i-1},x^i),
\end{equation}
where $y_0$ is considered a null symbol, i.e.,  $y_0=\emptyset$. We will often use $P_{Y^n \| X^n}$ as a convenient shorthand for the CC distribution given by \eqref{eq:cc_dist}, i.e., the mapping $(x^n,y^n) \mapsto P(y^n \| x^n)$ for all $(x^n,y^n) \in \cX^n \times \cY^n$.

The CC distribution is a properly defined probability distribution \cite{kramer1998directed}, i.e., we have
$$
P(y^n\|x^n)\geq0\quad \forall (x^n,y^n)\in \cX^n\times \cY^n,
$$
and for any $x^n \in \cX^n$, we have $\sum_{y^n \in \cY^n}P(y^n\|x^n)=1$.
Furthermore, the conditional distribution building blocks of the CC distribution uniquely determine its value \cite[Lemma~3]{permuter2009finite}.
This property was analyzed and used in \cite{permuter2009finite} to study the feedback capacity of a class of channels with memory.
The CC distribution is related to Marko's notion of free information $H_x$ \cite{marko1973bidirectional}, as noted by \cite{kramer1998directed}.

Another useful definition of CC considers conditioning up to the $(n-1)$-th time step, and is denoted $P(y^n\|x^{n-1}) \triangleq  \prod_{i=1}^n P(y_i|y^{i-1},x^{i-1})$, where we assume $(x_0,y_0)=\emptyset$. Again, $P_{Y^n \| X^{n-1}}$ is a convenient shorthand for this CC distribution. Consequently, we have the following decomposition of the joint distribution in terms of CC distributions
$$
P(x^n,y^n) = P(x^n\|y^{n-1})P(y^n\|x^n),
$$
or equivalently, $P(x^n,y^n)=P(x^n\|y^{n-1})P(y^n\|x^n)$. This is the causal counterpart of the traditional chain rule $P(x^n,y^n)=P(x^n)P(y^n|x^n)$.

\subsection{Causally Conditioned Entropy and Mutual Information}
Following \eqref{eq:cc_dist}, we define the CC entropy
\begin{align*}
    H(Y^n\| X^n) &= H(P_{Y^n\|X^n})\\ 
    &\triangleq   - \sum_{(x^n,y^n)\in \cX^n\times\cY^n}P(x^n,y^n)\log P(y^n\|x^n). 
\end{align*}
The CC entropy decomposes into conditional entropy terms as
\begin{equation}
    H(Y^n \| X^n) = \sum_{i=1}^n H(Y_i|X^i,Y^{i-1}).
\end{equation}
Note that, because conditioning reduces entropy, we have
\begin{equation}\label{eq:cc_vs_cond}
    H(Y^n \| X^n) \geq \sum_{i=1}^n H(Y_i|X^n,Y^{i-1}) = H(Y^n| X^n).
\end{equation}
Using the CC entropy, we can write DI as
\begin{equation}\label{eq:di_ents}
    I(X^n\to Y^n) = H(Y^n) - H(Y^n\| X^n).
\end{equation}
This expression opens a hatch through which we can better distinguish it from the classic MI, which we write similarly to \eqref{eq:di_ents} as
\begin{equation}
    I(X^n;Y^n) = H(Y^n) - H(Y^n | X^n).
\end{equation}
The above entropic expressions of MI and DI show that we have the relation
\begin{equation}\label{eq:di_leq_mi}
    I(X^n\to Y^n)\leq I(X^n;Y^n).
\end{equation}
In the next subsection, we elaborate on the inequality \eqref{eq:di_leq_mi}, and discuss an underlying law of information conservation.
Another property of DI, which is inherited from the classic MI, is the ability to bound its variation due to conditioning on an exogenous variable \cite[Lemma~4]{permuter2009finite}.
Formally, for any discrete variable $S$ defined over the alphabet $\cS$, we have
$$
\left|I(X^n\to Y^n)-I(X^n\to Y^n \mid S)\right|\leq H(S)\leq \log(|\cS|).
$$
This property was utilized to form achievability results on a class of channels with memory and time-invariant feedback.

It is common to identify the system of interest $(X^n,Y^n)$ with a sequence of transition kernels $(P(y_i|y^{i-1},x^i))_{i=1}^n$. 
This structural assumption implies that we treat \emph{causal} systems, i.e., systems in which the elements of $(X^n,Y^n)$ evolve sequentially, and can be affected only by past and present occurrences. For example, consider a system described by the Markov chain
\begin{equation}\label{eq:di_markov}
    Y_i - (Y^{i-1},X^i) - (Y_{i+1}^n,X_{i+1}^n),
\end{equation}
such as time series, a communication channel, or a biological system.

It is important to note that when the Markov relation \eqref{eq:di_markov} holds, the CC distribution $P(y^n\| x^n)$ is equivalent to the conditional distribution $P(y^n|x^n)$.
Thus, the sequential Markov setting in~\eqref{eq:di_markov}, which, as previously mentioned, often characterizes sequential systems, is the one under which DI emerges naturally.

\subsection{Causal Conditioned Directed Information and  Multivariate Variants}
The CC entropy, DI, and MI naturally generalize to more than two variables;
Consider three jointly distributed sequences $(X^n,Y^n,Z^n)$. 
We define the CC DI as 
\begin{align}\label{eq:cc_di}
    I(X^n\to Y^n \|Z^n) &= \sum_{i=1}^n I(X^i;Y_i|Y^{i-1},Z^{i}) \\
    &= H(Y^n\|Z^n) - H(Y^n\|X^n, Z^n).\nonumber
\end{align}
Similarly, we define conditional DI as
\begin{align}\label{eq:cond_di}
    I(X^n\to Y^n |Z^n) &= \sum_{i=1}^n I(X^i;Y_i|Y^{i-1},Z^n) \\
    &= H(Y^n|Z^n) - H(Y^n\|X^n| Z^n),\nonumber
\end{align}
where the conditioning is read from left to right, i.e., $Y^n$ is CC on $X^n$, and the pair $(X^n,Y^n)$ is conditioned on $Z^n$, and
$$
H(Y^n\|X^n| Z^n)\triangleq \sum_{i=1}^n H(Y_i|Y^{i-1},X^{i},Z^n).
$$
Having introduced a third sequence, the resulting joint DI measure adheres to the following chain rule decompositions \cite[Property~3.3]{kramer1998directed}
\begin{align*}
    I((X^n,Y^n)\to Z^n) &= I(X^n\to Z^n) + I(Y^n\to Z^n\|X^n)\\
    I(X^n\to (Y^n,Z^n)) &= I(X^n\to Y^n \| DZ^n) + I(X^n\to Z^n \|Y^n),
\end{align*}
where $DZ^n=(\emptyset, Z^{n-1})$ is a left concatenation of $Z^{n-1}$ with a null symbol.

\section{Properties and Decompositions}
This section further explores properties of DI and its relation with MI through decomposition formulas.
We note that DI is composed as a sum of conditional MI terms and inherits most of its functional properties.

\textbf{Non-negativity:} $I(X^n\to Y^n)\geq 0$ with equality if and only if (iff) all the conditional MI terms $(I(X^i;Y_i|Y^{i-1}))_{i=1}^n$ nullify, which happens iff the set of Markov relations $(X^i-Y^{i-1}-Y_i)$ hold for $i=1,2,\ldots,n$.

\textbf{KL-divergence representation:}
DI has the KL-divergence representation
\begin{align}
    I(X^n \to Y^n) &= \DKL(P_{X^n,Y^n} \, \big\| \, P_{X^n\|Y^{n-1}} P_{Y^n})\label{eq:di_dkl}
\end{align}
which can be thought of as a causal proxy of the MI representation 
$$
I(X^n ; Y^n) = \DKL(P_{X^n,Y^n} \, \| \, P_{X^n}P_{Y^n}).
$$
Writing $P_{X^n,Y^n}$ as $P_{X^n\|Y^{n-1}} P_{Y^n\|X^n}$, we see that \eqref{eq:di_dkl} measures how different the conditional distribution $P_{Y^n|X^n}$ is from the causal conditioned distribution $P_{Y^n\|X^n}$.


\textbf{Bounds:}
As shown in the previous section, we have the following bounds for DI
$$
0\leq I(X^n\to Y^n)\leq I(X^n;Y^n).
$$

\textbf{Convexity:} The authors of \cite{permuter2014capacity} show that DI is convex in $P_{X^n\|Y^{n-1}}$ when $\cX$ and $\cY$ are finite.
Since the set of CC distributions $P_{X^n\|Y^{n-1}}$ is a polyhedron, the problem
\begin{equation}\label{eq:max_di}
    \max_{P(X^n\|Y^{n-1})} I(X^n\to Y^n)
\end{equation}
is a convex optimization problem.
This is a helpful observation, as \eqref{eq:max_di} describes the feedback capacity of a class of communication channels, as will be further elaborated in Section~\ref{sec:FSC_defs}.
An analysis of the properties of DI as a functional defined in the space of distributions was performed in \cite{charalambous2016directed}.
\begin{remark}[DI on abstract spaces]
The authors of \cite{charalambous2016directed} generalize the convexity characterization for discrete distributions from \cite{permuter2014capacity} to Polish spaces, which are a class of metric spaces that generalize upon the notion of finite or continuous alphabets.
The authors also analyzed the compactness of the set of distributions and the continuity of the DI functional.
Beyond its theoretical significance, the mathematical analysis of the DI functional is used to construct the existence of a capacity-achieving input distribution under the DI optimization criterion.
We refer the reader to \cite{charalambous2016directed} for details.
\end{remark}

The relation between MI and DI is usually best understood through the lens of their decompositions, which connect them in several ways.
The characterization of MI in terms of DI was done by Massey, who suggested the following law of information conservation \cite{massey2005conservation}\
\begin{equation}\label{eq:massey_conservation}
    I(X^n;Y^n) = I(X^n\to Y^n) + I(DY^n\to X^n),
\end{equation}
where, recall that $DX^n=(\emptyset, X^{n-1})$ and therefore $I(DY^n\to X^n)=\sum_{i=1}^n I(Y^{i-1};X_i|X^{i-1})$.
The information conservation law shows us that, in a system $(X^n,Y^n)$, the total amount of information exchange in both directions is preserved as the MI of the two system components. 
Equation \eqref{eq:massey_conservation} also hints at a fundamental difference between DI and MI.
To gain further intuition on that difference, we provide an example in which the two differ.


    \begin{example}\label{di_example}
    The following example illustrates the relation and difference between MI and DI. Let $X^n$ be a binary, i.i.d. sequence that is distributed according to $\mathsf{Ber}(1/2)$ and define $Y_i=X_{i+1}$, for all $i=1,\ldots,n$ with $Y_{n}=\emptyset$. Since the current $Y_i$ reveals information about future $X_j$'s (the opposite is not true), the causal information flow in this example is unidirectional from $Y^n$ to $X^n$. This is captured by DI since $I(X^n\to Y^n) = 0$ while $I\big(DY^{n}\to X^n\big)=n-1$. MI is invariant to directionality and we have $I(X^n;Y^n)=n-1$.
    \end{example}

We can deepen our understanding of information flow and conservation by slightly modifying \eqref{eq:massey_conservation}, as proposed by \cite{amblard2011directed}:
\begin{equation}\label{eq:infor_conservation_2}
    I(X^n\to Y^n) + I(Y^n\to X^n) = I(X^n;Y^n) + I(X^n\to Y^n \| DX^{n}),
\end{equation}
where 
\begin{equation} \label{eq:inst_inf}   
    I(X^n\to Y^n \| DX^{n})=\sum_{i=1}^n I(X_i;Y_i|X^{i-1},Y^{i-1})
\end{equation}
is the \emph{instantaneous information}, which measures the information of $X_i$ and $Y_i$ given their joint history.
By introducing the notion of instantaneous information \eqref{eq:inst_inf} to the MI conservation law \eqref{eq:infor_conservation_2}, one can derive the following decomposition rule for DI:
\begin{equation}\label{eq:di_decomp}
    I(X^n\to Y^n) = I(DX^{n}\to Y^n) +  I(X^n\to Y^n \| DX^{n}),
\end{equation}
which naturally generalizes to CC-DI as
\begin{equation}
    I(X^n\to Y^n\|Z^n) = I(DX^{n}\to Y^n\|Z^n) +  I(X^n\to Y^n \| DX^{n}, Z^n),
\end{equation}
where $I(DX^{n}\to Y^n\|Z^n) = \sum_{i=1}^nI(X^{i-1};Y_i|Y^{i-1},Z^{i-1})$, in a similar sense as $I(DX^{n}\to Y^n\|DZ^n)$.

\textbf{Visualization of DI:} 
The above information conservation law and additional information decompositions can be obtained using a recently developed information matrix (InfoMat) \cite{tsur2024infomat}.
For an $m$-step jointly distributed sequence $(X^m,Y^m)$, 
the InfoMat arranges the $m^2$ conditional MI terms that describe the evolution of dependence along the temporal axis.
Formally, the InfoMat is an $m\times m$ matrix whose $(i,j)$th entry is $\rI^{XY}_{i,j}=I(X_i;Y_j|X^{i-1},Y^{j-1})$.
Using $\mathrm{I}^{XY}$, we can visualize the various decompositions of MI by coloring index subsets and taking the sum with respect to each index group in $\rI^{XY}$ as follows
\begin{equation}\label{eq:mi_mat}
\mathrm{I}^{XY} = 
\begin{pmatrix}
{\color{purple}\rI^{XY}_{1,1}} & {\color{purple}\rI^{XY}_{1,2}} & {\color{purple}\dots} & {\color{purple}\rI^{XY}_{1,n}}\\
{\color{teal}\rI^{XY}_{2,1}} & {\color{purple}\rI^{XY}_{2,2}} & {\color{purple}\ddots}& {\color{purple}\vdots}\\
{\color{teal}\vdots} &{\color{teal}\ddots} &{\color{purple}\ddots} & {\color{purple}\rI^{XY}_{n-1,n}}\\
{\color{teal}\rI^{XY}_{n,1}} &{\color{teal}\dots} & {\color{teal}\rI^{XY}_{n,n-1}} & {\color{purple}\rI^{XY}_{n,n}}
\end{pmatrix} 
=
\begin{pmatrix}
{\color{blue}\rI^{XY}_{1,1}} & {\color{red}\rI^{XY}_{1,2}} & {\color{red}\dots} & {\color{red}\rI^{XY}_{1,n}}\\
{\color{teal}\rI^{XY}_{2,1}} & {\color{blue}\rI^{XY}_{2,2}} & {\color{red}\ddots}& {\color{red}\vdots}\\
{\color{teal}\vdots} &{\color{teal}\ddots} &{\color{blue}\ddots} & {\color{red}\rI^{XY}_{n-1,n}}\\
{\color{teal}\rI^{XY}_{n,1}} &{\color{teal}\dots} & {\color{teal}\rI^{XY}_{n,n-1}} & {\color{blue}\rI^{XY}_{n,n}}
\end{pmatrix},
\end{equation}
The first equality visualizes Massey's law of information conservation \eqref{eq:massey_conservation}: DI corresponds to a triangular submatrix in $\rI^{XY}$, i.e., 
\begin{equation}
    I(X^n;Y^n) = {\color{purple}I(X^{n}\to Y^n)} + {\color{teal}I(Y^{n-1}\to X^n)}.
\end{equation}
Beyond Massey's law of information conservation, the second equality in \eqref{eq:mi_mat} visualizes a decomposition of MI into two time-shifted DI terms (which represent the effect of the past of one sequence on the present of another) and an \emph{instantaneous information exchange} term, which quantify the immediate effect of each sequence on the other, i.e.,
\begin{equation}
    I(X^n;Y^n)={\color{red}I(X^{n-1}\to Y^n)} + {\color{teal}I(Y^{n-1}\to X^n)} + {\color{blue}I_{\text{inst}}(X^n,Y^n)}\label{eq:conservation_mi_mat}
\end{equation}
To gain further intuition on the InfoMat, we refer to Example~\ref{di_example}, which demonstrated a setting where information transfer occurs only in a single direction.
In this example, the resulting InfoMat is a matrix of zeros outside a single off-diagonal that starts in the $(1,2)$-th entry, which represents the direct information transfer from $X_i$ to $Y_{i+1}$ for $i=1,\dots,n-1$.

Beyond its merit as a visualization tool for information decomposition laws, the InfoMat visualizes sequential information exchange in time series data.
For example, the larger the values of $\rI^{XY}$ on the diagonal, the larger the instantaneous information exchange between the sequences, and therefore the more dominant present interaction is compared to past interaction.
Consequently, the larger the values on the off-diagonal, the larger the effect of the past on the present interaction between the sequence elements.
The visual patterns in the InfoMat reflect dependence structures in the joint sequence. These connections help users analyze time-series data more effectively. By highlighting how interactions evolve, the InfoMat provides a fine-grained view of the underlying dynamics; see \cite{tsur2024infomat}.


\begin{figure}[t]
    \centering
    \subfigure[Oblivious coding scheme.]
    {
    \includegraphics[trim={30pt 30pt 30pt 1pt}, clip,width=0.35\linewidth]{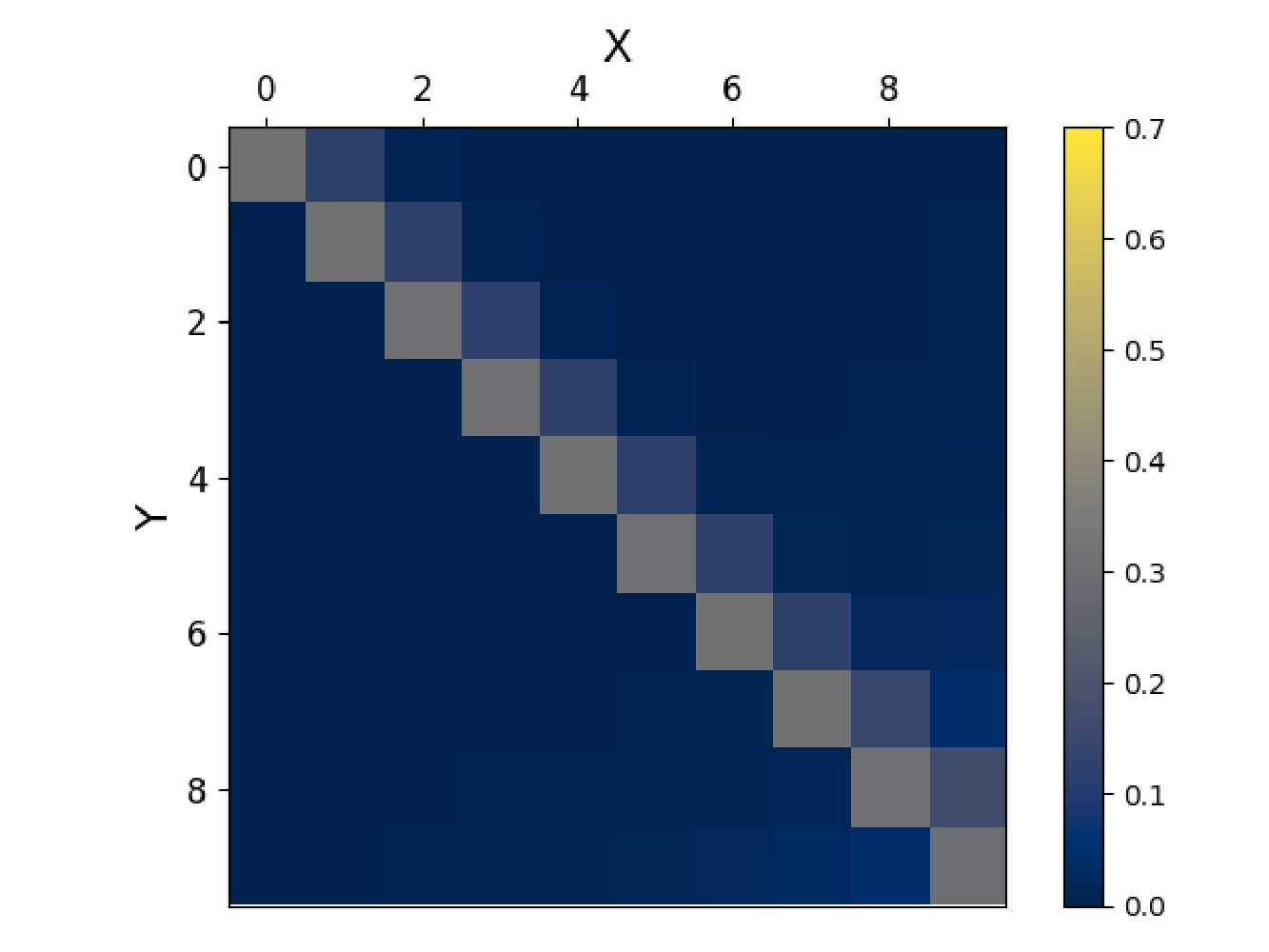}
    }
    \qquad\qquad
    \subfigure[Optimal coding scheme.]
    {
     \includegraphics[trim={30pt 30pt 30pt 1pt}, clip,width=0.35\linewidth]{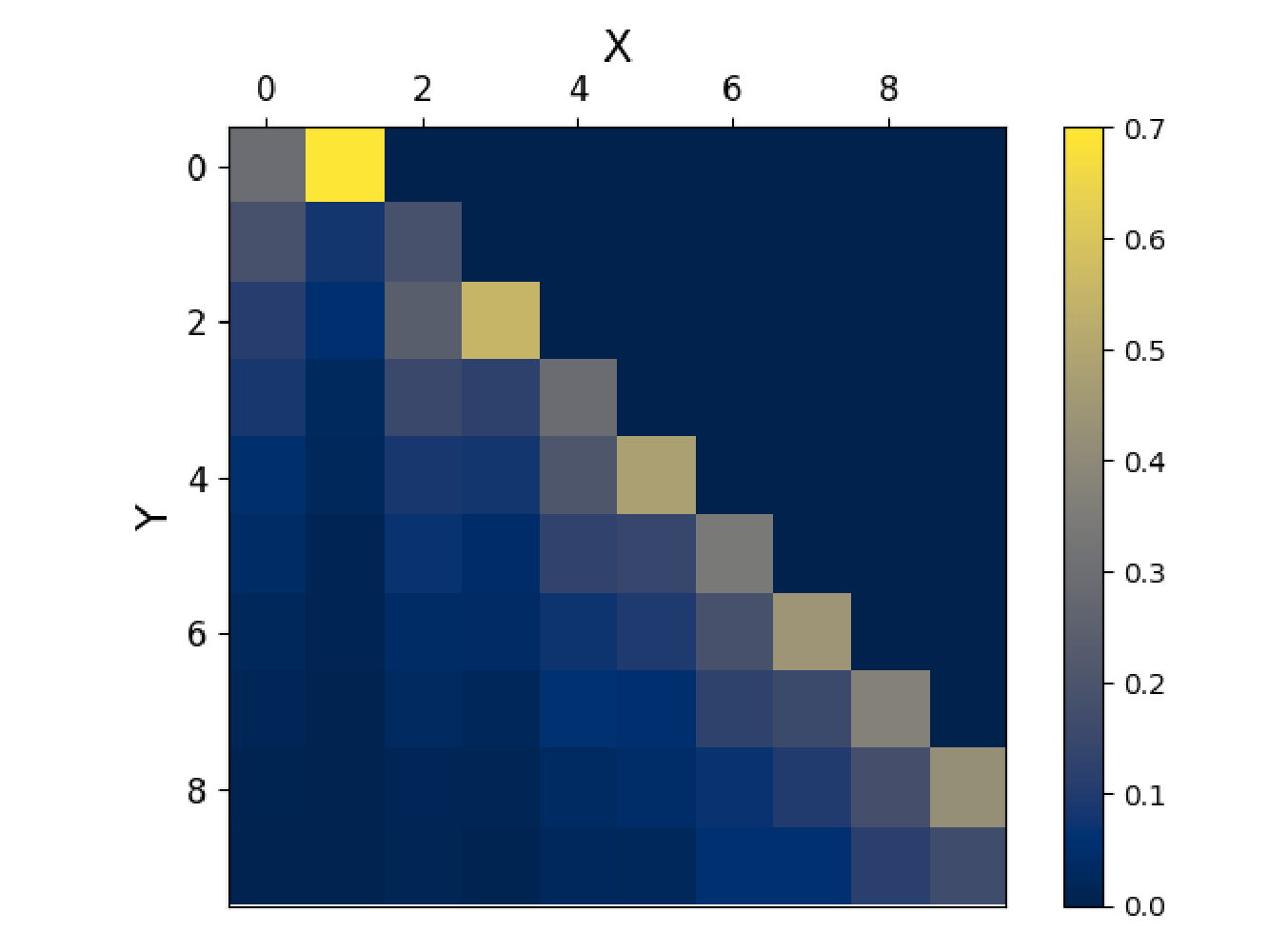}
    }
  \caption{InfoMat visualization on the Ising channel under (a) a channel-oblivious scheme and (b) the capacity-achieving scheme from \cite{ElishcoPermuter2014Ising}. DI corresponds to the sum of the upper triangular.}
  \label{fig:infomat_ising}
\end{figure}

\section{Directed Information Rate}
In many cases of interest, the sequence $(X^n,Y^n)$ is formed by the first $n$ terms of a jointly distributed stochastic process $(\XX,\YY)$.
In such cases, the analysis may require the notion of information rates \cite{CovThom06}.
Information rates measure an average of MI expressions, due to the observation of a growing horizon of the process.
Whenever the limit exists, the entropy rate of a stochastic process $\XX$ is
$$
H(\XX)\triangleq \lim_{n\to\infty}\frac{1}{n}H(X^n).
$$
For example, when $\XX$ is stationary, the limit is guaranteed to converge, and due to the properties of Cesàro sums, we have \cite{CovThom06}
$$
H(\XX) = \lim_{n\to\infty}H(X_n|X^{n-1}).
$$
Following the same reasoning, we define the CC entropy rate as
\begin{equation}\label{eq:cc_ent_rate}
    H(\YY\|\XX) = \lim_{n\to\infty}\frac{1}{n}H(Y^n\|X^n) .
\end{equation}
Stationarity implies
\begin{equation}\label{eq:cc_ent_rate_stat}
    H(\YY\|\XX) =  \lim_{n\to\infty}H(Y_n |X^n, Y^{n-1}).
\end{equation}
The analysis is often more comfortable when considering a fixed temporal index in the entropy term and an increasing conditioning horizon, i.e.
\begin{equation}\label{eq:fixed_time_ent_rate}
    H(\YY\|\XX) =\lim_{n\to\infty}H(Y_0 |X^0_{-n}, Y^{-1}_{-n}).
\end{equation}
This equivalence follows from the stationarity of the joint process.
The fixed time definition \eqref{eq:fixed_time_ent_rate} considers a fixed ``present'' symbol $X_0$, and the limit is translated into a growing horizon of history.

Having a notion of CC entropy rate, we can define the \emph{DI rate} as 
\begin{align}
    I(\XX\to\YY) &\triangleq  \lim_{n\to\infty} \frac{1}{n}I(X^n\to Y^n)\label{eq:di_rate}\\
    &= \lim_{n\to\infty}\frac{1}{n}H(Y^n) - \lim_{n\to\infty}\frac{1}{n}H(Y^n\|X^n)\label{eq:di_rate_as_ent_rates}\\
    &= H(\YY)-H(\YY\|\XX),
\end{align}
The DI rate (i.e., the convergence of the limit in its definition) exists whenever each of the limits in \eqref{eq:di_rate_as_ent_rates} exists, and under joint stationarity, we have
\begin{equation}\label{eq:di_rate_}
    I(\XX\to\YY) = \lim_{n\to\infty}I(X^n;Y_n|Y^{n-1}).
\end{equation}
As explained in Section \ref{chap:estimation}, most DI estimators focus on the DI rate, as they operate under the setting of jointly stochastic ergodic and stationary processes.


    \begin{remark}[DI in continuous time]
        The authors of \cite{weissman2012directed} define DI for continuous-time processes. Following the spirit of MI between continuous distributions, DI in continuous-time is defined as the infimum of discrete-time DI taken over all possible $n$-element discrete-time partitions of the considered time segments (this definition is in contrast to continuous-valued MI, which takes the \emph{supremum} over all quantizations). This line of work builds on the seminal results of Duncan \cite{duncan1970calculation}, who provided an explicit characterization of continuous-time mutual information for Gaussian channels in terms of filtering error, establishing a fundamental link between mutual information, stochastic calculus, and causal estimation. The continuous-time DI inherits favorable properties from its discrete counterpart, such as monotonicity, invariance to time dilations, and the conservation law \eqref{eq:massey_conservation}. The feedback capacity of a large class of channels (specifically, channels whose output is a function of the input and some stationary ergodic noise) can be characterized via continuous time DI. The definition above only applies if the time delay is positive. An alternative definition for continuous time DI has been proposed using approximation theory and sampling theorems, which applies to instantaneous feedback, i.e., zero time delay feedback \cite{liu2019continuous}.
\end{remark}

\section{Relation to Transfer Entropy and Granger Causality}
This section presents transfer entropy and Granger causality, which are popular measures of causal influence closely related to DI.

\subsection{Transfer Entropy}
Transfer entropy (TE) was proposed in \cite{schreiber2000measuring} to quantify the coherence between two time-evolving systems whose relations may be nonlinear.
It has seen popularity in physics \cite{battiston2021physics}, computational biology \cite{farahani2019application}, neuroscience \cite{wibral2014transfer}, and economics \cite{mikhaylov2020cryptocurrency}.
The TE was originally defined as
\begin{equation}\label{eq:TE}
    T^{X\to Y}_n(k,l) = \EE\left[\log\frac{P(Y_n|Y^{n-1}_{n-l}, X^{n-1}_{n-k})}{P(Y_n|Y^{n-1}_{n-l})}\right],
\end{equation}
which is equivalently expressed as a conditional MI:
\begin{equation}\label{eq:te_mi}
    T^{X\to Y}_n(k,l) = I(X^{n-1}_{n-k};Y_n|Y^{n-1}_{n-l}).
\end{equation}

TE depends on three parameters: $k$ and $l$, which determine the observation horizon of $X$ and $Y$, respectively, and $n$, which is the time at which TE is measured.
TE captures the entire dependence structure in $(X^n,Y^n)$ whenever $ Y_n $ depends solely on $ X^{n-1}_{n-k} $ and $ Y^{n-1}_{n-l}$, only under a joint $(k,l)$-Markov assumption on the model.
Also, the TE formula considers a single time step delay between $X^n$ and $Y^n$. It is therefore often interpreted as the effect of \emph{past} occurrences on present variables.
The resulting MI gap can be quantified through the instantaneous information element, i.e., 
$$
I(X^i;Y_i|Y^{i-1}) = T^{X\to Y}_i(i-1,i-1) + I(X_i;Y_i|X^{i-1},Y^{i-1}).
$$
This leads to the relation 
\begin{equation}\label{di_te}
    I(X^n\to Y^n) = \sum_{i=1}^n \left(T^{X\to Y}_i(i-1,i-1) + I(X_i;Y_i|X^{i-1},Y^{i-1})\right),
\end{equation}
which extends to information rates of stationary processes as
$$
I(\XX\to\YY)= T_{\infty} + I_{\infty}
$$
where 
$$
T_{\infty} = \lim_{n\to\infty}T^{X\to Y}_n(n-1,n-1),\quad I_{\infty} = \lim_{n\to\infty}I(X_n;Y_n|X^{n-1},Y^{n-1})
$$
are the TE rate and instantaneous information rate, respectively.
Finally, TE for $l=n$ can be visualized through the information matrix $\rI^{XY}$. Specifically, $T^{X\to Y}_n(n-1,n-1)$ is the sum of the $n-1$th column of the InfoMat.
TE can be interpreted as an approximate and limited memory version of DI. This simplifies estimation but loses precision in capturing long-range dependencies. Section \ref{chap:capacity} shows that one must consider random processes as variable-order Markov models for the channel outputs to achieve the feedback capacity. Describing the channel outputs with a finite history of the channel inputs and outputs gives the feedback capacity only in a few cases, such as memoryless and finite-state Markov channels.

\subsection{Granger Causality}\label{sec:gc}
Originating from Wiener's work on time-series prediction, Granger causality \cite{granger1969investigating} is a theoretical framework to assess directional dependencies between two time series based on conditional independence tests \cite{amblard2012relation}. The concept began in the field of econometrics, but has spread to become an interdisciplinary measure for causal effect testing in a wide range of areas, such as neuroscience and computational biology. As noted by \cite{amblard2012relation}, Granger causality can be defined either through the lens of prediction theory or probability. 

Consider a set of jointly distributed stationary processes $(\XX,\YY,\ZZ)$.
We say that $\XX$ does not \emph{Granger cause} $\YY$ relative to $(\XX,\YY,\ZZ)$ iff 
\begin{equation}\label{eq:granger_term}
    Y_{n}\indep X^{n-1} | (Y^{n-1},Z^{n-1}),\qquad \forall n\in\NN.
\end{equation}
Granger causality is thus defined through the notion of conditional independence.
We note that Granger causality is defined relative to the observation set, in the same sense that the measure of conditional independence further depends on the conditioned set.
Using the notion of DI, we can equivalently write the condition \eqref{eq:granger_term} as 
\begin{equation}\label{eq:granger_equivalent}
    I(DX^{n}\to Y^{n}\|DZ^n)=0,\qquad \forall n\geq 0.
\end{equation}

The authors of \cite{geweke1982measurement} fortify the connection between DI and Granger causality by addressing the functional definition of Granger causality. We draw this connection informally as follows.
We define the optimal MMSE risk of an estimator of $Y$ from $X$ as the risk of the optimal estimator, given by $E[Y|X]$. 
From a prediction-theoretic perspective, we say that $\XX$ Granger causes $\YY$ if the optimal risk of the estimator of $Y_n$  from $(Y^{n-1},Z^{n-1})$ improves when it also considers $X^{n-1}$, i.e., when $X^{n-1}$ does not possess significant additional information about $Y_n$, for any $n\in\NN$.
This definition is a special case of the conditional independence framework, given in \eqref{eq:granger_term}, as noted in \cite{amblard2012relation}.

The authors of \cite{geweke1982measurement,geweke1984measures,amblard2012relation} consider linear estimation and jointly Gaussian stationary processes with a quadratic loss and define causality indices that measure the causal effect of one process on another.
The causality indices measure the causal influence through the mean square error of estimating elements from one process to the other, with and without conditioning on one's past.
The indices are
\begin{align*}
    F_{X\leftrightarrow Y}&=\lim_{n\to\infty}\frac{R[Y_n|Y^{n-1},X^{n-1}]}{R[Y_n|Y^{n-1}]}\\
    F_{X\leftrightarrow Y\|Z}&=\lim_{n\to\infty}\frac{R[Y_n|Y^{n-1},X^{n-1},Z^n]}{R[Y_n|Y^{n-1},Z^n]}\\
    F_{X\rightarrow Y}&=\lim_{n\to\infty}\frac{R[Y_n|Y^{n-1}]}{R[Y_n|Y^{n-1},X^{n-1}]}\\
    F_{X\rightarrow Y\|Z}&=\lim_{n\to\infty}\frac{R[Y_n|Y^{n-1},Z^{n-1}]}{R[Y_n|Y^{n-1},X^{n-1},Z^{n-1}]},
\end{align*}
where $R[U|V]=\min\EE[(U-E[U|V])^2]$ is the aforementioned optimal risk. 
In the Gaussian case, distributions are fully characterized by the first two moments.
This leads to an equivalence between DI and Granger causality.
Specifically, the authors of \cite{amblard2012relation} show the following equivalence
\begin{align*}
    F_{X\rightarrow Y} &= I(\XX\to\YY)\\
    F_{X\rightarrow Y \| Z} &= I(\XX\to \YY\|\ZZ)\triangleq \lim_{n\to\infty}\frac{1}{n}I(X^n\to Y^n\|Z^n).
\end{align*}
For more information on the relation between Granger causality and DI, and a formal derivation of the prediction-theoretic perspective, we refer the reader to \cite{amblard2012relation}. The relation between DI, Granger causality, and TE makes DI relevant to numerous fields that relied on Granger causality and TE.

\section{Pearl's Causality and DI}\label{sec:causality}
Judea Pearl’s framework of causal analysis \cite{pearl2009causality,peters2017elements} provides formal semantics for reasoning about cause-and-effect relationships through structural causal models (SCMs). Pearl defines causality by asking how the distribution of variables would change under forced \emph{interventions} on the data-generating mechanisms, rather than by analyzing statistical dependence alone. This section follows the notation proposed in \cite{raginsky2011directed}.

An intervention that assigns a random variable $X$ the value $x$ is denoted by $\mathrm{do}(X=x)$ and is modeled by replacing the causal mechanism generating $X$ with a deterministic assignment, thereby decoupling $X$ from its original causes while leaving all other mechanisms unchanged. This formalism allows causal questions to be posed unambiguously and makes explicit the assumptions under which causal effects may or may not be identifiable from observational data.

A central motivation for intervention-based causality is that statistical dependence
need not correspond to a causal effect. Consider a \emph{confounding SCM} as a simple example, in which an exogenous variable $Z\sim P_Z$ and structural equations
\[
X \triangleq Z, \qquad Y \triangleq Z,
\]
so that $Z\to X$ and $Z\to Y$, but there is no causal arrow $X\to Y$. Observationally, $X$ and $Y$ are perfectly dependent. In contrast, intervening on $X$ breaks the link $Z\to X$ while leaving the mechanism generating $Y$ unchanged. Hence the distribution of $Y$ is unaffected by the intervention:
\[
P(Y\mid \mathrm{do}(X=x)) = P(Y) \qquad \text{for all } x\in\cX.
\]
This example highlights the distinction between dependence $P(Y\mid X=x)$
and causal effect $P(Y\mid \mathrm{do}(X=x))$.

\begin{figure}[!t]
    \centering
    \resizebox{0.65\linewidth}{!}{
    \begin{tikzpicture}
        \begin{scope}
            \node[draw,circle] (Z) at (0,2)  {$Z$};
            \node[draw,circle] (X) at (-2,0) {$X$};
            \node[draw,circle] (Y) at ( 2,0) {$Y$};

            \draw[->, line width=2pt] (Z) -- (X);
            \draw[->, line width=2pt] (Z) -- (Y);
            \draw[->, line width=2pt] (X) -- (Y);

            \node at (0,3) {\textbf{(a) Original DAG}};
        \end{scope}

        \begin{scope}[xshift=6cm]
            \node[draw,circle] (Z2) at (0,2)  {$Z$};
            \node[draw,circle] (X2) at (-2,0) {$X$};
            \node[draw,circle] (Y2) at ( 2,0) {$Y$};

            \draw[->, line width=2pt] (Z2) -- (Y2);
            \draw[->, line width=2pt] (X2) -- (Y2);

            \node[draw,rectangle] (do) at (-2,2) {$\mathrm{do}(X= x)$};
            \draw[->, line width=2pt] (do) -- (X2);

            \node at (-0.1,3) {\textbf{(b) Intervention }};
        \end{scope}
    \end{tikzpicture}
    }
    \caption{%
        (a) The original SCM: $Z$ confounds $X$ and $Y$, opening the back-door path $X \leftarrow Z \rightarrow Y$.  
        (b) After intervening with $\mathrm{do}(X{=}x)$, the incoming arrow to $X$ is removed and replaced by the intervention mechanism.}
    \label{fig:confounder_dag}
\end{figure}
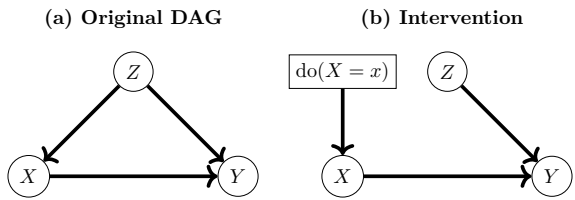

To formalize the intervention semantics, consider a collection of random variables $\{A_1,\dots,A_n\}$ associated with the nodes of a directed acyclic graph (DAG). For each variable $A_i$, let $\mathrm{pa}(A_i)\subseteq A^{i-1}$ denote its \emph{parent set}, i.e., the set of variables $A_j$ such that the DAG contains a directed edge
$A_j \to A_i$. The DAG induces the causal factorization of a joint distribution
\begin{equation}
P(A^n)=\prod_{i=1}^n P\bigl(A_i \mid \mathrm{pa}(A_i)\bigr).
\end{equation}
For some $j\in[n]$ (where $[n]\triangleq \{1,\dots,n\}$)
the atomic intervention $\mathrm{do}(A_j=a)$ is defined by replacing the conditional distribution $P(A_j\mid\mathrm{pa}(A_j))$ with its \textit{deterministic} assignment and replacing the conditioning on $A_j$ with $A_j=a$ wherever it appears as a parent. Formally, the joint distribution under such intervention is given by
\begin{equation}
P(A^n \mid \mathrm{do}(A_j=a))
\;\triangleq\;
\mathbf{1}\{A_j=a\}
\prod_{i\neq j} P\bigl(A_i \mid \mathrm{pa}^{(j,a)}(A_i)\bigr),
\end{equation}
where the conditioning set is 
\[
\mathrm{pa}^{(j,a)}(A_i)\triangleq
\begin{cases}
\bigl(\mathrm{pa}(A_i)\setminus\{A_j\},\,A_j=a\bigr), & A_j\in \mathrm{pa}(A_i),\\
\mathrm{pa}(A_i), & A_j\notin \mathrm{pa}(A_i).
\end{cases}
\]
The atomic intervention extends naturally to simultaneous interventions on a set of
variables: For $S\subseteq[n]$ the intervention $\mathrm{do}(A^S=a^S)$ replaces the
mechanisms generating the variables in $S$ by the deterministic assignments $A^S=a^S$. All other mechanisms are left as they are, except that whenever a variable from $S$ appears as a parent in the mechanism of some other node, it is fixed to its intervened value. Figure \ref{fig:confounder_dag} illustrates a confounding variable and the effect of an intervention in a DAG.

DI admits an interpretation within Pearl’s intervention-based semantics \cite{raginsky2011directed}. Consider the two-process setting $(X^n,Y^n)$ and embed it into a time-ordered SCM by interleaving the variables as
\begin{align}
A^{2n}=(X_1,Y_1,\ldots,X_n,Y_n),    
\end{align}
with a topological order in which each $A_i$ has parents among the past variables,
$\mathrm{pa}(A_i)\subseteq A^{i-1}$. In this time-ordered setting, note that intervening on the entire input trajectory yields the causal conditional law:
\begin{align}\label{eq:pearl_do_is_CC}
P(Y^n \mid \mathrm{do}(X^n=x^n)) \;=\; P(Y^n \Vert X^n = x^n).    
\end{align}
Thus, causal conditioning corresponds to Pearl’s notion of intervention applied to the input process. 

A natural ``intervention-vs-observation'' discrepancy measure is the KL divergence between the observational conditional law and the interventional law \cite{raginsky2011directed}. For a fixed $X^n=x^n$, we write it as
\begin{align}
\DKL\!\left(P_{Y^n\mid X^n=x^n}\,\big\|\,P_{Y^n\mid \mathrm{do}(X^n=x^n)}\right).
\end{align}
Averaging this divergence over $P_{X^n}$ gives the conditional KL divergence
\begin{align}
\EE_{X^n}\left[\DKL\!\left(P_{Y^n\mid X^n}\,\big\|\,P_{Y^n\mid \mathrm{do}(X^n)}\right)\right]
&= \DKL\!\left(P_{Y^n\mid X^n}\,\big\|\,P_{Y^n\Vert X^n}\,\big|\,P_{X^n}\right) \nonumber\\
&= \mathbb{E}\!\left[\log\frac{P_{Y^n\mid X^n}}{P_{Y^n\Vert X^n}}\right] \nonumber\\
&= \mathbb{E}\!\left[\log\frac{P_{X^n\Vert Y^{n-1}} }{P_{X^n}}\right] \nonumber\\
&= I(Y^{n-1}\to X^n),
\end{align}
where we used \eqref{eq:pearl_do_is_CC} and $P(X^n,Y^n)=P(X^n\Vert Y^{n-1})\,P(Y^n\Vert X^n)$ in the third equality. This shows that the average gap between conditioning on $X^n$ and intervening on $X^n$ is exactly the feedback/back-door information flow $I(Y^{n-1}\!\to\!X^n)$; by the DI decomposition $I(X^n;Y^n)=I(X^n\!\to\!Y^n)+I(Y^{n-1}\!\to\!X^n)$, it is the portion of mutual information attributable to  the influence of $Y^n$ on $X^n$.

\subsection{Setwise directed information and the back-door criterion}\label{subsec:setwise-di-backdoor}
The preceding $A^{2n}=(X_1,Y_1,\ldots,X_n,Y_n)$ calculation is a special case of a general phenomenon: the discrepancy between \emph{observing} and \emph{intervening} on a set of variables. Let $A^n=\{A_1,\ldots,A_n\}$ be the variables of an SCM/DAG as defined above, and for any index set $S\subseteq[n]$ write $A^S\triangleq (A_i:i\in S)$. For disjoint sets $S,T\subseteq[n]$, define the \emph{setwise DI} from $A^T$ to $A^S$ by the conditional KL divergence
\begin{align}\label{eq:setwise-di}
I(A^T\to A^S)
&\triangleq
\DKL\!\left(P_{A^T\mid A^S}\,\big\|\,P_{A^T\mid \mathrm{do}(A^S)}\,\big|\,P_{A^S}\right) \nonumber\\
&= \mathbb{E}\!\left[\log\frac{P_{A^T\mid A^S}}{P_{A^T\mid \mathrm{do}(A^S)}}\right].
\end{align}
In the special case when $(T,S)$ constitutes a partition of $[n]$ (i.e., $T=S^c$ and $S \cup T = [n]$), the setwise DI admits the chain rule
\begin{align}
I(A^{S^c}\!\to\!A^{S})=\sum_{i\in S} I\!\left(A^{S^c_{<i}};\,A_i \,\middle|\, A^{S_{<i}}\right),    
\end{align}
where $U_{<i}\triangleq \,U\cap\{1,\ldots,i-1\}$. More generally, for pairwise disjoint sets $S,T,Z\subseteq[n]$, the \emph{conditional} setwise DI is defined as
\begin{align}\label{eq:cond-setwise-di}
I(A^T\to A^S\mid A^Z)
&\triangleq
\DKL\!\left(P_{A^T\mid A^S,A^Z}\,\big\|\,P_{A^T\mid \mathrm{do}(A^S),A^Z}\,\big|\,P_{A^S,A^Z}\right) \nonumber\\
&= \mathbb{E}\!\left[\log\frac{P_{A^T\mid A^S,A^Z}}{P_{A^T\mid \mathrm{do}(A^S),A^Z}}\right],
\end{align}
where $P_{A^T\mid \mathrm{do}(A^S),A^Z}$ is obtained from the joint interventional law $P(A^T,A^Z\mid \mathrm{do}(A^S=a^S))$ by conditioning on $A^Z$.

Equations \eqref{eq:setwise-di}--\eqref{eq:cond-setwise-di} make the ``intervention--vs--observation'' gap explicit: by nonnegativity of KL divergence,
\begin{align}\label{eq:kl-gap-zero}
I(A^T\to A^S\mid A^Z)=0
\quad\Longleftrightarrow\quad
P_{A^T\mid A^S,A^Z}=P_{A^T\mid \mathrm{do}(A^S),A^Z},
\end{align}
for $P_{A^S,A^Z}$-a.e.\ $(a^S,a^Z)$.

In Pearl’s framework, the causal effect of an intervention is defined through the
interventional distribution $P(A^T\mid \mathrm{do}(A^S=a^S))$. In applications, however, we only observe the joint distribution of measured variables (e.g., $P(A^T,A^S,A^{S_d})$). The problem is therefore to determine when $P(A^T\mid \mathrm{do}(A^S=a^S))$ can be computed from observational data and how. 

To compute $P(A^T\mid \mathrm{do}(A^S))$, Pearl introduced the back-door criterion as a graphical sufficient condition. The underlying idea is to condition on variables outside of $S$ and $T$ to block all paths through which variables outside $S$, which are connected to $T$, may also have causal influence on $S$. 

Raginsky \cite{raginsky2011directed} provides an information-theoretic characterization of the back-door criterion. Specifically, the vanishing of the conditional setwise DI
$I(A^T\!\to\!A^S\mid A^{S_d})$ is equivalent to
\[
P(A^T\mid A^S,A^{S_d}) \;=\; P(A^T\mid \mathrm{do}(A^S),A^{S_d}),
\]
meaning that, given $A^{S_d}$, conditioning on $A^S$ matches intervening on $A^S$ for
predicting $A^T$. If, in addition, $S_d$ is unaffected by intervening on $S$ (in
particular, if $S_d$ contains no descendants of $S$), then averaging over $A^{S_d}$
using its observational distribution yields a Bayesian adjustment formula
\cite{raginsky2011directed} that can be computed from observed variables.
\begin{theorem}[Back-door criterion via directed information {\cite{raginsky2011directed}}]\label{thm:backdoor}
Let $S,T,S_d\subseteq\{1,\dots,n\}$ be pairwise disjoint sets of an SCM.
If $S_d$ consists only of non-descendants of $S$ and $I(A^T\to A^S\mid A^{S_d})=0$, then the causal effect of intervening on $S$ is computable by adjustment:
\begin{align}\label{eq:backdoor}
P(A^T \mid \mathrm{do}(A^S)) = \sum_{a^{S_d}} P(A^T \mid A^S, A^{S_d})\,P(A^{S_d}).
\end{align}
\end{theorem}

The work \cite{raginsky2011directed} aims at identifiability of the SCM, i.e., the process of classifying confounders for further adjustment of the SCM. This work is complemented by \cite{wieczorek2019information}, which propose an information-theoretic measure to quantify the strength of causal effect. Specifically, once the presence of a confounder is adjusted for (via  Theorem \ref{thm:backdoor}), the authors of \cite{wieczorek2019information} quantify the strength of causal relations along the DAG. This results in an information-theoretic two-step procedure for causality: We begin with DI and the back-door criterion to identify and adjust for confounders, and then we calculate conditional mutual information terms to quantify the strength on causal relations in the adjusted SCM.

\section{Optimizing Directed Information}\label{sec:BA_DI}
DI plays a central role in communication and control when models involve processes with memory.
It characterizes the fundamental limit of core problems in information theory, both in the context of channel capacity \cite{yang2005feedback,Tatikonda2009feedbakcapacity,kim2008coding} and rate distortion \cite{venkataramanan2007source}.
It is therefore used as the objective of optimization problems when a closed-form solution is not tractable. DI may simplify to a single-letter objective when the channel has a specific structure, e.g., a finite number of states, as we elaborate in Section \ref{chap:capacity}.

For discrete alphabets, a sequential generalization of the Blahut-Arimoto (BA) algorithm \cite{blahut1972computation,arimoto1972algorithm} for the optimization of DI was proposed \cite{naiss2012extension}. We consider a fixed channel $r_{Y^n\|X^n}$, and we alternately optimize a conditional and a CC distribution, denoted $q_{X^n|Y^n}$ and $r_{X^n\|Y^{n-1}}$.
Note that $r_{Y^n\|X^n}$ and $q_{X^n|Y^n}$ are fully characterized by the sets $\{r(y_i|y^{i-1},x^i)\}_{i=1}^n$ and $\{r(x_i|y^{i-1},x^{i-1})\}_{i=1}^n$, respectively.
We define the following bounds on the DI
\begin{align}
    I_\mathrm{U} &\triangleq \frac{1}{n} \max_{x_1}\sum_{y_1}\dots\max_{x_n}\sum_{y_n}P(y^n\|x^n)\log\frac{P(y^n\|x^n)}{\sum_{z^n}P(y^n\|z^n)r(z^n\|y^{n-1})} \\
    I_\mathrm{L} &\triangleq \frac{1}{n}\sum_{x^n,y^n}P(y^n\|x^n)r(x^n\|y^{n-1})\log\frac{q(x^n|y^n)}{r(x^n\|y^{n-1})}
\end{align}
such that 
$$
I_\mathrm{L}\leq \frac{1}{n}I(X^n\to Y^n) \leq I_\mathrm{U}.
$$
The iterative procedure alternated between optimizing $I_\mathrm{L}$ over $r_{X^n\|Y^{n-1}}$ and $I_\mathrm{U}$ over $q_{X^n|Y^n}$.
Using Lagrange multipliers, a closed-form expression is derived for $r_{X^n\|Y^{n-1}}$ and $q_{X^n|Y^n}$ from $r'_{X^n\|Y^{n-1}}$, which is obtained in the previous iteration. That is, on the $k$th iteration, we have
\begin{align}
&E(r_{k-1}(x^n\|y^{n-1})) = P(y_i|x^{i},y^{i-1})\prod_{j=i+1}^nr_{k-1}(x_j|x^{j-1},y^{j-1})P(y_j|x^j,y^{j-1})\nonumber\\
    &r_k(x_i|x^{i-1},y^{i-1})=\prod_{x_{i+1}^n, y_i^n}\left[\frac{q_{k-1}(x^n|y^n)}{\prod_{j=i+1}^n r_{k-1}(x_j|x^{j-1},y^{j-1})} \right]^{E(r_{k-1}(x^n\|y^{n-1}))},\label{eq:di_ba_r}\\
    &q_k(x^n|y^n) = \frac{r_k(x^n\| y^{n-1})P(y^n\|x^n)}{\sum_{x^n}r_k(x^n\|y^{n-1})P(y^n\|x^n)}\label{eq:di_ba_q}.
\end{align}
The complete iterative alternating optimization is as follows:
After initializing $q(x^n|y^n)$ and a tolerance parameter $\epsilon>0$, we perform an iterative optimization, where the $k$th iteration consists of the following actions: 
\begin{enumerate}
    \item Calculate $r_k(x_i|x^{i-1},y^{i-1})$ for $i=n,n-1,\dots,1$ according to \eqref{eq:di_ba_r}
    \item Calculate $q_k(x^n|y^n)$ according to \eqref{eq:di_ba_q}.
    \item Calculate $\Delta_k = I_\mathrm{U}-I_\mathrm{L}$
    \item If $\Delta_k>\epsilon$ set $k\leftarrow k+1$ and repeat, Otherwise, set $\hat{C}=I_\mathrm{L}$.
\end{enumerate}
Note that the update of the elements of $r_{X^n\|Y^{n-1}}$ is performed in the reverse order, i.e., from $i=n$ to $i=1$.
By an appropriate generalization of \eqref{eq:di_ba_r} and \eqref{eq:di_ba_q}, the authors of \cite{naiss2012extension} generalize the above method to scenarios in which the feedback is given by a general time-invariant function of $Y^{i-1}$. This setting also includes delayed feedback schemes.

The BA-driven algorithm optimizing DI was applied in \cite{naiss2012computable} to compute the rate distortion function of stationary ergodic sources with delayed feedforward from the encoder to the decoder.
DI optimization can also be performed when the channel is unknown if an estimate of $P_{Y^n\|X^n}$ is available. We describe an alternative optimization scheme in Section \ref{chap:estimation} if such an estimate is unavailable.

\chapter{Estimating Directed Information}\label{chap:estimation}

This section provides an overview of methods to estimate DI. The algorithms have various levels of complexity and universality. 
In general, the fewer assumptions on the joint distribution $P(x^n,y^n)$ an estimator requires, the higher its computational and sample complexity, and the fewer formal guarantees it offers. 
We assume a DI estimator receives $n$ samples, $(x^n,y^n)$, from the underlying joint distribution (or in some cases, $k$ instances of $(x^n,y^n)$).
When additional structure on the joint process is known, one can simplify the estimators.
We note that $I(X^n\to Y^n)$ is in general a function of $n$, but most works estimate the rate $n^{-1}I(X^n\to Y^n)$ which is assumed to converge to \eqref{eq:di_rate} in the limit $n\to\infty$.
Such estimators quantify the steady-state behavior of the joint process and not transient phenomena, which decay with $n$ under appropriate assumptions.


\section{Classic Estimators}
This section describes popular DI estimation methods that do not rely on contemporary neural network methodologies. We discuss both parametric and non-parametric methods. Section \ref{sec:est_summary} briefly discusses which estimator should be used in which setting.

\subsection{Plug-In Estimators}
The simplest form of DI estimation is a plug-in estimator. The idea is to use the DI definition as an expectation with respect to a population distribution, and ``plug in'' the empirical distribution obtained from a $k$-fold sample of sequences of length $n$, ${\bigl(x^n(i),y^n(i)\bigr)}_{i=1}^k$. When stationarity is assumed, the plug-in estimator can be defined as
$$
\hat{I}(\XX\to\YY)=I(\hat{X}^n;\hat{Y}_n|\hat{Y}^{n-1})
$$
for $(\hat{X}^n,\hat{Y}^n)\sim \hat{P}$, where $\hat{P}$ is the empirical distribution given by 
$$
\hat{P}(x^n,y^n) = \frac{1}{k}\sum_{i=1}^k \mathbbm{1}_{\{(x^n(i),y^n(i)) = (x^n,y^n)\}}.
$$
Alternatively, we can use the decomposition of the DI rate into entropy rates and define the plug-in estimator as 
$$
\hat{I}(\XX\to\YY)=\hat{H}(\YY)-\hat{H}(\YY\|\XX).
$$
This representation allows the use of entropy rate estimation tools. As they rely on the empirical distribution, plug-in estimators assume that $\cX$ and $\cY$ are finite.
When some Markov order $\ell$ is assumed, we can extract a bigger sample size from ${\bigl(x^n(i),y^n(i)\bigr)}_{i=1}^k$ by dividing it into sub-sequences of length $\ell$ \cite{quinn2015directed}.

The plug-in DI estimator performance was analyzed in \cite{quinn2015directed} for a collection of $m$ processes $(\XX_1,\dots,\XX_m)$, which covers the situation of two processes $(\XX,\YY)$ as a special case. The random variables constituting the $i$-th process $\XX_i$ are denoted by $X_{i,t}$, $t = 1,2,3,\ldots$.
The authors of \cite{quinn2015directed} assume the processes on a network are \emph{strictly causal}, i.e., no instantaneous information transfer occurs. Therefore, the joint distribution over $n$ samples from each of the $m$ processes --- $x^{1,n},\dots,x^{m,n}$, with $x^{i,n}$ denoting $n$ samples from $\mathbb{X}_i$ --- decomposes as
\begin{align*}
    P(x^{1,n},\dots,x^{m,n}) &= \prod_{i=1}^m P(x^{i,n}\|x^{1,n-1},\dots,x^{i-1,n-1})\\
    &=\prod_{i=1}^m \prod_{j=1}^n P(x^{i,j}|x^{1,n-1},\dots,x^{i-1,n-1},x^{i,j-1})
\end{align*}
The joint process $(\XX_1,\dots,\XX_m)$ is assumed to be stationary, ergodic and Markov of order $\ell$. Furthermore, each pair $(\XX_i,\XX_j)$ is also assumed to be Markov of order $\ell$.
Ergodicity of the joint Markov chain means that the Markov chain is irreducible and aperiodic, so that it has a unique stationary distribution.
As noted in \cite{quinn2015directed}, these assumptions are made to simplify the setting and to avoid degeneracies. Specifically, strict causality is reasonable when the system's sampling rate is high enough, and Markovity, which is necessary for ergodicity, is often a modeling choice of the system. 

Next, we state the DI plug-in estimator sample complexity bounds from \cite{quinn2015directed}.
To that end, for $\delta>0$, we denote $B_\delta$ as the indicator of the $\delta$ confidence interval of the estimator around the DI rate, i.e., 
$$
B_\delta = \mathbbm{1}_{\{ \hat{I}(\XX_i\to\XX_j)\in [I(\XX_i\to\XX_j)-\delta,I(\XX_i\to\XX_j)+\delta],\forall i,j\}}.
$$
Under the above assumptions, the authors of \cite{quinn2015directed} analyze the sample complexity of the plug-in estimator and characterize the tradeoff between accuracy quantified by $\delta$ and the number of samples $n$.
Specifically, they provide a bound that depends on the joint Markov process mixing. For $\lambda\in(0,1]$ and a $d$-step transition parameter, $d\geq 2$, the process mixing can be defined between the first state, $V_1$, and the $d$-th state, $V_d$, of the joint Markov chain. Here, $V_t$ denotes the state of the joint Markov chain at time $t$, i.e., $V_t=(X_{1,t},\dots,X_{m,t})$. The mixing condition is then given by
$$
P_{V_d|V_1}(v|v')\geq \lambda \pi(v),
$$
where $(v,v')$ are elements in the state space generated by the joint Markov process and $\pi$ is the stationary distribution; see \cite{quinn2015directed} for more details. The bound is given as follows:
\begin{theorem}\label{thm:quinn_sample_complex}
    Let $\lambda\in(0,1]$ and $d\geq 2$ be mixing parameters of the collections of pairwise Markov chains and let $\delta>0$.
    We have $\EE[B_\delta]\geq 1-\rho$ for 
    $\epsilon$ such that $\delta = -4|\cX|^{2\ell+1}\epsilon\log\epsilon$ and 
    $$
    \rho=8m(m-1)|\cX|^{2\ell+1}\exp\left( -\frac{(n\epsilon-2d/\lambda)^2}{2nd^2/\lambda^2} \right).
    $$
    Furthermore, for any $\epsilon'>0$, the sample complexity of jointly estimating all pairwise DI rate terms is $\delta=O(n^{-1/2+\epsilon'})$ for a fixed number $m$ of processes. Conversely, for a fixed $\delta$, we require $n=O(\log m)$.
\end{theorem}

The authors of \cite{kontoyiannis2016estimating} deepen the discussion of the plug-in estimator and study its asymptotic distribution.
Specifically, the plug-in DI estimator converges to a distribution that depends on the ground truth DI and the estimator variance. 
If $I(\XX\to\YY)>0$ (i.e., $\XX$ has a positive causal influence on $\YY$), the plug-in estimator error converges to a zero-mean Gaussian with rate $O(n^{-1/2})$.
Otherwise, the error converges to a $\chi^2$ distribution with $|\cX|^k(|\cY|^{k+1}-1)(|\cX|-1)$ degrees of freedom 
The convergence depends on the alphabet sizes $(|\cX|,|\cY|)$ with rate $O(n^{-1})$.
These results extend to almost sure (a.s.) and $L_1$ convergence as follows
\begin{align*}
    &\hat{I}(\XX\to\YY)-I(\XX\to\YY) = \cO\left(\sqrt{\frac{\log\log n}{n}}\right),\quad P\text{-a.s.}\\
    &\EE\left[\left|\hat{I}(\XX\to\YY)-I(\XX\to\YY)\right|\right] = O(n^{-1/2}).
\end{align*}

Furthermore, the authors of \cite{kontoyiannis2016estimating} connect the plug-in estimator with the \textit{causal influence hypothesis test}.
Suppose we test for a causal influence of the elements of the process $\XX$ on the elements of $\YY$. 
The null hypothesis corresponds to $Y_i\indep X^{i-1}$ given $Y^{i-1}$ for any $i>k$, i.e., $I(\XX\to\YY)>0$, and the alternate hypothesis corresponds to $I(\XX\to\YY)=0$.
In such a setting, the likelihood ratio test statistic is given by the difference of log-likelihoods under the hypotheses and amounts to $2n\hat{I}(\XX\to\YY)$.

Following \cite{kontoyiannis2016estimating}, the authors of \cite{ferreira2021concentration} propose concentration bounds for ergodic processes with \textit{unbounded} memory.
To state the result, for any $\epsilon\in(0,1)$, let $(k_i)_{i\in\NN}$ satisfy $k_i\leq \frac{\epsilon}{2\log|\cX\times\cY|}\log(i)$ and let
$$
\theta(k_i) = \EE\left[\frac{\hat{I}(X^{k_i}\to Y^{k_i})}{k_i}\right].
$$
be the normalized DI estimate over $k_i$ elements.
For this setting, \cite{ferreira2021concentration} showed that for any $t>0$ we have
$$
\Pr\left(\left| \frac{\hat{I}(X^{k_i}\to Y^{k_i})}{k_i} - \theta(k_i) \right|\geq t \right)\leq 2\exp\left( -\frac{8\log^2|\cX\times\cY|n^{1-\epsilon}t^2}{9\Gamma(p)^{-2}\epsilon^2\log^4n} \right),
$$
where $\Gamma(z)$ is the Gamma function and $p$ is the infinite-order Markov chain transition kernel.

\subsection{Bias of Plug-In DI estimators}
The authors of \cite{schamberg2019bias} study the bias of plug-in DI estimators through the lens of $d$-separation and the Markov-based bounds on DI.
Specifically, they show that while the mentioned DI estimators operate under the assumptions that both the joint process $(\XX,\YY)$ and the `receiving' process $\YY$ are Markov with a finite order, the latter assumption on $\YY$ does not necessarily hold.
This results in an estimation gap due to the Markov assumption, which is quantified by \cite{schamberg2019bias}.

To this end, for a set of Markov processes $(\XX,\YY,\ZZ)$ which are jointly stationary and have Markov order $d$, they define the \textit{truncated} and \textit{partial} DI respectively as
\begin{align*}
    I_T^{(k)}(X^n\to Y^n\|Z^n) &\triangleq  \sum_{i=1}^n I(X^i_{i-k};Y_i|Y^{i-1}_{i-k},Z^{i}_{i-k})\\
    I_P^{(k)}(X^n\to Y^n\|Z^n) &\triangleq  \sum_{i=1}^n I(X^i_{i-k};Y_i|Y^{i-1},Z^{i}_{i-k},X^{i-k-1}).
\end{align*}
Let $k_1\geq 1$ and $k_2\geq d$. The truncated and partial DI measures are used to construct the following DI bounds for processes of Markov order $d$:
$$
\lim_{n\to\infty}\frac{1}{n}I_P^{(k_1)}(X^n\to Y^n \| Z^n) \leq I(\XX\to \YY \| \ZZ) \leq \lim_{n\to\infty}\frac{1}{n}I_T^{(k_2)}(X^n\to Y^n \| Z^n).
$$

\subsection{Parametric Maximum-Likelihood Estimation}
Another variation of plug-in DI estimation takes a parametric approach \cite{quinn2015directed}.
Instead of treating an empirical distribution, consider a parametrized distribution for $Q$ parameters $\theta\in\Theta$, where $\Theta\subset\RR^{Q}$ is compact.
Denote the corresponding conditional log-likelihood by
$$L_n(\theta)=\log P(y_n|y^{n-1},x^n;\theta).
$$
The authors of \cite{quinn2015directed} propose an analysis of the maximum-likelihood estimator under the following assumptions:
\begin{enumerate}
    \item $L_n(\theta)$ is continuously differentiable with respect to $\theta\in\Theta$.
    \item The expected log-likelihood $\EE[L_n(\theta)]$ has a unique maximizer.
    \item The expected Hessian, given by $A_n(\theta)\triangleq \EE\left[-\frac{\partial^2L_n(\theta)}{\partial\theta_{i}\partial\theta_{j}}\middle|_{\theta_{i,j}}\right]$, is positive definite with finite entries.
\end{enumerate}
The above assumptions
allow to calculate the maximum-likelihood estimator sample complexity, resulting in the error $\delta=O(n^{-1/2})$.
This sample complexity follows from the sample complexity of the density maximum-likelihood estimator.

\subsection{Intensity-Function-Based Estimator}
DI is used in neuroscience to analyze neural spike-train data.
The joint distribution of neural spike data follows a specific structure that motivated a parametric DI estimator in \cite{quinn2011estimating}.
In statistical inference of neural data, well-behaved processes can be fully characterized via the conditional intensity function (CIF) \cite{brown2003likelihood}, which, for the process $\YY$, is defined as\footnote{We adapted our notation for consistency.} \cite{daley2003introduction}
$$
\lambda(t\|x^t,y^t)\triangleq \lim_{\Delta\to0}\frac{P(y_{t+\Delta}-y_t=1|x^t,y^t)}{\Delta},
$$
where $t$ is a continuous time step.
By discretizing the time interval to $n$ samples and taking $\Delta\ll 1$, we can use the CIF to approximate the log-probabilities defined by the discretized conditional process. Specifically, by using a Bernoulli we have
\begin{equation}\label{eq:logf_cif}
    \log f_\lambda(y^n\|x^n) = -\sum_{i=1}^n\log\lambda(i\|x^i,y^i)\dd y_i + \lambda(i\|x^i,y^i)\Delta.
\end{equation}
The formula \eqref{eq:logf_cif} stems from the Bernoulli-spike model \cite{brown2003likelihood}. The time index $t$ is replaced with a discrete index $i\in\{1,\dots,n\}$ and on each step we calculate the CIF on time $i$ given $(x^{i-1},y^{i-1})$.
The quantity $dy_i$ quantifies the increment of spike count, and often boils down to a binary variable indicating if a spike occurred. The log-probability $\log f_{\lambda}(y^n)$ is similarly defined with a CIF that evolves conditioned on the history of $\YY$ alone.

The construction of the CIF often employs generalized linear models (GLMs) \cite{truccolo2005point} to model the CIF.  Formally, for functions $h=(h_1,\dots,h_k)$ the set $\mathrm{GLM}_{J,K}(h)$ is given by:
\begin{align*}
    &\mathrm{GLM}_{J,K}(h)\triangleq \\
    &\left\{ \lambda:\log\lambda(i\|x^i,y^i) = \alpha_0+\sum_{j=1}^J \alpha_j\dd y_{i-j}+\sum_{k=1}^K \beta_kh_k(x_{i-(k-1)})\right\},
\end{align*}
and we define $\mathrm{GLM}(h)=\bigcup_{J,K\in\NN}\mathrm{GLM}_{J,K}(h)$.
The choice of the function $h$ is not specified in \cite{quinn2011estimating}. Popular choices are the identity function, piecewise constant functions, or parametric models (e.g. splines \cite{truccolo2005point}).
In traditional GLM terms, $\mathrm{GLM}(h)$ is a \emph{log-linear Poisson GLM with the canonical log link}. The spike-history and stimulus-basis terms comprise the predictor vector, and the log-intensity is a linear function of these predictors.

The CIF-based estimator operates under the following assumptions:
\begin{enumerate}
\item For any point processes $(\XX,\YY)$ and a pre-specified set of functions $h=\{h_k:k\geq0\}$, the CIF has a generalized linear model (GLM) representation, i.e. $\lambda(i\|x^i,y^i)\in\mathrm{GLM}(h)$ for all $i$.
    \item The joint distribution has a finite alphabet, i.e. $|\cX|,|\cY|<\infty$.
    \item The joint process $(\XX,\YY)$ is stationary and ergodic.
    \item The joint distribution follows a finite-order Markov chain structure.
    
\end{enumerate}
As discussed in \cite{quinn2011estimating}, when neural spike trains are discretized (e.g., $\Delta\!=\!1$\,ms) and modeled as point-process GLMs with finite history windows $(J,K)$, they satisfy Assumptions (1)–(3). 

Under the above assumptions, one can estimate the DI rate as follows.
First, the DI rate is decomposed into the entropy rates via
$I(\XX\to\YY)=H(\YY)-H(\YY\|\XX)$, such that each is separately estimated.
To estimate $H(\YY)$, \cite{quinn2011estimating} proposes to use a provably consistent existing estimator such as a Lempel-Ziv universal estimator \cite{ziv1977universal} or a Burroughs-Wheeler transform-based estimator \cite{cai2004universal}.

The estimate of $H(\YY\|\XX)$ follows by combining a maximum-likelihood estimate and minimum description length-based model order selection. Under the considered assumptions, the optimization objective is
\begin{equation}\label{eq:di_quinn_opt}
    \min_{J,K}\min_{\theta\in\RR^{J+K}}-\frac{1}{n}\sum_{i=1}^n g_\theta(y^i_{i-J},x^i_{i-K}) + \frac{J+K}{2n}\log n
\end{equation}
where $g_\theta$ is the CIF-based log density approximation under the GLM assumption and $(J,K)$ are the model orders.
Note that for every pair $(J,K)$ the optimization over $\theta\in\RR^{J+K}$ should be solved, and the optimal model order $(\hat{J},\hat{K})$ is chosen to minimize over the range of $(J,K)$.
Note that $\hat{K}=0$ corresponds to no causal effect from $\XX$ to $\YY$, which therefore implies that $H(\YY)=H(\YY\|\XX)$.
The optimization $\eqref{eq:di_quinn_opt}$ is solved for $\theta$ using a GLM optimizer, which is a convex optimization scheme with a built-in implementation in popular software (e.g. \texttt{glmfit} in MatLab).
The authors of \cite{quinn2011estimating} show that the resulting DI estimator is a strongly consistent estimator of the DI rate, i.e., we have
$$
\hat{I}(\XX\to\YY)\stackrel{a.s.}{\to}I(\XX\to\YY).
$$

\subsection{Context Tree Weighting-Based Estimator}
The authors of \cite{jiao2013universal} develop a class of DI estimators using universal probability assignments and the context-tree-weighting (CTW) algorithm.
The estimators are designed for stationary and ergodic stochastic processes over finite alphabets, assuming that the joint process memory does not exceed the maximal CTW tree depth.

Given data from a distribution, the CTW algorithm learns a graphical data distribution model in the form of a weighted tree. To explain the CTW algorithm, we consider the distribution $P_{Y^n,X^n}$.
Each node on the tree corresponds to a different possible sequence, such that each node has $|\cX|\times|\cY|$ edges, each labeled with a distinct possible following item in the sequence.
The tree path that leads to a particular node corresponds to a specific context, i.e., a sequence of previous symbols.
Under the assumption of a memory $l$, a tree with depth $D\geq l$ captures all possible distributions of memory up to $l$. When $D\leq l$, the CTW is generally considered a good approximation of the joint distribution.

\begin{figure}[!t]
\centering
\resizebox{0.7\linewidth}{!}{
\begin{tikzpicture}[
    node distance=1.5cm and 2cm,
    every node/.style={circle, draw, minimum size=0.6cm},
    level/.style={sibling distance=60mm/#1}
]

\node (root) at (8,0) {root};
\node[draw=none] at (8,0.8) {(4,4)};

\node (n1) at (5,1.5) {1};
\node[draw=none] at (5,2.2) {(3,1)};

\node (n0_1) at (5,-1.5) {0};
\node[draw=none] at (5,-0.9) {(1,3)};

\node (n1_2) at (2,2.2) {1};
\node[draw=none] at (2,2.8) {(2,1)};

\node (n0_2) at (2,0.8) {0};
\node[draw=none] at (2,1.4) {(1,1)};

\node (n1_3) at (2,-0.8) {1};
\node[draw=none] at (2,-0.1) {(1,1)};

\node (n0_3) at (2,-2.2) {0};
\node[draw=none] at (2,-1.6) {(0,2)};

\node (l1) at (-1,2.6) {1};
\node[draw=none] at (-1.9,2.6) {(0,0)};

\node (l0_1) at (-1,1.8) {0};
\node[draw=none] at (-1.9,1.8) {(1,0)};

\node (l1_2) at (-1,1.0) {1};
\node[draw=none] at (-1.9,1.0) {(1,0)};

\node (l0_2) at (-1,0.2) {0};
\node[draw=none] at (-1.9,0.2) {(1,1)};

\node (l1_3) at (-1,-0.6) {1};
\node[draw=none] at (-1.9,-0.6) {(0,1)};

\node (l0_3) at (-1,-1.4) {0};
\node[draw=none] at (-1.9,-1.4) {(1,0)};

\node (l1_4) at (-1,-2.2) {1};
\node[draw=none] at (-1.9,-2.2) {(0,1)};

\node (l0_4) at (-1,-3.0) {0};
\node[draw=none] at (-1.9,-3.0) {(0,1)};

\draw (root) -- (n1);
\draw (root) -- (n0_1);

\draw (n1) -- (n1_2);
\draw (n1) -- (n0_2);

\draw (n0_1) -- (n1_3);
\draw (n0_1) -- (n0_3);

\draw (n1_2) -- (l1);
\draw (n1_2) -- (l0_1);

\draw (n0_2) -- (l1_2);
\draw (n0_2) -- (l0_2);

\draw (n1_3) -- (l1_3);
\draw (n1_3) -- (l0_3);

\draw (n0_3) -- (l1_4);
\draw (n0_3) -- (l0_4);

\end{tikzpicture}
}
\caption{The Context Tree Weighting (CTW) algorithm for depth $D=3$ based on the example in \cite{jiao2013universal} with the sequence $(x_{-2}, x_{-1}, \ldots, x_{8}) = 000110100010$. Each leaf represents a context, and internal nodes recursively aggregate predictions. Edge labels denote observed binary outcomes. The value inside each node is the prediction based on weighted probability estimates. Tuples such as $(2,1)$ represent symbol counts during the recursive update. The root node combines all subtree estimates into the final prediction.}
\label{fig:ctw}
\end{figure}
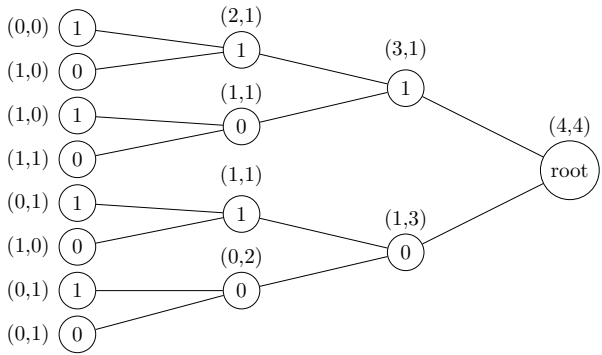

Each node has a weight value corresponding to the occurrence probability of the node's sequence.
At each step, the current observed sequence (symbol and context) corresponds to a node on the graph.
The values of the nodes along the node's path are updated according to the current tree weights.
By processing a sequence of $n$ symbols $(x^n,y^n)$, the CTW iteratively updates the tree weights, yielding a consistent estimate of the process PMF.
For tree depth $D$, the CTW algorithm has time complexity $O(nD)$ and space complexity $(|\cX|\times|\cY|)^D$.
See Figure \ref{fig:ctw} for a visualization of the CTW algorithm.

To obtain DI estimates, the CTW approximates the marginal and joint distributions $(P_{Y^n}, P_{X^n}, P_{X^n,Y^n})$ and utilizes common entropy decompositions of DI (as presented in Chapter \ref{chap:di}). We denote the CTW-based approximation of $P$ with $\hat{P}$ in this subsection.
Using the CTW algorithm, four estimators of DI are proposed.

The first estimator follows directly from the asymptotic equi-partition property (AEP) for entropy rates \cite{CovThom06} and CC entropy rates \cite{venkataramanan2007source}.
Therefore, it is constructed as the difference of two AEP-based entropy estimates, and is therefore given by 
\begin{equation}\label{eq:di_ctw_1}
    \hat{I}_1(x^n,y^n)=-\frac{1}{n}\log\hat{P}(y^n) + \frac{1}{n}\log\hat{P}(y^n\|x^n).
\end{equation}

The second estimator considers the conditional entropy functional of the probability assignment $\hat{P}$ instead of the AEP term in each MC term \eqref{eq:di_ctw_1}.
To this end, the second estimator is given by
\begin{equation}\label{eq:di_ctw_2}
    \hat{I}_2(x^n,y^n) \triangleq  \hat{H}_2(y^n) - \hat{H}_2(y^n \| ^n)
\end{equation}
where
\begin{align*}
    \hat{H}_2(y^n) &\triangleq  -\frac{1}{n}\sum_{i=1}^n\sum_{y_{i+1}}\hat{P}(y_{i+1}|y^i)\log\hat{P}(y_{i+1}|y^i)\\
    \hat{H}_2(y^n\|x^n) &\triangleq  -\frac{1}{n}\sum_{i=1}^n\sum_{x_{i+1},y_{i+1}}\hat{P}(x_{i+1},y_{i+1}|x^i,y^i)\log\hat{P}(y_{i+1}|y^i,x^i,x_{i+1}).
\end{align*}
An advantage of $\hat{I}_2$ over $\hat{I}_1$ is that its outputs are bounded within the interval $[-\log|\cY|,\log|\cY|]$, while the outputs are $\hat{I}_1$ are unbounded in general.
However, as we will discuss, $\hat{I}_1$ benefits from a stronger convergence rate.

A disadvantage of both $\hat{I}_1$ and $\hat{I}_2$ is that they are constructed as the difference of two estimators, each corresponding to a different entropy estimate. Consequently, they can produce negative outputs. To circumvent this, two additional estimators are proposed.
The first estimator uses the DI representation via KL divergence and calculates a Monte-Carlo average of the KL divergence functional between two CTW-based distributions.
It is as follows
\begin{equation}
    \hat{I}_3(x^n,y^n)\triangleq \frac{1}{n}\sum_{i=1}^n\DKL(\hat{P}(y_{i+1}|x^{i},y^{i})\|\hat{P}(y_{i+1}|y^{i})),
\end{equation}
The final estimator is similar to $\hat{I}_3$, but the KL divergence term considers the joint distribution and a CTW-based estimate of the CC distribution $P_{X^n\|Y^n}$. It is given by
\begin{equation}\label{eq:di_ctw_4}
    \hat{I}_4(x^n,y^n)=\frac{1}{n}\sum_{i=1}^n\DKL(\hat{P}(x_{i+1},y_{i+1}|x^{i},y^{i})\|\hat{P}(y_{i+1}|y^i)\hat{P}(x_{i+1}|x^i,y^i)).
\end{equation}

All of these estimators were shown to be consistent estimators of the DI rate, in both the $L_1$ sense and almost surely.
The consistency statements assume that the joint and marginal processes of $\YY$ are stationary, ergodic, aperiodic Markov chains with a finite alphabet. Error convergence rates were also constructed. Specifically, $\hat{I}_1$ was show to converge in the $L_1$ and almost sure senses with rates $O(n^{-1/2}\log n)$ and $O(n^{-1/2}(\log n)^{5/2})$, respectively, while $\hat{I}_2$ was shown to converge in the $L_1$ sense with rate $O(n^{-1/2}\log n^{3/2})$.
It is also shown that no estimate of DI can obtain a convergence rate that is sharper than $O(n^{-1/2})$, which emphasizes the tightness of the proposed error bounds \cite{jiao2013universal}.



\subsection{$k$-Nearest Neighbors}
The authors of \cite{murin2017k} proposed to use the $k$-nearest neighbors (kNN) algorithm to estimate DI.
The kNN algorithm is a supervised learning algorithm that utilizes a labeled dataset $(x^n, y^n)$ with samples $X^n$ and corresponding labels $Y^n$ as follows.
Given a new sample $x_{n+1}$, a label $\hat{y}_{n+1}$ is determined according to the $k$ nearest neighbors under some predetermined metric $\sd$.\footnote{In the context of density estimation $L_p$ spaces are considered.}
A modification of the celebrated Kraskov, St\"{o}gbauer, Grassberger (KSG) estimator \cite{kraskov2004estimating} is proposed in \cite{murin2017k}.

The KSG MI estimator can be constructed either directly for MI or for differential entropies, which, in turn, yield an MI estimate.
Let $\psi(x)=\Gamma(x)(d\Gamma(x)/dx)$ be the digamma function and $\rho_{k,i,p}$ the distance of $i$th sample $x_i$ from its $k$th nearest neighbor under the $\ell_p$ metric.
Define the number of samples with distance $\rho_{k,i,p}$ or less from $x_i$ as 
$$
n_{x_i,p}=\sum_{j=1, j\neq i}^n\mathbbm{1}_{\{\|x_i-x_j\|_p\leq \rho_{k,i,p}\}}.
$$
Consequently, KSG MI estimator is\footnote{A similar estimator can be derived using the Euclidean norm instead of the sup norm; see \cite{gao2018demystifying}.} 
$$
\hat{I}_{\mathsf{KSG}}(X;Y)=\psi(k)+\log(n)-\frac{1}{n}\sum_{i=1}^n\left( \psi(n_{x_i,\infty})-\psi(n_{y_i,\infty}) \right).
$$
To arrive at an MI estimate from differential entropies, we  define the KSG differential entropy as
$$
\hat{h}_{\mathsf{KSG}}(X)=\log n + \log c_{d_x,p}+\frac{1}{n}\sum_{i=1}^n(d_x\log\rho_{k,i,p}-\psi(n_{x,i,p}+1)),
$$
where $c_{d,p}$ is volume of the $d$-dimensional unit ball in the $\ell_p$ space.

The authors of \cite{murin2017k} leverage the KSG estimator and propose a DI rate estimator for Markov processes.
Suppose the joint process $(\XX,\YY)$ is a stationary Markov process of order $m$. The DI rate can be written as 
\begin{align*}
I(\XX\to\YY)&=I(X^m;Y_m|Y^{m-1}) \\
&= h(Y^m)-h(Y^{m-1})-h(Y^m,X^{m-1})+h(Y^{m-1},X^{m-1}).
\end{align*}

When estimating DI, we deal with sequences rather than a single sample each time. For sequences of length $m$, those samples are given from the set of $(n-m)$ overlapping sequences from $(x^n,y^m)$.
Consequently, for a sequence e.g. $x^{i-1}_{i-m}$, its corresponding set, from which we evaluate the $k$-neighbors is the set $(x^{j-1}_{j-m})_{j\neq i}$.
Consequently, the KSG estimator can be adapted to give
\begin{align*}
    \hat{I}_{\mathsf{KSG}}(\XX\to\YY)&=\psi(k)+\frac{1}{n-m}\sum_{i=m+1}^m\psi(n_{y^{i-1}_{i-m},\infty}+1)\\
    &-\psi(n_{y^{i}_{i-m+1},\infty}+1)-\psi(n_{(x^{i-1}_{i-m},y^{i-1}_{i-m}),\infty}+1).
\end{align*}
The KSG estimator is widely used to estimate MI and is a consistent MI estimator under smoothness conditions of the joint density. 

\subsection{Kernel-Density Estimation}
Another work that estimates the DI when alphabets are continuous via kernel density estimation \cite{izenman1991review} is established in \cite{malladi2016identifying}.
The main idea is that the joint process evolution is modeled with a Gaussian linear autoregressive model of the form
$$
Y_i = \sum_{j=1}^J\alpha_jY_{i-j} + \sum_{k=1}^K\beta_kX_{i-k+1} + Z_i,
$$
where the model parameters are $\theta=((\alpha_j)_{j=1}^J,(\beta_k)_{k=1}^K)$ and $(Z_i)_{i\in\NN}$ is a white zero-mean Gaussian process.
The resulting conditional distribution $P(y_i|y^{i-1},x^{i})$ is Gaussian, which gives a closed-form formula of the normalized log-likelihoods $1/n\log P(y^n\|x^n;\theta)$ and $1/n\log P(y^n;\theta)$, which are then optimized with respect to $\theta$, resulting in a maximum-likelihood estimate of DI.
The authors of \cite{malladi2016identifying} suggest coupling the optimization with minimum-descriptive-length-based optimization for model order selection \cite{grunwald2007minimum}, i.e., the choice of optimal $K$ and $J$ values.





\section{Neural Estimation}
With the rising popularity of neural networks in most scientific fields due to their computational and expressive powers, a recent line of work has emerged, in which deep learning is applied to tackle various problems in information theory.
Specifically, neural estimation, which is the process of estimating some functional by optimizing neural networks, gained increased popularity since it was proposed for MI.
Employing neural estimation to DI, the authors of \cite{tsur2023neural} developed a provably consistent estimator, called the DI neural estimator (DINE).

The DINE represents the DI rate as the solution of a tractable optimization problem which can be solved by gradient descent methods over a class of neural networks.
As a first step, by leveraging a decomposition of entropy to cross entropy and KL divergence terms with respect to a reference uniform measure, the DI rate can be decomposed into the difference \cite[Lemma~4]{tsur2023neural}
\begin{equation}\label{eq:di_rate_kls}
    I(\XX\to\YY)=\DKL(P_{Y_0|X^{0}_{-\infty},Y^{-1}_{-\infty}}\|P_{\tilde{Y}})-\DKL(P_{Y_0|Y^{-1}_{-\infty}}\|P_{\tilde{Y}}),
\end{equation}
where $P_{\tilde{Y}}$ is a reference uniform distribution over $\cY$.\footnote{When $\cY$ is unbounded we can either take a bounding box over the dataset $Y^n$ or replace the uniform distribution with a Gaussian one \cite{aharoni2022density}.}
Then, each KL term is represented using the Donsker-Varadhan (DV) variational formula \cite{donsker1983asymptotic}.
For a pair $P,Q$, the DV formula is
\begin{equation}\label{eq:dv}
    \DKL(P\|Q)=\sup_{f\in\cF}\EE\left[f(X_1)\right]-\log\EE\left[\exp(f(X_2))\right],
\end{equation}
where $X_1\sim P$ and $X_2\sim Q$, and $\cF$ is the class of all Borel-measurable functions with finite expectations.
Next, we apply the DV formula to the KL terms in \eqref{eq:di_rate_kls}, replace the optimization feasible set with a class of neural networks, and replace expectations with sample means.
Doing so results in the following estimator of the DI rate from a sample from $P_{X^n,Y^n}$, $(x^n,y^n)$:
\begin{equation}\label{eq:dine}
    \hat{I}_{\mathsf{DINE}}(x^n,y^n)
    \triangleq \sup_{\theta_{xy}}\hat{\sD}_{Y\|X}(x^n,y^n,g_{\theta_{xy}})-\sup_{\theta_y}\hat{\sD}_{Y}(y^n,g_{\theta_y})
\end{equation}
where $\hat{\sD}_{Y}(y^n, g_{\theta_y}), \hat{\sD}_{Y\|X}(x^n,y^n,g_{\theta_{xy}})$ are given by
\begin{subequations}
\begin{align}
   \hat{\sD}_Y(y^n, g_{\theta_y}) 
   &\triangleq \frac{1}{n}\sum_{i=1}^n{g_{\theta_y}}\left(y^i\right)
   -\log\left(\frac{1}{n}\sum_{i=1}^n e^{g_{\theta_y}\left(\widetilde{y}_i,y^{i-1}\right)}\right)\\
   \hat{\sD}_{Y\|X}(x^n,y^n,g_{\theta_{xy}}) 
   &\triangleq \frac{1}{n}\sum_{i=1}^n g_{\theta_{xy}}\left(y^i,x^i\right)-\log\left(\frac{1}{n}\sum_{i=1}^n e^{g_{\theta_{xy}}\left(\widetilde{y}_i,y^{i-1},x^i\right)}\right)
\end{align}\label{eq:DINE_KL_est_main}%
\end{subequations}
and $\tilde{y}^n$ are i.i.d. samples from the uniform distribution on $\cY$, $\mathsf{Unif}(\cY)$. The parametric models $g_{\theta_y}$, $g_{\theta_{xy}}$ can be realized with a modification of either recurrent neural networks \cite{tsur2023neural} or with Transformer models \cite{luxembourg2024treet}, to maintain a causal dependence structure.

To estimate the DI rate, one must solve the optimization \eqref{eq:dine}.
Note that, on the $i$th step, the required network operates on two different inputs $(y_i,\tilde{y}_i)$, and a single shared history ($(y^{i-1})$ for $g_{\theta_y}$ and $(y^{i-1}, x^i)$ for $g_{\theta_{xy}}$).
To this end, the authors of \cite{tsur2023neural} construct a novel recurrent neural network architecture, which is used for the selective accumulation of memory in the network according to the elements in each KL mean estimate term.
The DINE is a consistent estimator of the DI rate for a class of jointly stationary and ergodic processes \cite[Theorem~2]{tsur2023neural}. However, the DINE does not have finite-sample guarantees on estimation. This is due to two main reasons: 1)~It operates under relatively general assumptions on the data (joint stationarity and ergodicity), for which only asymptotic analysis is feasible, and 2)~its performance is bottlenecked by the performance of recurrent neural networks in dynamic system approximation, which is currently purely asymptotic.

Recently, the DINE methodology was extended to state-of-the-art neural network architectures.
Specifically, the authors of \cite{luxembourg2024treet} propose an estimator of transfer entropy that builds upon masked attention transformer models.

As a neural estimator, the DINE applies to contemporary data-driven tasks and is independent of whether the alphabets are continuous or discrete.
When alphabets are continuous, the DINE shows good scalability compared to existing estimators, as it utilizes neural network optimization.
However, optimizing neural networks often requires hyperparameter optimization and architecture tuning. This may induce unnecessary computational overhead when the underlying joint distribution has a simple structure that simpler estimators can capture.

\begin{remark}[Curse of dimensionality]
    While, as presented above, various potential estimators are available in continuous spaces, they all share a common issue; they suffer from the curse of dimensionality; see e.g. \cite{gao2018demystifying}.
    That is, the convergence rate of the proposed estimators does not scale well with dimension.
    This is because such estimators generalize MI estimation methodologies that suffer from this issue whenever high-dimensional continuous variables are involved. One promising method to overcome the curse of dimensionality is by considering sliced information measures \cite{goldfeld2021sliced,tsur2023max}, where we estimate information between low-dimensional projections of the variables.
\end{remark}

\section{Summary of Presented Estimators}\label{sec:est_summary}

In this section, we aim to provide guidance on which DI estimator fits which setting, and examine the differences between the DI estimators presented in this section.

\begin{enumerate}
  \item \textbf{Plug-in estimator} %
        \emph{Use when:} alphabets are very small, discrete and a finite Markov order~$\ell$ is known.  
        \textbf{Pros:} easy to implement, asymptotically unbiased.  
        \textbf{Cons:} sample and memory requirements scale as $|\mathcal X|^{2\ell+1}$, assumes data is discrete.

  \item \textbf{MLE} %
        \emph{Use when:} the data is low-dimensional and the distribution comes from an exponential model.  
        \textbf{Pros:} $\sqrt n$ convergence guarantees.
        \textbf{Cons:} sensitive to misspecification.

  \item \textbf{CTW} %
        \emph{Use when:} discrete data with \emph{unknown} but finite memory.  
        \textbf{Pros:} finite-sample error bounds, adaptivity to tree depth, universality.  
        \textbf{Cons:} bad scalability with alphabet size and tree depth (memory).

  \item \textbf{kNN / KSG} %
        \emph{Use when:} Low dimensional continuous data.  
        \textbf{Pros:} non-parametric, no binning.  
        \textbf{Cons:} error grows fast with dimension; kd-trees degrade beyond \(\sim\!10\) D.

  \item \textbf{GLM} %
        \emph{Use when:} point-process data.  
        \textbf{Pros:} convex optimization setting. 
        \textbf{Cons:} number of parameters scale as \(p^{2}(J\!+\!K)\); assumes sparsity or low-rank penalties.

  \item \textbf{Neural (DINE)} %
        \emph{Use when:} high-dimensional or mixed-type data, unknown distribution.
        \textbf{Pros:} universal, differentiable, scalable with contemporary optimization schemes.  
        \textbf{Cons:} compute-heavy; relies on network optimization and hyperparameter setting.
\end{enumerate}

\section{Optimizing Estimated Directed Information}
In some cases, the user requires optimizing the DI estimate with respect to some of the involved distributions.
For example, calculating channel capacity involves optimizing the average DI with respect to the channel input distribution. Thus, coupling a DI estimator with an optimizer can yield a capacity estimation mechanism.
We optimize an estimate of $I(X^n\to Y^n)$ with respect to the distribution of $X^n$, assuming the conditional distribution $P_{Y^n\|X^n}$ is fixed.

When the DI estimator at hand calculates an estimate of the channel distribution $P_{Y^n\|X^n}$ as an intermediate stage, we can turn to ``model-based'' DI optimization schemes \cite{naiss2012extension}, and plug in the estimated channel distribution.
When the channel distribution is unknown, we distinguish between discrete and continuous input distributions, i.e., if the optimization of estimated DI is performed over continuous or discrete spaces $\cX$.
In what follows, we focus on the construction of the optimizers, outlining the loss function derivation and alternating optimization with the DI estimator. For information on the asymptotic convergence guarantees of the optimizer to a local optimum, we refer the reader to \cite{tsur2023neural,tsur2022optimizing}.

Next, we describe the proposed optimization scheme under each assumption on the input distribution domain.

\underline{Continuous spaces:} When the $\cX\subset\RR^d$ is continuous, the authors of \cite{tsur2023neural} proposed an optimization methodology for the DINE that optimizes an auxiliary generative model of the input distribution $P_{X^n}$.
Formally, they propose the neural distribution transformer (NDT), which is a parametric model that sequentially generates a sequence of channel inputs $X^n$ from a given noise sequence $U^n$ through the relation
\begin{equation}\label{eq:frl}
    X_i = h_\phi(Z_{i-1},U_i),
\end{equation}
where $Z_i=(X_i,Y_i)$ when feedback is considered and $Z_i=X_i$ otherwise.
This relation can be implemented by realizing $h_\phi$ with an autoregressive parametric model, such as recurrent neural networks.
The formulation \eqref{eq:frl} follows from the functional representation lemma \cite{li2018strong}.

To optimize the NDT parameters $\phi$, we exploit the differentiability of the DINE network $(g_{\theta_{y}},g_{\theta_{xy}})$.
Thus, the gradients of the DINE objective \eqref{eq:dine} with respect to $\phi$ can be derived via a chain rule.
This allows minibatch gradient optimization of the NDT parameters to maximize the downstream estimated DI.
Practical implementation is based on automatic differentiation tools
The overall framework consists of alternating optimization of the DINE and NDT parameters.

The same actions are performed on each iteration, but a different model is optimized.  The proposed method can be applied to any parametric estimator of DI that is differentiable with respect to its inputs.
The proposed method was demonstrated to estimate both the feedback and non-feedback channel capacity in \cite{tsur2023neural}, and was recently modified to estimate capacity regions of multi-user channels with memory in \cite{huleihel2023neural}.

\underline{Discrete spaces:} When $\cX$ is discrete and $|\cX|$ is finite, an NDT of the form \eqref{eq:frl} is no longer applicable, as the differentiability of the combined model is lost.
To this end, an optimization scheme that focuses on the inputs PMF was proposed in \cite{tsur2022optimizing}.
Specifically, the \textit{discrete} NDT is a model whose output on each step is a probability on $\cX$, i.e., an $|\cX|$-dimensional probability vector.
Formally, on each step $t$, the discrete NDT, which has parameters $\phi$, receives a history $(x^{t-1},y^{t-1})$ and produces the conditional probability of the next element, which we denote with $p_t^\phi$. The probability vector $p_t^\phi$ is thus given as\footnote{In practice, we realize the NDT with a recurrent mechanism with some intrinsic state, and the output sequentially evolves as $p_t^\phi\triangleq  h_\phi(x_{t-1}, y_{t-1},p^\phi_{t-1})$}
$$
p_t^\phi=p^\phi(\cdot|x^{t-1},y^{t-1}),\qquad t\geq 1.
$$

To optimize the parameters of $h_\phi$, gradients of the corresponding DI rate should be derived with respect to $\phi$.
To that end, an alternative optimization for gradient-based estimation of $h_\phi$ should be derived. 
The alternative objective is constructed such that its gradients approximate the DI rate gradients with respect to $\phi$.
To derive such an objective, the DI rate optimization with respect to the input PMF is formulated as an infinite-horizon average-reward Markov decision process (MDP) (For more information on MDPs, we refer the reader to Section~\ref{sec:MDP}.), allowing us to borrow tools of reinforcement learning.
Specifically, \cite{tsur2022optimizing} combine the policy gradient theorem \cite{sutton1999policy} with expectation estimation and function approximation to derive the alternative optimization criteria $\hat{\sJ}_{\theta,T}(x^n,y^n,\phi)$, which is calculated from samples $(x^n,y^n)$ and whose purpose is, again, to optimize the discrete NDT parameter $\phi$. The objective is given by:
\begin{equation}\label{eq:hat_j}
     \hat{\sJ}_{\theta,T}(x^n,y^n,\phi)\triangleq \frac{1}{n}\sum_{t=1}^{n}\log \left(p^\phi_t(x_{t})\right)\hat{\sQ}^\theta_{T,t}(x^n,y^n),
\end{equation}
where:
\begin{enumerate}
    \item The term:
    \begin{equation}\label{eq:est_q_func}
    \hat{\sQ}^\theta_{T,t}(x^n,y^n) \triangleq  \sum_{i=t}^{t+T-1}
     \hat{r}_\theta(y^{i},x^{i})-\hat{I}(x^n,y^n,\theta)
\end{equation}
is approximates the so-called $Q$-function.
    \item $\hat{I}(x^n,y^n,\theta)$ is a parametric estimator of DI with parameters $\theta$, which can be obtained via, e.g., DINE.
    The term $\hat{r}_{\theta} \triangleq  g_{\theta_{xy}} - g_{\theta_y}$, which are, in turn the optimized DINE neural networks.
    \item $p^\phi_t(x_{t})$ is the discrete NDT output in time $t$, evaluated on $x_t$.
\end{enumerate}

The calculation of the objective \eqref{eq:hat_j} from $n$ samples $(x^n,y^n)$ that come from an $n$-step interaction between the discrete NDT and the channel model, i.e., $(x^n,y^n)\sim\prod_{i=1}^n p^\phi_i p(y_i|x^i,y^{i-1})$ can be understood as follows:
for each pair $(x_i,y_i)\sim p^\phi_i p(y_i|x^i,y^{i-1})$, we use the DINE networks $(g_{\theta_y},g_{\theta_{xy}})$ to calculate an approximation of the accumulated MDP reward over $T$ steps, as given in \eqref{eq:est_q_func}, where $T$ is a predetermined horizon size.
Denoting by $P^\phi$ the distribution induced by the probability vector $p^\phi$, the alternative objective is constructed as a Monte Carlo estimate of the expectation $\EE\left[\log P^\phi_{X_0|X^{-1}_{-\infty},Y^{-1}_{-\infty}}\hat{\sQ}^\theta_{T}\right]$, whose gradient w.r.t $\phi$ is approximated by $\nabla_\phi I(\XX\to\YY)$ due to the policy gradients theorem.

As noted in \cite{tsur2022optimizing}, the proposed framework is not limited to the DINE as the underlying DI estimator.
Specifically, every estimator of DI that yields an estimate of the log-likelihood ratio $\log\frac{p_{Y_0|Y^{-1}_{-i}}}{p_{Y_0|X^0_{-N}Y^{-1}_{-N}}}$ for $i=1,\dots,n$, as it can be plugged to replace the difference of DINE networks in $r_t$. 
These elements can be obtained by the DI estimators above that rely on PMF estimation, such as the plug-in or CTW estimators.
This renders the proposed optimizer compatible with most existing DI estimators.

\chapter{Feedback Capacity}\label{chap:capacity}
This section studies communication channels with memory and their close connection with  DI. We show that the feedback capacity is given by a multi-letter expression in terms of DI. We then present tools and methods from control and learning theories, such as dynamic programming and reinforcement learning, that can be utilized to compute multi-letter capacity expressions. These tools offer significant computational advantages for numerical evaluation, and in some cases, analytical solutions to the feedback capacity can be deduced or are inspired by their numerical estimates.

In his seminal work, Shannon considered memoryless channels defined by a transition probability kernel, $P_{Y|X}$, from the channel input to the channel output \cite{shannon1948mathematical}. The memoryless property implies that at each time $i$, the following Markov condition holds
\begin{align}\label{eq:ch_memoryless}
    P(y_i|y^{i-1},x^i,m)&= P_{Y|X}(y_i|x_i),
\end{align}
where $m$ denotes the random message to be transmitted. That is, the channel output does not depend on past channel outputs, inputs and the message, when conditioned on the current channel input. Shannon showed that the capacity of a memoryless channel is given by the single-letter expression
\begin{align}\label{eq:cap_memoryless}
    \mathrm{C} &= \max_{P(x)} I(X;Y).
\end{align}

For memoryless channels, Shannon also showed that feedback does not increase the channel capacity \cite{shannon1956zero}. That is, the channel capacity of a memoryless channel with feedback is 
\begin{align}\label{eq:cap_fb_memoryless}
    \mathrm{C}^\mathrm{fb} &= \max_{P(x)} I(X;Y)
\end{align}
whether the encoder has access to past channel outputs or not. The fact that feedback does not increase the capacity of a memoryless channel implies that the standard random coding construction of a codebook with i.i.d.\ channel inputs (that do not depend on the feedback) achieves the capacity. 

While feedback does not increase the capacity of memoryless channels, it does have significant advantages. For example, in the variable-length coding regime, Burnashev \cite{burnashev1976data} characterized the optimal error exponent that improves upon the non-feedback error exponent. Feedback can also be used to construct simple and explicit coding schemes. Examples of such coding schemes include that for the BSC by Horstein \cite{horstein1963sequential} (also, \cite{burnashev1974interval}), for the Gaussian case by Schalkwijk and Kailath \cite{schalkwijk1966coding} (which achieves a doubly-exponential decay of the probability of error), and the posterior matching principle that unifies these schemes for memoryless channels \cite{shayevitz2011optimal}. 

For a fixed $P_{Y|X}$, the MI $I(X;Y)$ is a concave function of the input distribution $P(x)$, so the problem of computing the capacity in \eqref{eq:cap_memoryless} or \eqref{eq:cap_fb_memoryless} is a convex optimization problem. However, except in a few specific instances of channels, it is not generally possible to obtain a closed-form expression for capacity. Many good numerical tools are available, both in theory and in practice, for solving generic convex optimization problems, and these can be used for the numerical computation of capacity for a given memoryless channel. In practice, however, it is more convenient to use an iterative algorithm called the Blahut-Arimoto algorithm \cite{blahut1972computation},\cite{arimoto1972algorithm}. The Blahut-Arimoto algorithm is a particular instance of an MM algorithm (``MM'' here stands for ``minorization-maximization'') \cite{Lange_MMAlg}, and its iterates are guaranteed to converge to the channel capacity while also producing a capacity-achieving input distribution -- see e.g., \cite[Chapter~9]{Yeung_book}. Thus, from a computational perspective, it is relatively easy to determine the capacity, with or without feedback, of a memoryless channel. The situation is rather different for a channel with memory, as we will see in Section~\ref{sec:Cfb_DI}.

A general channel with memory can be described by a sequence of channel output probabilities
\begin{align*}
P(y_i|y^{i-1},x^i)
\end{align*}
that can be encapsulated by the causal conditioning (CC), which was defined earlier in Eq.~\eqref{eq:cc_dist} as
\begin{align*}
    P(y^n\|x^{n})\triangleq  \prod_{i=1}^n P(y_i|y^{i-1},x^{i}).
\end{align*}
This class of channels is too general if one is interested in computing the channel capacity or constructing simple capacity-achieving coding schemes. Thus, a common model to capture the influence of past events with a finite-dimensional dependency is a \emph{channel with state}. The idea of a channel state is that it captures the information \emph{relevant} to the current channel output, $y_i$, from the history $(y^{i-1},x^i)$ in a variable whose dimension (alphabet) does not grow with time. In what follows, we present a model of finite-state channels that represents a sufficiently rich class of channels with memory. We remark that the study of channels with memory is inherently more challenging than the memoryless case since the optimal channel input and output processes have memory too. 

\section{Definitions} \label{sec:FSC_defs}

\subsection{Finite-State Channels}
A Finite-State Channel (FSC) is defined by a time-invariant transition probability kernel, $P_{S^+,Y|X,S}$, where $X$, $Y$, $S$, $S^+$ denote the channel input, output, and state before and after a single transmission, respectively. Inputs, outputs and states take values in the sets $\mathcal{X}$, $\mathcal{Y}$ and $\mathcal{S}$, respectively; We assume throughout the current section that $|\mathcal{X}|,|\mathcal{Y}|,|\mathcal{S}|<\infty$ although many of the results can be extended to infinite input and output alphabets as well. Similarly to the memoryless channel property, the FSC property implies that the channel output and next channel state depend only on the previous state and the current channel input. Formally, an FSC has the Markov property
\begin{align}\label{eq:ch_FSC}
    P(s_t,y_t|x^t,y^{t-1},s^{t-1},m) = P_{S^+,Y|X,S}(s_t,y_t|x_t,s_{t-1}), \ \ t\ge1
\end{align}
with $s_0\in\mathcal S$ being the initial state and $m$ the message.

FSCs constitute a rich family of channels that contains several well-studied classes of channels, some of which we describe in the examples below.
\begin{example}[Memoryless channels]
Memoryless channels, characterized in \eqref{eq:ch_memoryless} by $P_{Y|X}$, are a special case of FSCs obtained by taking $|\mathcal{S}|=1$, so that the state is the same at all times.
\end{example}
\begin{example}[Intersymbol interference]
In wireless communication, the transmitted waves may arrive at different times, leading to a delayed effect of transmitted symbols on channel outputs. In this case, the channel output may depend on the current and past channel inputs. This is known as intersymbol interference (ISI), which can be modeled as 
\begin{align}
    P(y_i|y^{i-1},x^i,m)&= P(y_i|x_i,x_{i-T}^{i-1})
\end{align}
for some $T$ that determines the number of past input symbols that affect the current channel output.
A channel with ISI is a special case of an FSC, obtained by choosing $s_{i-1} = x_{i-T}^{i-1}$.
\end{example}
\begin{example}[Independent states]
Another well-known example of FSCs is the class of state-dependent channels whose state is distributed as an i.i.d. sequence with distribution $P_S = P_{S^+}$. In this example, the state is memoryless, in the sense that it does not depend on past states, and channel inputs and outputs. This is a special case of FSC, obtained by choosing 
 \begin{align}
     P_{S^+,Y|X,S} = P_{S^+}P_{Y|X,S}.
 \end{align}
 \end{example}
\begin{example}[Markovian states]
    A generalization of the case of independent states is when the states form a Markov process \cite{goldsmith1996capacity}. This corresponds to a scenario in wireless communication, for example, wherein the present channel condition depends on the previous channel condition. The state-dependent channel with Markovian states is an FSC, obtained by setting
    $$P_{S^+,Y|X,S} = P_{S^+|S}P_{Y|X,S}.$$ 
\end{example}

\begin{example}[Markov channels]\label{ex:markov}
    Finite-state Markov channels constitute a family of FSCs whose state is the last channel output. In particular, the channel is given by
     \begin{align}\label{eq:def_Markov_channel}
        P_{S^+,Y|X,S}&= \mathbbm{1}\{S^+ = Y\} P_{Y|X,S}.
     \end{align}
     A key property of this channel is that the state is available to the decoder. If the channel outputs are fed back to the encoder, then the state also becomes available to the encoder. Markov channels are also known as the Previous Output is STate (POST) channels \cite{permuter2014capacity}. 

     We remark that the property of the state being available to the decoder (and to the encoder, via output feedback) holds for any channel whose state at time $t$ is a deterministic function of a bounded number, $L \ge 1$, of past channel outputs, $y_{t-L}^{t-1}$. In fact, any FSC whose state is known to the decoder (and to the encoder, via output feedback) can be modeled as a finite-state Markov channel as in \eqref{eq:def_Markov_channel}, by the simple device of augmenting the channel output to include the state as well: $\tilde{Y}_t \triangleq  (S_t,Y_t)$.
\end{example}

\begin{example}[Finite-memory state channels] The channel state depends on a finite number of past inputs and outputs. Specifically, 
       \begin{align}\label{eq:input_def}
        P(s_i,y_i|x^i,y^{i-1},s^{i-1},m) &= P(s_i|x_{i-k_1}^i,y_{i-k_2}^{i})P_{Y|X,S}(y_i|x_i,s_{i-1}) \nonumber\\
        &\triangleq  P_{S|\tilde{S}}(s_i|\tilde{s}_i)P_{Y|X,S}(y_i|x_i,s_{i-1}),
    \end{align}
where $k_1$, $k_2$ are arbitrary non-negative finite integers, and we defined a new channel state $\tilde{s}_i = (x_{i-k_1}^i,y_{i-k_2}^i)$. This channel captures both ISI channels and Markov channels as special cases. Although the state transition is not deterministic, one can convert \eqref{eq:input_def} to a FSC whose state is $\tilde{S}_t$, which evolves deterministically, and the channel is given by 
\begin{align}
  P_{Y|X,\tilde{S}}(y_i|x_i,\tilde{s}_{i-1}) &= \sum_{s_{i-1}} P_{S|\tilde{S}}(s_{i-1}|\tilde{s}_{i-1}) P_{Y|X,S}(y_i|x_i,s_{i-1}).
\end{align}
If the channel state depends only on channel outputs, this is called the Noisy Output is the STate (NOST) channel \cite{Shemuel2022nost}.
\end{example}



\begin{figure}
    \centering
    \includegraphics[width=\linewidth]{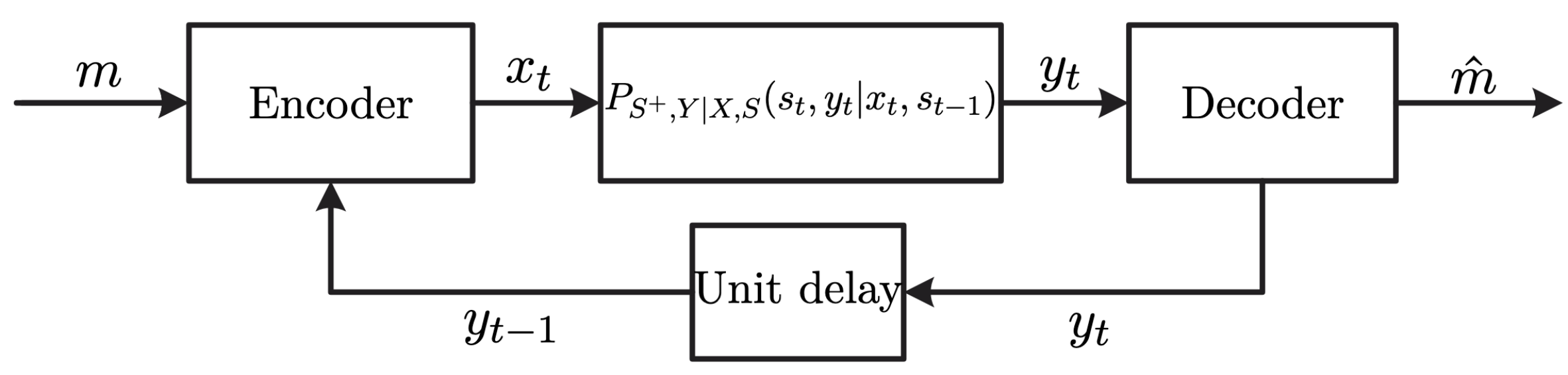}
    \caption{Finite-state channel with feedback.}
    \label{fig:FSC}
 \end{figure}

\subsection{Feedback Capacity} The notion of feedback capacity arises in the communication setup depicted in Fig. \ref{fig:FSC}. A message $m$ is uniformly distributed over the integer set $[1:2^{nR}]$, where $n$ is the blocklength and $R$ is the rate. The encoder has access to the message and the previous channel outputs, so its mapping is defined by the sequence
\begin{align}
    \phi_i: \mathcal M \times \mathcal Y^{i-1}\to\mathcal X, \ \ i = 1,\dots,n.
\end{align}
The channel input $x_i = \phi_i(m,y^{i-1})$ is fed to the channel and the new channel output $y_i$ is made available to the decoder and the encoder (via feedback). For communication over a blocklength of size $n$, the decoder has access to the channel output sequence $y^n$ and it outputs a message estimate $\hat{M}$ using a mapping $\psi: \mathcal Y^n \to \mathcal M$. For a fixed code, i.e., an encoder and a decoder mapping, the average probability of error is $P_e^{(n)} = \Pr (M\neq \psi(Y^n))$. The feedback capacity of a FSC is defined as the supremum over all rates such that there exists a feedback code of rate $R$ with $P_e^{(n)}\to 0$. We denote the feedback capacity as $\mathrm{C}^\mathrm{fb}$. With a slight abuse of notation, we omit the dependence of the feedback capacity on the initial state $s_0$ (or its distribution) due to mild regularity conditions that make this dependence redundant \cite{Permuter2008Trapdoor}.

\section{Feedback Capacity and Directed Information} \label{sec:Cfb_DI}
The following theorem relates the feedback capacity of an FSC to an optimization problem to maximize DI \cite{permuter2009finite,Tatikonda2009feedbakcapacity}.

\begin{theorem}\label{FSC_feedback_Capacity}
The feedback capacity of a strongly connected\footnote{An FSC is strongly connected if for any pair of states $s,s^{\prime}\in\mathcal{S}$, there exists an integer $T$ and an input distribution $\{P_{X_i|S_{i-1}}\}_{i=1}^{T}$ such that $\sum_{i=1}^T P_{S_i|S_0}(s|s^{\prime})>0$.} FSC is
\begin{align}
	\mathrm{C}^\mathrm{fb} = \lim_{n\to\infty}\frac{1}{n}\max_{P(x^n\|y^{n-1})}I(X^n\rightarrow Y^n).
\end{align}
\end{theorem}

In comparison, the capacity \emph{without feedback} for a large class of channels with memory is given by~\cite{gallager1968information}
\begin{equation}\label{e_cap_mut}
\mathrm{C} = \lim_{n\to \infty} \max_{P(x^n)} \frac1{n} I (X^n;Y^n).
\end{equation}    
Interestingly, the only difference between the capacity with and without feedback is the optimization domain since, in the absence of feedback, we have $I(X^n;Y^n) = I(X^n \to Y^n)$. 

The capacity formulae for the cases with and without feedback are both \emph{multi-letter expressions}, as they involve a maximization over an increasing number of random variables. Such capacity expressions typically cannot be computed directly since they involve a limit and a domain of maximization that grows exponentially in $n$ \cite{permuter2009finite,DaboraGoldIndecomposable,Tatikonda2009feedbakcapacity}. 

To reveal the inherent challenges in the computation of the channel capacity, or equivalently, in the maximization of DI, we will present several special cases based on the availability of the channel state, or lack thereof, to the encoder and the decoder. In the general case of an FSC, the channel state is not available to the encoder or the decoder. The feedback, however, induces a degraded situation where the encoder can only know more than the decoder. We start by studying the first scenario in which the channel state is known by all parties. 

\subsection{Markov Channels} The first scenario is Markov channels, introduced in Example \ref{ex:markov}, wherein the channel state is available to the decoder and encoder, via output feedback. Such a channel can be described by a probability kernel of the form
\begin{align}\label{eq:fsc_sc_sy}
    P_{Y|X,Y^-}: \mathcal X \times \mathcal Y\to \mathcal P(\mathcal Y).
\end{align}
That is, we omit the state variable and simply denote the state by $Y^-$, the last channel output.

The feedback capacity of Markov channels is given by a single-letter expression. 
\begin{theorem}\label{th:cap_sy}
    The feedback capacity of an FSC $P_{Y|X,Y^-}$ is given by 
    \begin{equation}
        \mathrm{C}^{\mathrm{fb}} = \max_{P(x|y^-)}I(X;Y|Y^-),
    \end{equation}
    where the joint distribution is $\pi(y) P(x|y^-)P_{Y|X,Y^-}(y|x,y^-)$, and $\pi(\cdot)$ is the stationary distribution computed from the Markov transition probability
    \begin{equation}
        P(y|y^-) = \sum_x P(y|x,y^-)P(x|y^-).
    \end{equation}
\end{theorem}
The feedback capacity in Theorem \ref{th:cap_sy} shows that an optimal encoder needs only to depend on the last channel output. That is, at time $t$, while the encoder has access to $(x^{t-1}, y^{t-1})$, it need only use $y_{t-1}$. This result also implies that the optimal channel output distribution is Markov. We will see later in Section \ref{sec:Qgraph} that the structure of the optimal output distribution plays a crucial role in capacity computation. 

The feedback capacity is not a convex function of the conditional input distribution $P(x|y^-)$, but it can be formulated as a convex optimization on the domain of the joint distribution $P(y^-,x,y)$ \cite{Shemuel2022nost}. The capacity formula in Theorem \ref{th:cap_sy} was derived under some assumptions on the state evolution called strong connectivity \cite{Chen2005Markov}. We show here a simple derivation that utilizes the fact that the DI can be written as a sequential convex optimization problem (SCOP) \cite{Shemuel2022nost,SabagKostinaHassibi24Gaussian} also leading to relaxation of the strong connectivity assumption. We also remark that for particular POST channels, it was shown that feedback does not increase the channel capacity.

As a first step of the SCOP derivation, it is easy to show (see e.g. \cite{Chen2005Markov,Shemuel2022nost})
\begin{align}\label{eq:markov_proof}
    \max_{\CCD} \DI&= \max_{\{P(x_t|y_{t-1})\}} \sum_{i=1}^n I(X_i;Y_i|Y_{i-1}),
\end{align}
but the objective on the right-hand side is not a concave function of the input distributions $P(x_t|y_{t-1})$. Thus, it is difficult to derive a single-letter upper bound on the right-hand side of \eqref{eq:markov_proof}. 

A simple converse proof that relaxes the strong connectivity assumption utilizes the sequential form of the optimization. The idea is to exchange the maximization over the sequence of conditional distributions $P(x_t|y_{t-1})$ in \eqref{eq:markov_proof} with an optimization over the sequence of joint distributions ${\{P(x_t,y_{t-1})\}}_t$ subject to 
\begin{align}
  \sum_{x_{t+1}} P(x_{t+1},y_{t}) = \sum_{x_t,y_{t-1}} P_{Y|X,Y^-}(y_t|x_t,y_{t-1}) &P(x_t,y_{t-1}), \notag \\  & t=2,\dots,n. \label{eq:markov_proof_condition}
\end{align} 
The condition \eqref{eq:markov_proof_condition} ensures that the distribution of $P(y_t)$ is consistent whether it is computed as one-step forward of $P(x_t,y_{t-1})$ or directly as the marginal of $P(x_{t+1},y_{t})$. The objective in \eqref{eq:markov_proof} is a convex function of the new optimization variables, so Jensen's inequality can be used to conclude the single-letter objective function. For the optimization domain, it can be shown that the uniform convex combination of ${\{P(x_t,y_{t-1})\}}_{t=1}^n$ asymptotically satisfies the constraint in \eqref{eq:markov_proof_condition}, which converges to the standard stationarity constraint in Theorem~\ref{th:cap_sy}.  

Theorem \ref{th:cap_sy} can be extended to channels where the channel output depends on a bounded number, $L \ge 1$, of past channel outputs $P(y_i|x_i,y^{i-1}_{i-L})$. Indeed, as we noted in Example~\ref{ex:markov}, any channel whose state is causally available to the encoder and the decoder can be modeled as a Markov channel \cite{Chen2005Markov}. Finally, if the state is generated by a stochastic kernel $P_{S^+|Y}$, the feedback capacity can still be computed as a single-letter expression with an auxiliary random variable \cite{Shemuel2022nost}. 

\subsection{Unifilar FSCs}\label{sec:unifilar_fsc}
In the rest of this section, we focus on the scenario where the state evolution can be described by a time-invariant function. 
\begin{definition}\label{ex:unifilar}
An FSC is called \emph{unifilar} if the state evolution can be described by a time-invariant function, $f:\cS \times \cX \times \cY \to \cS$, such that 
\begin{align}
   \label{eq:def_unifilar}
s_i = f(s_{i-1},x_i,y_i).
\end{align}
\end{definition}
\noindent The unifilar FSC with feedback is a special case of the scenario in which the state is causally available to the encoder but not necessarily to the decoder. If the initial state is available to the encoder, it can compute using $(x^{i-1},y^{i-1})$ the channel states up to time $i-1$ by recursively applying the mapping \eqref{eq:def_unifilar}. Note that even though the state-evolution function $f$ is known to both the encoder and decoder, the decoder may not be able to accurately track the channel state evolution since it does not know the true input sequence.


As an example of a unifilar FSC, we introduce the Ising channel, which will serve as a running example in this section to illustrate the various techniques developed to analyze feedback capacity. 
\begin{example}\label{ex:ising}
The Ising channel was defined by Berger and Bonomi in 1990 \cite{Berger1990Ising}, as an instance of an intersymbol interference channel. The channel input, output and state alphabets are all equal to $(0,\dots,q-1)$. The channel output can be succinctly described as follows: 
\begin{align}\label{eq:def_Ising}
Y_i = \begin{cases}
    X_i & \text{ w.p.} \ \frac1{2} \\
    S_{i-1} & \text{ w.p.} \ \frac1{2}
\end{cases}
\end{align}
The new state equals the current channel input, i.e., $S_i = X_i$. This channel represents a simple scenario of memory where the previous channel input may affect the current channel input. In particular, if the current channel input is equal to the previous channel input then the channel is clean. Otherwise, the channel output is randomly chosen, and the channel input serves as the future channel state. We will see that even for this example, the solutions are not trivial due to the memory between consecutive time instances. 

Clearly, the channel is unifilar, with the function $f$ that governs the state evolution being $x_i = f(s_{i-1},x_i,y_i)$. For most of this section, we focus on the special case of the \emph{binary Ising channel} with $q=2$. A possible extension of the defined channel is where the probability in \eqref{eq:def_Ising} is not equal to $\frac1{2}$ (see \cite{Ising_artyom_IT,sabag2019graph}).
\end{example}

For unifilar FSCs, the DI in the feedback capacity can be simplified as follows \cite{Permuter2008Trapdoor,Tatikonda2009feedbakcapacity}:
\begin{theorem}
\label{th:unifilar}
The feedback capacity of a strongly connected unifilar FSC, when the initial state $s_0$ is known at the encoder and decoder, is given by
\begin{align}\label{eq:intro_cap}
\mathrm{C}^\mathrm{fb} &= \lim_{N\rightarrow \infty} \sup_{\{P(x_t|s_{t-1},y^{t-1})\}_{t=1}^N} \frac1{N} \sum_{i=1}^N I(X_i,S_{i-1};Y_i|Y^{i-1}),
\end{align}
and does not depend on the initial state $s_0$.
\end{theorem}
The encoder's perspective is reflected in the maximization domain. For general FSCs, the encoder needed to maximize over the causal conditioned channel, i.e., conditioned on all past channel inputs and outputs. However, in unifilar FSCs, the channel state, $s_{t-1}$, is available to the encoder and can replace past channel inputs both in the maximization and the objective function. This simplification is significant from a control perspective: unifilar channels admit a Markov decision process formulation of feedback capacity, as we show in the following sections. 

\begin{remark}[Delayed feedback]
The unifilar setting extends to the scenario where only delayed feedback is available to the encoder: the encoder receives the output feedback with a delay of $d$ time units. When the feedback has a delay of $d$ time units, the maximization over the DI in Theorem \ref{FSC_feedback_Capacity} is performed over $P(x^n\|y^{n-d})$ instead of over $P(x^n\|y^{n-1})$ \cite{permuter2009finite}. The special case of instantaneous feedback, i.e., $d=1$, is the one that most of the literature focuses on, with the exceptions of \cite{huleihel2023capacity,viswanathan1999capacity,yang2005feedback}. Delayed feedback is useful to obtain upper bounds on the non-feedback capacity since feedback techniques can still be applied \cite{huleihel2023capacity}.
\end{remark}

\section{Markov Decision Processes}\label{sec:MDP}
Starting from the feedback capacity expression in Theorem~\ref{th:unifilar} above, it was shown in that the feedback capacity of a strongly connected unifilar FSC, when the initial channel state $s_0$ is known at the encoder and decoder, is also expressible as 
\begin{equation}
\mathrm{C}^{\mathrm{fb}} = \sup \liminf_{n \to \infty} \frac{1}{n} \sum_{t=1}^n I(X_t,S_{t-1} ; Y_t \mid Y^{t-1}).
\label{eq:Cfb}
\end{equation}
Here, the supremum is over all choices of conditional probability distributions $P(x_t \mid s_{t-1},y^{t-1})$, $t = 1,2,3,\ldots$. The $\mathrm{C}^{\mathrm{fb}}$ expression in \eqref{eq:Cfb} can be interpreted as the optimal average reward of a Markov decision process \cite{tatikonda_phd,Tatikonda2009feedbakcapacity,Permuter2008Trapdoor}, as we describe in this section. 

We present the family of Markov decision processes (MDP) problems for an infinite-horizon and an average reward. We follow the exposition in \cite{arapostathis_etal}.

\subsection{Infinite-Horizon, Average-Reward MDP}
An instance of an infinite-horizon MDP is specified by the following elements:
\begin{itemize}
\item a space $\cZ$ of \emph{DP states}, which must be a Borel space;
\item a space $\cU$ of \emph{actions}, which must be a compact subset of a Borel space;
\item a space $\cW$ of \emph{disturbances}, which must be a measurable space;
\item a function $F:\cZ \times \cU \times \cW \to \cZ$, which specifies the evolution of MDP states $z_t$ in discrete time $t \in \ZZ_+$ according to the equation $z_t = F(z_{t-1},u_t,w_t)$, where $u_t$ and $w_t$ are the action and disturbance, respectively, at time $t$;
\item a distribution $P(z_0)$ on the initial MDP state $z_0$;
\item a conditional probability distribution $P_{W|Z,U}$, which specifies how the disturbance $w_t$ is drawn: $w_t \sim P_{W|Z,U}(w_t|z_{t-1},u_t)$;
\item actions $u_t$ chosen according to a \emph{policy} $\pi = (\mu_1,\mu_2,\mu_3,\ldots)$, such that $u_t = \mu_t(z_0,w^{t-1})$ for all $t \ge 1$;
\item a bounded reward function $g: \cZ \times \cU \to \RR$.
\end{itemize}

Note that, given the \emph{history} $(z_0,w^{t-1})$ at time $t$, the entire trajectory of MDP states $z_1,z_2,\ldots,z_{t-1}$, and all actions $u_1,u_2,\ldots,u_t$ up to (and including) time $t$ are completely determined by the policy $\pi$ and the MDP state evolution $F$. A policy is called \emph{stationary} if there is a function $\mu: \mathcal{Z} \to \mathcal{U}$ such that $\mu_t(z_0,w^{t-1}) = \mu(z_{t-1})$ for all $t \geq 1$. 

The \emph{average reward} for a given policy $\pi = {(\mu_t)}_{t \ge 1}$ is defined as 
\begin{equation}
\rho_{\pi} \triangleq  \liminf_{n \to \infty} \frac{1}{n} \EE\left[\sum_{t=0}^{n-1}g\left(Z_t,\mu_{t+1}(Z_0,W^t)\right)\right],
\label{eq:rho}
\end{equation}
where $Z_0$, $Z_t$ and $W^t$ are random variables representing, respectively, the initial MDP state, the MDP state at time $t$, and the sequence of disturbances up to time $t$. The optimal average reward $\rho^* = \sup_\pi \rho_{\pi}$ is obtained by optimizing over all policies $\pi$. 

\medskip

Comparing the expressions in \eqref{eq:Cfb} and \eqref{eq:rho}, it should come as no surprise that the feedback capacity of a strongly connected unifilar FSC can be viewed as the optimal average reward of an appropriately formulated MDP. The MDP formulation was explicitly laid out in \cite{Permuter2008Trapdoor}, and we describe the same in the following section. 

\subsection{MDP Formulation of Feedback Capacity} \label{sec:DP_Cfb}

Throughout this exposition, we assume that the initial channel state $s_0$ is arbitrary, but fixed, and is known to the encoder and decoder. Thus, in the various entities involved in the MDP formulation in this section, there is a dependence on $s_0$. However, we will typically suppress the dependence on $s_0$ from the notation we use for these entities. For example, the probability distribution $\beta_t \triangleq  P_{S_t|Y^t=y^t}$, which we will define below, has a dependence on $s_0$, which we suppress from the notation. The same is the case with the MI $I(X_t,S_{t-1};Y_t \mid \beta_{t-1},u_t)$, which is the MDP reward function.

\begin{table}
\begin{center}
\begin{tabular}{|l|l|} \hline
\ \ \ MDP element & \ \ \ \ Instantiation in $\mathrm{C}^\mathrm{fb}$ set-up \\ \hline\hline
$z_0$, initial MDP state & $\beta_0 \triangleq  \delta_{s_0}$ \\ \hline
$P_Z$, distribution on $z_0$ & the Dirac measure $\delta_{\beta_0}$ \\ \hline
$z_t$, MDP state at time $t$ & the prob.\ distrib.\ $\beta_t \triangleq  P_{S_t|Y^t=y^t}$ \\
& on the channel state at time $t$ \\ \hline
$u_t$, action at time $t$ & an $|\cS| \times |\cX|$ stochastic matrix $\Pi_t$ \\ & that describes $P_{X_t | S_{t-1},Y^{t-1}=y^{t-1}}$ \\ \hline
$w_t$, disturbance at time $t$ & $y_t$, channel output at time $t$ \\ \hline
$P_{W| Z,U}(w_t| z_{t-1},u_t)$, & a distribution $P(y_t| \beta_{t-1},\Pi_t)$ on $\cY$ \\ 
the disturbance distribution & given by Eq.~\eqref{eq:PY}  \\ \hline
$z_t = F(z_{t-1},u_t,w_t)$ & Eq.~\eqref{eq:state_evol} \\ \hline
$g(z_{t-1},u_t)$, the reward function & $I(X_t,S_{t-1};Y_t \mid \beta_{t-1},\Pi_t)$ \\ \hline
\end{tabular}
\end{center}
\caption{The MDP formulation of feedback capacity for a unifilar FSC}
\label{table:DP}
\end{table}

The MDP formulation of the feedback capacity computation problem, summarized in Table~\ref{table:DP}, involves specifying each of the elements that together constitute the MDP. The MDP state, $z_t$, at time $t$ is taken to be the channel state's belief at the decoder, that is, the probability distribution of $S_t$, given that the channel outputs up to time $t$ are observed to be $y^t$: $z_t = P_{S_t|Y^t=y^t}$. We denote this probability distribution by $\beta_t$; thus, $z_t = \beta_t \triangleq  P_{S_t|Y^t=y^t}$, and $\beta_t(s) = P_{S_t|Y^t}(s|y^t)$ for all $s \in \cS$. With this, the MDP state space $\cZ$ is the $|\cS|$-dimensional simplex of probability distributions on $\cS$. Note that the distribution $\beta_t$ depends on $y^t$ as well as the initial channel state $s_0$. The MDP state is computable at the receiver as well as at the transmitter, by virtue of the channel output feedback and the common knowledge of $s_0$. 

The initial MDP state $z_0$ corresponds to the prior distribution $\beta_0$ from which the initial channel state $s_0$ is drawn. In our set-up, $s_0$ is fixed, so $\beta_0$ is simply $\delta_{s_0}$, the probability measure that puts a mass of $1$ on $s_0$. Thus, the probability distribution $P_Z$ from which $z_0$ is drawn is the Dirac measure $\delta_{\beta_0}$ on the simplex of probability distributions on $\cS$.

Recall that, in MDPs, the action $u_t$ is a function $\mu_t$ of $(z_0,w^{t-1})$, which in the present formulation, corresponds to $(\delta_{\beta_0},y^{t-1})$. In keeping with this, an action $u_t$ in our MDP formulation is a choice of an $|\cS| \times |\cX|$ stochastic matrix $\Pi_t$ that describes the conditional distribution $P_{X_t | S_{t-1},Y^{t-1}=y^{t-1}}$. To be precise, the rows and columns of $\Pi_t$ are indexed by $\cS$ and $\cX$, respectively, and for $s \in \cS$ and $x \in \cX$, the $(s,x)$-the entry of $\Pi_t$, denoted by $\Pi_t(s,x)$, specifies the probability $P_{X_t|S_{t-1},Y^{t-1}}(x \mid s, y^{t-1})$. The space of actions $\mathcal{U}$, therefore, is the set of all $|\cS| \times |\cX|$ stochastic matrices, and for $t \ge 1$, the assignments $y^{t-1} \mapsto \Pi_t$ constitute the functions $\mu_t$ that make up the policy $\pi$. Thus, the choice of a policy $\pi$ is equivalent to a choice of the conditional distributions $P(x_t \mid s_{t-1},y^{t-1})$, $t = 1,2,3,\ldots$, over which the supremum in \eqref{eq:Cfb} is taken.

The disturbance $w_t$ is taken to be the channel output $y_t$, so that the disturbance space $\cW$ is the channel output alphabet $\cY$. The disturbance $w_t$ is drawn from a distribution $P_{W|Z,U}(w_t \mid z_{t-1},u_t)$, which in our case is the probability distribution on the output alphabet $\cY$ given, for $y_t \in \cY$, by
\begin{equation} 
\label{eq:PY}
P(y_t \mid \beta_{t-1},\Pi_t) \triangleq  \sum_{x_t,s_{t-1}} \beta_{t-1}(s_{t-1}) \, \Pi_t(s_{t-1},x_t) \, P_{Y|X,S}(y_t \mid x_t,s_{t-1}),
\end{equation}
where $P_{Y|X,S}$ arises from the channel law $P_{S^+,Y|X,S}$ --- see Section~\ref{sec:FSC_defs}.

The MDP state evolution $z_t = F(z_{t-1},u_t,w_t)$ can be derived as follows. For $s \in \cS$,
\begin{align}
\beta_t(s) &= P(S_t = s \mid Y^t = y^t) \notag \\
&= \sum_{s',x} P(S_{t-1}=s', S_t = s, X_t = x \mid Y^t = y^t) \notag  \\
&= \sum_{s',x} \frac{P(S_{t-1}=s', S_t = s, X_t = x, Y_t = y_t \mid Y^{t-1} = y^{t-1})}{P(Y_t = y_t \mid Y^{t-1} = y^{t-1})} \notag \\
&= \frac{\sum_{s',x} P(S_{t-1}=s', S_t = s, X_t = x, Y_t = y_t \mid Y^{t-1} = y^{t-1})}{\sum_{s',s'',x} P(S_{t-1}=s', S_t = s'', X_t = x, Y_t = y_t \mid Y^{t-1} = y^{t-1})} \notag \\
&= \frac{\sum_{s',x} \beta_{t-1}(s') \Pi_t(s',x) P_{S^+,Y|X,S}(s,y_t|x,s')}{\sum_{s',s'',x} \beta_{t-1}(s') \Pi_t(s',x) P_{S^+,Y|X,S}(s'',y_t|x,s')} \label{eq:state_evol} \\
&=: F_s(\beta_{t-1},\Pi_t,y_t)\nonumber
\end{align}
The state evolution function $F$ is then the collection $(F_s)_{s \in \mathcal{S}}$.
We further note that, for a unifilar channel (see Example~\ref{ex:unifilar}), the channel law $P_{S^+,Y|X,S}$ factors as
$$
P_{S^+,Y|X,S} = \mathbbm{1}\{S^+ = f(S,X,Y)\} P_{Y|X,S},
$$
which can be incorporated into Eq.~\eqref{eq:state_evol}.

Finally, we take the reward to be $I(X_t,S_{t-1};Y_t \mid Y^{t-1} = y^{t-1})$. While this appears, at first glance, to be a function of $y^{t-1}$, we argue that it depends on $y^{t-1}$ only through $\beta_{t-1}$ and $\Pi_t$, so that the reward is a function only of $(z_{t-1},u_t)$, as required in MDPs. To see this, we note that the joint distribution of $S_{t-1}$, $X_t$ and $Y_t$, given $y^{t-1}$, is expressible as 
\begin{align*}
P(s_{t-1},x_t,y_t | y^{t-1}) &= P(s_{t-1} | y^{t-1}) P(x_t | s_{t-1},y^{t-1}) P(y_t | x_t,s_{t-1},y^{t-1}) \\
&= \beta_{t-1}(s_{t-1}) \, \Pi_t(s_{t-1},x_t) \, P_{Y|X,S}(y_t \mid x_t,s_{t-1})
\end{align*}
Thus, the joint distribution of $S_{t-1}$, $X_t$ and $Y_t$, given $y^{t-1}$, is completely determined by $\beta_{t-1}$ and $\Pi_t$, and hence, so is the reward $I(X_t,S_{t-1};Y_t \mid Y^{t-1} = y^{t-1})$. We can then alternatively express the reward function as $g(\beta_{t-1},\Pi_t)$ $\triangleq  I(X_t,S_{t-1};Y_t \mid \beta_{t-1}, \Pi_t)$, which is what we have done in Table~\ref{table:DP}. 

With this as our definition of reward, the average reward $\rho_\pi$ in \eqref{eq:rho} becomes 
$$
\rho_\pi = \liminf_{n\to\infty} \frac{1}{n} \EE\left[\sum_{t=1}^n I(X_t,S_{t-1};Y_t \mid Y^{t-1})\right]
$$
and taking the supremum over all policies $\pi$ yields the optimal average reward $\rho^*$. As already noted above, a policy $\pi$ is a collection of conditional distributions $P(x_t \mid s_{t-1},y^{t-1})$, $t = 1,2,\ldots$; so comparing $\rho^*$ with the expression for $\mathrm{C}^{\mathrm{fb}}$ in \eqref{eq:Cfb}, we see that $\rho^* =\, \mathrm{C}^{\mathrm{fb}}$.

The MDP formulation of feedback capacity allows us to attack the problem of computing $\mathrm{C}^{\mathrm{fb}}$ using a variety of tools that have been developed over the years for solving MDPs, starting with the classic Bellman equation.

\subsection{The Bellman Equation}
Given a policy $\pi$, the average reward attained by $\pi$ can usually be estimated using Monte Carlo simulations. Numerical methods such as policy iteration \cite{Bertsekas05} can be used to nudge policies in a direction of increasing average rewards. The Bellman equation is then a useful tool for determining whether an average reward obtained through some such procedure is optimal. 

\begin{theorem}[Bellman equation]
If, for some $\rho \in \mathbb{R}$, there is a bounded function $h: \mathcal{Z} \to \mathbb{R}$ that satisfies the equation
\begin{equation}
    \rho + h(z) = \sup_{u \in \mathcal{U}} \biggl[g(z,u) + \int_{\mathcal{W}} h(F(z,u,w)) P_{W|Z,U}(dw | z,u) \biggr] \, \, \text{ for all } z \in \mathcal{Z},
    \label{eq:Bellman}
\end{equation}
then $\rho = \rho^*$. Moreover, if there is a function $\mu:\mathcal{Z} \to \mathcal{U}$ such that $\mu(z)$ attains the supremum in \eqref{eq:Bellman} for all $z \in \mathcal{Z}$, then the stationary policy $\pi = {(\mu_t)}_{t \ge 1}$ defined, for all $t \ge 1$, by $\mu_t(z_0,w^{t-1}) = \mu(z_{t-1})$ attains the optimal average reward, i.e., $\rho_{\pi} = \rho^*$.
\end{theorem}

The theorem above is a consequence of Theorem~6.2 in \cite{arapostathis_etal}. Its utility lies in its assertion that the optimality of a given average reward $\rho$ can be established by presenting a witness, namely, a \emph{value function} $h:\mathcal{Z} \to \mathbb{R}$ that satisfies the Bellman equation \eqref{eq:Bellman}. For the specific case of the MDP formulation of feedback capacity described in Section~\ref{sec:DP_Cfb}, a solution to the Bellman equation is a bounded value function $h(\cdot)$ that assigns a real number to each probability distribution on the channel state space $\mathcal{S}$ such that, for some $\rho \in \mathbb{R}$,
\begin{equation}
\rho + h(\beta) = \sup_{\Pi} \biggl[g(\beta,\Pi) + \sum_{y \in \mathcal{Y}} h(F(\beta,\Pi,y)) P(y | \beta,\Pi) \biggr]
\label{eq:Bellman_Cfb}    
\end{equation}
for all probability distributions $\beta$ on $\mathcal{S}$. The supremum above is taken over all $|\mathcal{S}| \times |\mathcal{X}|$ stochastic matrices $\Pi$. The real number $\rho$ in this case is the feedback capacity, $\mathrm{C}^{\mathrm{fb}}$. 

\subsection{Value Iteration} \label{sec:VI}
Explicitly finding a value function that satisfies the Bellman equation is normally a difficult task, but numerical estimates can be obtained using the method of \emph{value iteration} \cite{Bertsekas05}. To describe value iteration, it is useful to define a \emph{DP operator}, $T$, as follows: for any bounded function $h: \mathcal{Z} \to \mathbb{R}$, 
\begin{equation}
    (Th)(z) \triangleq  \sup_{u \in \mathcal{U}} \biggl[g(z,u) + \int_{\mathcal{W}} h(F(z,u,w)) P_{W|Z,U}(dw | z,u) \biggr]
\label{DP_op}
\end{equation}
for all $z \in \mathcal{Z}$.

Value iteration consists of repeated application of the Bellman operator $T$, starting with an initial bounded function $h_0: \mathcal{Z} \to \mathbb{R}$:
\begin{equation} \label{eq:VI}
    h_{k+1} = Th_{k}, \ \ k = 0,1,2,\ldots.
\end{equation}
Often, $h_0$ is simply chosen to be the identically-zero function, i.e., $h_0(z) = 0$ for all $z \in \mathcal{Z}$. Under suitable conditions (for instance, when the state, action and disturbance spaces are finite \cite{Bertsekas05}), it holds that $\lim_{k \to \infty} \frac{1}{k} h_k(z) = \rho^*$ for all $z \in \mathcal{Z}$. This implies that for large $k$, we should expect $h_k(z) \approx k \rho^*$, and hence, $h_{k+1}(z) - h_k(z) = Th_k(z) - h_k(z) \approx \rho^*$ for all $z \in \mathcal{Z}$. In other words, we may expect that for large $k$, $h_k$ is a good approximation of a solution to the Bellman equation \eqref{eq:Bellman}. Stating general conditions under which this conclusion can be rigorously drawn is beyond the scope of this monograph. Our aim is only to use the value iteration method to guide us in making informed ``guesses'' for the form taken by the solution to the Bellman equation. We illustrate this next, using the example of the Ising channel defined in Example~\ref{ex:ising}.

In the feedback capacity setting, the Bellman operator $T$ is given by the expression on the right-hand side of \eqref{eq:Bellman_Cfb}:
\begin{equation}
    (Th)(\beta) \triangleq  \sup_{\Pi} \biggl[g(\beta,\Pi) + \sum_{y \in \mathcal{Y}} h(F(\beta,\Pi,y)) P(y | \beta,\Pi) \biggr]
\label{DP_op_fb}
\end{equation}
for all probability distributions $\beta$ on the channel state space $\mathcal{S}$. We further specialize this to the case of the binary Ising channel. 

\bigskip

\textbf{Example~\ref{ex:ising} (cont.)} The MDP state at time $t$, $\beta_t$,  is a probability distribution on $\mathcal{S} = \{0,1\}$; it can be parameterized by a single real number, $z_t \triangleq  \beta_t(0) = P_{S_t|Y^t}(0 \mid y^t)$, since $\beta_t(0)+\beta_t(1) = 1$. The action at time $t$ is a $2 \times 2$ stochastic matrix $\Pi_t$, with entries $\Pi_t(s,x)$, $s,x \in \{0,1\}$. Since the matrix $\Pi_t$ can be determined from two of its entries, for example, $\Pi_t(0,0)$ and $\Pi_t(1,1)$, we will, following \cite{ElishcoPermuter2014Ising}, parametrize the action at time $t$ using the two parameters
\begin{align}
    \gamma_t &= (1-z_{t-1}) \, \Pi_t(1,1), \notag \\
    \delta_t &= z_{t-1} \, \Pi_t(0,0). \notag
\end{align}
With this, the MDP state evolution equation \eqref{eq:state_evol} can be shown to reduce to \cite[Lemma~2]{ElishcoPermuter2014Ising}
\begin{equation} \label{eq:simple_state_evol}
    z_t= \left\{ \begin{array}{ll}
1+\frac{\delta_t -z_{t-1}}{1+\delta_t -\gamma_t},& \text{if } y_t=0\\
\frac{1-z_{t-1}-\gamma_t}{1+\gamma_t -\delta_t},& \text{if } y_t=1.
\end{array} \right.
\end{equation}

The Bellman operator $T$ can also be expressed in terms of the action parameters $\gamma$ and $\delta$.
For this, let us define the binary entropy function $H_2(x) \triangleq  -x \log_2 x - (1-x) \log_2(1-x)$, for $x \in [0,1]$. Then, the Bellman operator, $T$, for the Ising channel feedback capacity problem is expressible as follows \cite[Lemma~2]{ElishcoPermuter2014Ising}: for any bounded function $h:[0,1] \to \mathbb{R}$,
\begin{align}
    (Th)(z) = \sup_{\substack{0 \leq \delta \le z, \\ 0 \le \gamma \le 1-z}} 
    \biggl[ 
    & H_2\biggl(\frac12 + \frac{\delta-\gamma}{2}\biggr) + \delta + \gamma - 1 \ + \notag \\
    & \frac{1+\delta-\gamma}{2} \, h\biggl(1 + \frac{\delta-z}{\delta+1-\gamma}\biggr) \ + \notag \\
    & \frac{1-\delta+\gamma}{2} \, h\biggl(\frac{1-z-\gamma}{1+\gamma-\delta}\biggr)
    \biggr]
\label{DP_op_Ising}
\end{align}
for all $z \in [0,1]$.

\begin{figure}[!t]
\begin{center}
\includegraphics[width=12cm]{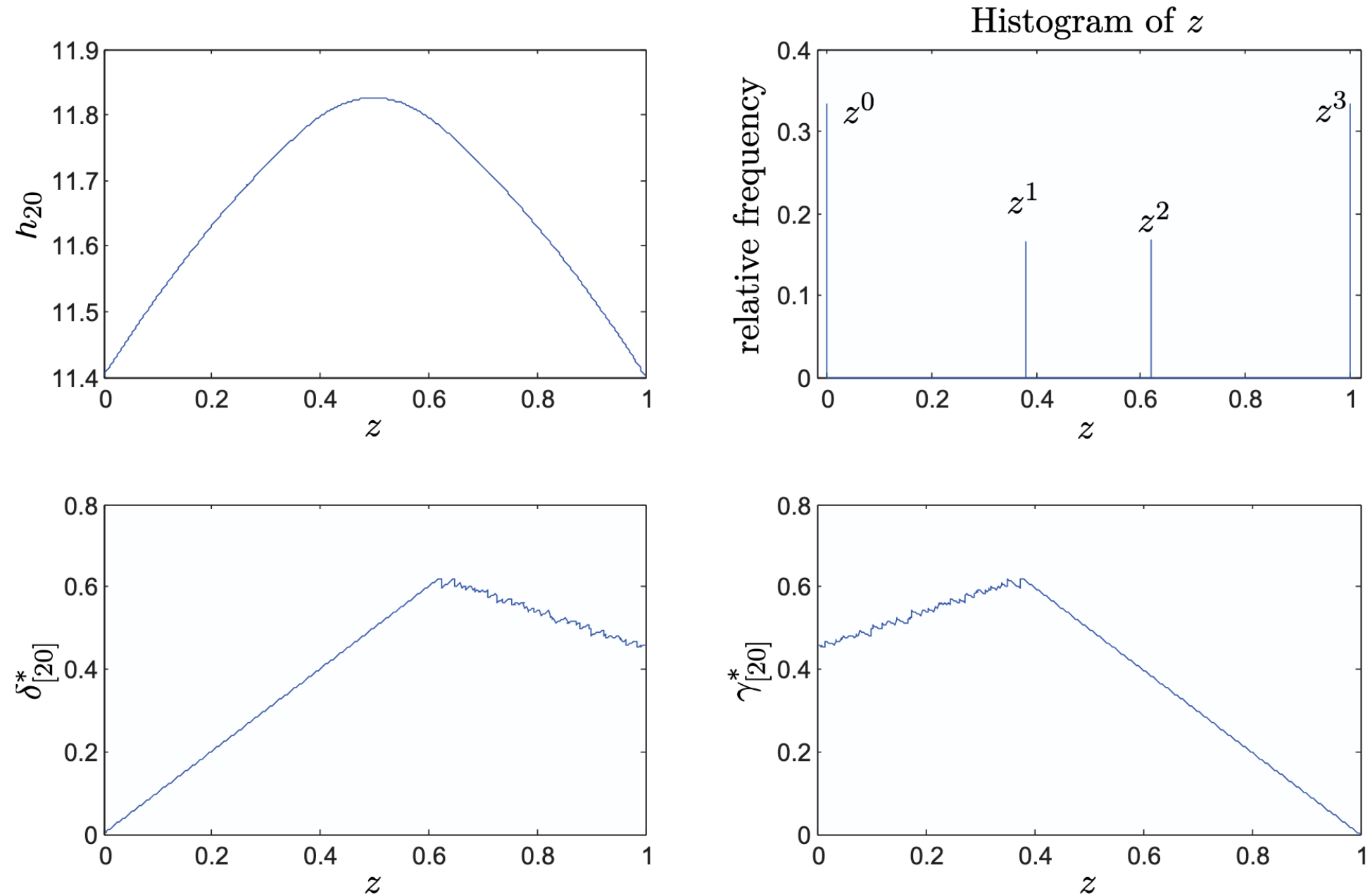}
\caption{This figure, borrowed from \cite{ElishcoPermuter2014Ising}, plots the results obtained after $20$ iterations of \eqref{DP_op_Ising}, starting with $h_0 \equiv 0$. The top-left plot shows the estimate for $h_{20}$ obtained numerically. The bottom two plots show the action parameters, $\delta^*_{[20]}$ and $\gamma^*_{[20]}$, as functions of $z$, that attain the supremum on the right-hand side of \eqref{DP_op_Ising} for $h_{20}(z) = (Th_{19})(z)$. 
The histogram on the top-right shows the (approximate) relative frequencies with which various states of the state space $\mathcal{Z} = [0,1]$ are visited under the MDP state evolution \eqref{eq:simple_state_evol}, applied with $\delta_t = \delta^*_{[20]}(z_{t-1})$ and $\gamma_t = \gamma^*_{[20]}(z_{t-1})$. The disturbance $y_t$ is sampled from the distribution in \eqref{eq:PY}.} 
\label{fig:Ising_VI}
\end{center}
\end{figure}

The feedback capacity of the binary Ising channel can be obtained by solving the Bellman equation for the corresponding MDP, but finding a solution is a highly non-trivial task. As shown in \cite{ElishcoPermuter2014Ising}, value iteration can be used to obtain a handle on the solution. This involves iteratively applying the Bellman operator in \eqref{DP_op_Ising} starting with $h_0 \equiv 0$. For $k=0,1,2,\ldots$, the iteration $h_k(z) = (Th_{k-1})(z)$ was evaluated for each value of $z \in [0,1]$. This was done by discretizing the state space $\mathcal{Z} = [0,1]$ and numerically evaluating the supremum in \eqref{DP_op_Ising}. The results obtained after $20$ iterations are shown in Figure~\ref{fig:Ising_VI}, with the caption of the figure providing a succinct summary of how the various plots were obtained.

Remarkably, based on intuition gained from Figure~\ref{fig:Ising_VI}, the authors of \cite{ElishcoPermuter2014Ising} found a closed-form solution, $h^*(z)$, for the Bellman equation. Briefly, from the shape of the plots for $\delta_{[20]}$ and $\gamma_{[20]}$, they guessed that the parameters $\delta^*$ and $\gamma^*$ for the optimal action must be piecewise linear functions of $z$, with $\gamma^*(z) = \delta^*(1-z)$. Also, from the symmetric, concave form of $h_{20}$ reminiscent of the plot of the binary entropy function, they guessed that the Bellman solution $h^*$ may arise from the binary entropy function $H_2$. The crucial insight was obtained from the histogram in Figure~\ref{fig:Ising_VI}: it seemed that, under the optimal policy, the MDP state evolution eventually settled into trajectories that only visited $4$ MDP states (labelled $z^0,z^1,z^2,z^3$) out of the entire state space $\mathcal{Z} = [0,1]$. These insights were pooled together to obtain an explicit function $h^*(z)$ that could be analytically verified to solve the Bellman equation \cite[Lemmas~3 and 4]{ElishcoPermuter2014Ising}:
\begin{equation}
\label{Ising_hstar}
h^*(z)= \begin{cases}
\eta(1-z), & \text{if }  z \in [0,\frac{1-a}{1+a}] \\
H_2(z), & \text{if } z\in [\frac{1-a}{1+a},\frac{2a}{1+a}] \\
\eta(z), & \text{if } z\in [\frac{2a}{1+a},1],
\end{cases}
\end{equation}
where $a\approx 0.4503$ is the only solution in $[0,1]$ to the equation $x^3 = (1-x)^4$, 
the function $\eta(z)$ is given by 
$$\textstyle \left( \frac{1}{1-a} \right) H_2\left( \frac{2a+(1-a)z}{2} \right) -z - \frac{4a+(1-a)z}{2(1-a)}\rho^* + \frac{2a+(1-a)z}{2(1-a)}H_2\left( \frac{2a}{2a + (1-a)z}\right),$$
and\footnote{In \cite{ElishcoPermuter2014Ising}, $\rho^*$ is expressed as  $\frac{2H_2(a)}{3+a}$, but it is straightforward to check, using $a^3 = (1-a)^4$, that this simplifies to $\rho^* = -\frac12 \log_2 a$.}
$$
\rho^* = -\frac12 \log_2 a \approx 0.5755.
$$
It can then be shown \cite[Theorem~1]{ElishcoPermuter2014Ising} that $(Th^*)(z) = h^*(z) + \rho^*$ for all $z \in [0,1]$, from which we infer that the feedback capacity of the binary Ising channel is equal to $\rho^* \approx 0.5755$.

\bigskip

As noted in the above example, in the case of the MDP associated with the binary Ising channel feedback capacity, all sample paths of the MDP state evolution under the optimal policy eventually ended up in a finite set (in this case, $4$ in number) of sink states. This phenomenon played a crucial role in finding an explicit solution to the Bellman equation. 
This behaviour of sample paths settling into a finite set of sink states is highly atypical for the Markov decision processes that arise in general stochastic control settings. But it seems to be a feature of the MDPs that are associated with the feedback capacity problems that we can solve exactly --- see \cite{Permuter2008Trapdoor,SabagPermuterKashyap2016BEC,Sabag_BIBO_IT,Ising_artyom_IT,trapdoor_generalized,ElishcoPermuter2014Ising,sabag2019graph,sabag2017single}. As these works show, identifying a finite set of sink states under the optimal policy also allows us to design simple capacity-achieving feedback coding schemes.

Nonetheless, as the above example amply illustrates, it is extremely challenging to exactly solve the Bellman equation for average-reward scenarios, and in particular, when the reward depends on a belief (i.e., a posterior probability distribution) on the channel states, as in the Ising channel case. This motivates us to find other, more indirect ways, of arriving at the feedback capacity. In the following sections, our focus is on methods to bound the feedback capacity from above and below. In some cases, we can get the upper and lower bounds to match, which results in an exact determination of the feedback capacity.

\section{Capacity Bounds via Q-Graphs}\label{sec:Qgraph}
In this section, we present a methodology to compute simple bounds on the feedback capacity of unifilar FSCs. The challenge in optimizing the DI for channels with memory, including FSCs, is that the memory of the channel implies that the solution has memory as well. In particular, the optimal channel input and output processes have memory. Indeed, even if one considers memoryless (i.i.d.) channel inputs, the channel output process will still have memory due to the dynamical behavior of the channel state. 
In what follows, we impose particular structures of memory both on the channel inputs and outputs that result in computable upper and lower bounds on the channel capacity.

The structure that we will impose is inspired by the solutions to the Bellman equation for a wide range of channels, including the Trapdoor and Ising channels, and the Binary Erasure and Binary Symmetric channels with input constraints \cite{SabagPermuterKashyap2016BEC,Sabag_BIBO_IT,Ising_artyom_IT,trapdoor_generalized,Permuter2008Trapdoor,ElishcoPermuter2014Ising}. In all cases of channels where the associated Bellman equation for the feedback capacity MDP has been explicitly solved, the unique optimal output distribution is a variable-order Markov process (also called a hidden Markov model) \cite{sabag2017single}. That is, there exists a finite, directed graph whose labeled edges can describe the optimal channel output process. We proceed to show how to leverage this insight to obtain useful bounds that can eliminate the need to find an explicit solution to the Bellman equation.

Recall that the DI for FSCs can be written as 
\begin{align} \label{eq:DI-FSC}
    \sum_{i=1}^n I(X_i,S_{i-1};Y_i|Y^{i-1})&= \sum_{i=1}^n H(Y_i|Y^{i-1}) - H(Y_i|X_i,S_{i-1}).
\end{align}
The main challenge in translating such expressions into a single-letter expression is the dependence of the channel output $Y_i$ on \emph{all} previous channel outputs $Y^{i-1}$. An idea that was found to work well in tackling this challenge is a \emph{quantization} of the past outputs into \emph{finitely} many bins. Specifically, one can leverage the following simple and general bound:
\begin{align}\label{eq:Q_derivation2}
H(Y^n) = \sum_{i=1}^n H(Y_i|Y^{i-1}) &= \sum_{i=1}^n H(Y_i|Y^{i-1}, \Phi_{i-1}(Y^{i-1}))\nonumber\\
&\le \sum_{i=1}^n H(Y_i|\Phi_{i-1}(Y^{i-1})),
\end{align}
the inequality clearly holding for any sequence of deterministic mappings $\{\Phi_i\}_{i\ge1}$. The choice of mappings $\Phi_i$, $i=1,2,3,\ldots$, from $\mathcal{Y}^i$ to a set gives us a means of imposing a structure on the output process, and a careful choice of these mappings results in tight bounds.

For example, a tight upper bound for memoryless channels can be obtained by choosing $\Phi_{i}(\cdot)$ as a constant function, so that \eqref{eq:Q_derivation2} becomes $H(Y^n) \le \sum_i H(Y_i)$. For general FSCs, though, this choice will not result in a tight upper bound since the optimal output process has memory. Another useful choice of mapping is $\Phi_{i-1}(y^{i-1}) = y_{i-1}$, which imposes a Markov structure. It turns out, however, that a more general notion of a variable-order Markov structure is required to achieve tight upper bounds. This is described in terms of the $Q$-graphs that we introduce next.

\subsection{$Q$-Graphs and the $Q$-Graph Upper Bound}\label{subsec:Qgraph}
\begin{figure}
    \centering
\centering
    \psfrag{Q}[][][1]{$y=1$}
    \psfrag{E}[][][1]{$y=0/?$}
    \psfrag{F}[][][1]{$y=?$}
    \psfrag{O}[][][1]{$y=0/?/1$}
    \psfrag{L}[][][1]{$q=2$}
    \psfrag{H}[][][1]{$q=1$}
    \includegraphics[scale = 1]{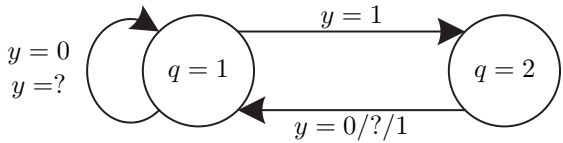}
    \caption{An example of a $Q$-graph with $|\mathcal{Q}|=2$, and $\mathcal{Y}=\{0,1,?\}$. Each node $q$ has three outgoing edges, which are labelled by the distinct symbols from $\mathcal Y$. In the figure, the self-loop shown at $q=1$ is a compact representation of two parallel edges, one labeled `$0$' and the other labeled `$?$'. Similarly, the edge depicted from $q=2$ to $q=1$ represents three parallel edges, each labelled with a different symbol from $\mathcal{Y}$.}
    \label{fig:Q_graph_example}
 \end{figure}

A $Q$-graph is a directed, irreducible\footnote{A directed graph is \emph{irreducible} if for each ordered pair of its vertices $(u,v)$, there is a directed path from $u$ to $v$.}, edge-labeled graph of constant out-degree $|\mathcal{Y}|$ that is used to define a sequence of mappings $\{\Phi_i\}_{i \ge 1}$ as discussed in the previous subsection. Specifically, the vertex set, $\mathcal{Q}$, of a $Q$-graph is a finite set, and each vertex $q \in \mathcal{Q}$ has $\mathcal{Y}$ outgoing edges, with distinct outgoing edges from $q$ being labeled by distinct symbols from the channel output alphabet $\mathcal Y$. An example of a $Q$-graph is shown in Fig.~\ref{fig:Q_graph_example}. A $Q$-graph should be viewed as a graphical representation of a mapping $\Phi:\mathcal{Q} \times \mathcal{Y} \to \mathcal{Q}$, where $\Phi(q,y) = q'$ iff there is an edge labeled $y$ from $q$ to $q'$. The mappings (binnings) $\Phi_i: \mathcal{Y}^i \to \mathcal{Q}$ are then recursively defined as follows: 
$$
\Phi_i(y^i) = \Phi\bigl(\Phi_{i-1}(y^{i-1}),y_i\bigr), \text{ for all } i \ge 1,
$$
initialized by $\Phi_0(\varepsilon) = q_0$, where $\varepsilon$ is the empty string and $q_0 \in \mathcal{Q}$ is a distinguished initial vertex of the $Q$-graph. Thus, a $Q$-graph is indeed a ``$Q$-quantization'' of the set of all finite-length channel output sequences into a finite set, $\mathcal Q$.

\begin{example}
    Consider the $Q$-graph shown in Fig.~\ref{fig:Q_graph_example}. The graph has two vertices, marked as $1$ and $2$, which represent the two quantization bins into which finite-length sequences over the alphabet $\cY = \{0,1,?\}$ are mapped. The self-loop at vertex $1$ shown in the figure represents two parallel self-loops at that vertex --- one labeled `0' and the other labeled `1'. Similarly, the edge shown from vertex 2 to vertex 1 represents three parallel edges from vertex 2 to vertex 1, each labeled by a distinct symbol from $\cY$. 
    
    Now, fix $q_0$ to be one of the two vertices; this is the distinguished initial vertex from which walks on the graph always start. Each walk along the directed edges of the $Q$-graph generates a sequence of symbols from $\cY$ obtained by concatenating the labels of the edges along that walk. For each $y^t \in \cY^t$, there is a unique walk starting at the vertex $q_0$, whose edge labels form the sequence $y^t$. The vertex at which this walk terminates is the quantization bin for $y^t$, which in our notation is written as $\Phi_t(y^t)$. For instance, starting at $q_0 = 1$, the walk on the graph that generates the sequence $y^{10} = (0,1,?,?,1,?,0,1,1,1)$ ends at vertex $2$, so $\Phi_{10}(y^{10}) = 2$.
\end{example}

\bigskip
We use $Q$-graphs to simplify the description of input distributions that an encoder may follow. Given a $Q$-graph (along with a distinguished initial vertex $q_0$) or each $(s,q) \in \mathcal{S} \times \mathcal{Q}$, we specify distributions ${\bigl(P(x | s,q)\bigr)}_{x \in \mathcal{X}}$, from which we define
\begin{equation} \label{eq:Phi_i}
P(x_t \mid s_{t-1},y^{t-1}) \triangleq  P(x_t \mid s_{t-1},q_{t-1}) \text{ with } q_{t-1} = \Phi_{t-1}(y^{t-1}).
\end{equation}
We collectively refer to the distributions ${\bigl(P(x | s,q)\bigr)}_{x \in \mathcal{X}}$ as an input distribution. Any such input distribution defines a Markov chain on $\mathcal{S} \times \mathcal{Q}$, having the transition probability kernel 
\begin{align} \label{SQ_Markov} 
  P(s',q' \mid s,q) &= \sum_{y: \Phi(q,y) = q'} \sum_{x: f(s,x,y)=s'} P(x|s,q) P(y|x,s).
\end{align}
Under some technical conditions on $P(x|s,q)$, which we omit for brevity (see \cite{sabag2017single}), this Markov chain admits a unique stationary distribution $\pi(s,q)$. We will always choose input distributions $P(x|s,q)$ that induce Markov chains with this property. We are now in a position to state our $Q$-graph-based upper bound on feedback capacity.



\begin{theorem}\label{th:Q_UB}
Given any $Q$-graph, the feedback capacity is bounded by 
\begin{align}\label{eq:th_Q}
\mathrm{C}^\mathrm{fb}\leq \sup_{P(x|s,q)}I(X,S;Y|Q)
\end{align}
and the joint distribution of $(S,Q,X,Y)$ decomposes as $P(s,q,x,y) = \pi(s,q) P(x|s,q) P(y|x,s)$.
\end{theorem}

The derivation of the upper bound follows from the steps in \eqref{eq:Q_derivation2} by choosing the sequence of mappings to be described by a $Q$-graph. 
\begin{align*}
\sum_{i=1}^n I(X_i,S_{i-1};Y_i|Y^{i-1})&\stackrel{(a)}\leq \sum_{i=1}^n H(Y_i|\Phi_{i-1}(Y^{i-1}))- H(Y_i|X_i,S_{i-1})\\
&\stackrel{(b)}{=} \sum_{i=1}^n I(X_i,S_{i-1};Y_i|Q_{i-1})\\
&\stackrel{(c)}{\le} \max_{\{P(x_t|s_{t-1},q_{t-1})\}_{t=1}^n} \sum_{i=1}^n I(X_i,S_{i-1};Y_i|Q_{i-1}),
\end{align*}
where $(a)$ follows from \eqref{eq:DI-FSC} and \eqref{eq:Q_derivation2}, $(b)$ holds because we define $Q_{i-1} \triangleq  \Phi_{i-1}(Y^{i-1})$, and $(c)$ follows from the fact that the conditional distributions ${\{P(x_t|s_{t-1},q_{t-1})\}}_{t=1}^n$ uniquely determine the objective function being maximized (proof omitted). The remaining step of the proof is to transform the $n$-letter upper bound to the single-letter upper bound of Theorem \ref{th:Q_UB} --- see \cite{sabag2017single}.

The advantage of the upper bound in Theorem~\ref{th:Q_UB} compared to solving the Bellman equation in \eqref{eq:Bellman_Cfb} lies in its computability. To solve the Bellman equation, we are faced with the challenging task of finding a value function $h(\beta)$ defined on the continuous MDP state space of all possible probability distributions $\beta$ on the channel state alphabet $\mathcal{S}$. On the other hand, as we shall see in Theorem~\ref{theorem:UB_convex}, the $Q$-graph upper bound can be formulated as a convex optimization problem, which can be solved using standard solvers. 

Theorem~\ref{th:Q_UB} shows that each choice of $Q$-graph leads to an upper bound on the feedback capacity. This leads naturally to the question of how to choose $Q$-graphs that result in good upper bounds, which we discuss at the end of this section. Indeed, if one can show that there always exists an optimal $Q$-graph that achieves the bound in \eqref{eq:th_Q} with equality, then minimizing the upper bound over all possible $Q$-graphs would transform \eqref{eq:th_Q} into a min-max single-letter capacity formula for the feedback capacity of unifilar FSCs.

We illustrate the efficacy of the upper bound of Theorem~\ref{th:Q_UB} by applying it to our running example of the Ising channel.

\begin{figure}
    \centering
    \includegraphics[scale=1, trim=0 0 0 0, clip]{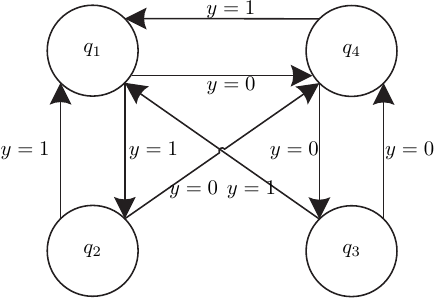}
    \caption{A $Q$-graph for the Ising channel with $\mathcal{Y}=\{0,1\}$, denoted by $Q_1$.}
    \label{fig:Q_Ising}
\end{figure}

\textbf{Example~\ref{ex:ising} (cont.)} 
    For the binary Ising channel, using the $Q$-graph in Fig.~\ref{fig:Q_Ising}, the upper bound in Theorem \ref{th:Q_UB} can be computed as 
    \begin{align} \label{ex:Ising_Q_upbnd}
    \mathrm{C}^\mathrm{fb} \le \frac{2H_2(a)}{a+3} = -0.5\log_2(a),
    \end{align}
    where $a$ is the solution in $[0,1]$ for the equation $a^3 = (1-a)^4$.
Comparing with the exact feedback capacity of the binary Ising channel derived in Section~\ref{sec:VI}, we see that the upper bound of Theorem~\ref{th:Q_UB} provides a tight bound on the feedback capacity of this channel. The computation of the bound in \eqref{ex:Ising_Q_upbnd} requires explicitly determining the supremum in \eqref{eq:th_Q}. This can be done by formulating the problem as a convex optimization problem, as summarized in Theorem~\ref{theorem:UB_convex} below, and analytically solving the Karush-Kuhn-Tucker (KKT) conditions. We omit the technical details.


\subsubsection{Computing the bound via convex optimization}
The feedback capacity upper bound in \eqref{eq:th_Q} is not a concave function of $P(x|s,q)$. Thus, a direct computation of the bound is feasible only for small alphabets. In this section, we show that the bound can be formulated as a convex optimization problem, which makes its efficient computation possible even for channels with large alphabets. This formulation also allows us to use the Karush-Kuhn-Tucker (KKT) conditions to compute the bound analytically for specific examples.

The main idea in making the upper bound a convex optimization problem is to enlarge the optimization domain from the space of conditional distributions $P(x|s,q)$ to the space of joint distributions $P_{S,Q,X,Y} \in \mathcal{P}(\cS\times \cQ\times \cX\times \cY)$. While this makes the objective function a concave function of the optimization variables, it also requires adding some new constraints. These new constraints are needed to ensure that optimizing the joint distribution aligns with maximizing the upper bound as stated in Theorem \ref{th:Q_UB}.

\textbf{1. Stationarity constraint:} 
Let $(S,Q)$ and $(S^+,Q^+)$ respectively denote the current and next states of the Markov chain on $\mathcal{S} \times \mathcal{Q}$ defined by \eqref{SQ_Markov}. Theorem~\ref{th:Q_UB} specifies a joint distribution on $(S,Q,X,Y)$, from which we obtain, via \eqref{SQ_Markov}, the next-state distribution
\begin{align}\label{eq:constraint_original}
  &P_{S^+,Q^+}(s',q') \nn\\
  &= \sum_{s,q,x,y}P_{S,Q,X,Y}(s,q,x,y)\mathbbm{1}\{s' = f(s,x,y)\}\mathbbm{1}\{q' = \Phi(q,y)\}.
\end{align}
Stationarity means that $P_{S^+,Q^+}$ is identical to the marginal onto $(S,Q)$ of the joint distribution $P_{S,Q,X,Y}$. To capture this, we define a linear operator, $\Lambda_{\mathrm{st}}$, that maps probability distributions $P_{S,Q,X,Y} \in \mathcal{P}(\cS\times \cQ\times \cX\times \cY)$ to a real-valued function on $\mathcal{S} \times \mathcal{Q}$ given by
$$
(s,q) \ \longmapsto \ P_{S^+,Q^+}(s,q) - \sum_{x,y} P_{S,Q,X,Y}(s,q,x,y).
$$
With this, the stationarity requirement is equivalent to mandating that $\Lambda_{\mathrm{st}}(P_{S,Q,X,Y}) = \mathbf{0}$, i.e., that the image of $P_{S,Q,X,Y}$ under the operator $\Lambda_{\mathrm{st}}$ is the identically-zero function on $\mathcal{S} \times \mathcal{Q}$. 


\textbf{2. Channel law constraint:} We next enforce the requirement that the joint distribution $P_{S,Q,X,Y}$ is consistent with the channel law $P_{Y|X,S}$ and satisfies the Markov chain $Y-(X,S)-Q$. That is, $P(y|x,s) = \frac{ P_{S,Q,X,Y}(s,q,x,y)}{\sum_{y'}P_{S,Q,X,Y}(s,q,x,y')}$. Define a linear operator, $\Lambda_{\mathrm{ch}}$, that maps $P_{S,Q,X,Y} \in \mathcal{P}(\cS\times \cQ\times \cX\times \cY)$ to a real-valued function on $\mathcal{S} \times \mathcal{Q} \times \mathcal{X} \times \mathcal{Y}$ given by
$$
(s,q,x,y) \ \longmapsto \ P_{S,Q,X,Y}(s,q,x,y) - P(y|x,s) \cdot \sum_{y'}P_{S,Q,X,Y}(s,q,x,y'),
$$
(Note that the coefficients $P(y|x,s)$ are specified by the channel law.) With this, the channel law and Markov chain requirement on $P_{S,Q,X,Y}$ is enforced by setting $\Lambda_{\mathrm{ch}}(P_{S,Q,X,Y}) = \mathbf{0}$.

The new formulation for the upper bound in Theorem \ref{th:Q_UB} can be summarized as the following optimization problem.
\begin{empheq}[box=\fbox]{align}
\label{opt:UB} 
& \underset{P_{S,Q,X,Y}\in \Delta}{\text{minimize}} & f(P_{S,Q,X,Y}) \triangleq  -I(X,S;Y|Q)\nn\\
& \text{subject to} &\Lambda_{\mathrm{st}}(P_{S,Q,X,Y}) = \mathbf{0}, \ \Lambda_{\mathrm{ch}}(P_{S,Q,X,Y}) & = \mathbf{0}.
\end{empheq}


\bigskip

It is straightforward to check that \eqref{opt:UB} is a convex optimization problem.
\begin{theorem}[\cite{sabag2019graph}]\label{theorem:UB_convex}
For any $Q$-graph, the problem in \eqref{opt:UB} is a convex optimization problem. That is, $f(P_{S,Q,X,Y})$ is a convex function, while $\Lambda_{\mathrm{st}}$ and $\Lambda_{\mathrm{ch}}$ are linear functions of $P_{S,Q,X,Y}$.
\end{theorem}

The following section provides a sufficient condition for the optimality of the $Q$-graph upper bound.

\subsection{The $Q$-Graph Lower Bound}\label{subsec:Qgraph-lobnd}
The idea of the $Q$-graph lower bound parallels that of the upper bound in that it enforces a structured distribution on the channel inputs, and hence, on the channel outputs. As in Section~\ref{subsec:Qgraph}, we consider input distributions $P(x|s,q)$ that are defined via a $Q$-graph. But we now enforce an additional condition on $P(x|s,q)$ to ensure that the distribution induced on the channel outputs admits the same $Q$-graph structure. 

The additional condition we impose is related to the posterior distribution (often termed as the set of \emph{beliefs}) ${\bigl(P(s_t | y^t)\bigr)}_{s_t \in \mathcal{S}}$ of channel states at the decoder. Using Bayes' rule, the belief $P(s_t | y^t)$ can be written recursively as
\begin{align}\label{eq:belief_update}
P(s_t|y^t)&= \frac{\sum_{x_{t},s_{t-1}} P(s_t,y_t,x_{t},s_{t-1}|y^{t-1})}{\sum_{s'_t,x_{t},s_{t-1}} P(s'_t,y_t,x_{t},s_{t-1}|y^{t-1})}.
\end{align}
This equation, known as the Bahl–Cocke–Jelinek–Raviv (BCJR) forward equation in channel coding, is also the MDP state evolution in \eqref{eq:state_evol}.

For a given choice of $Q$-graph and input distribution $P(x|s,q)$, the BCJR equation \eqref{eq:belief_update} is expressible as
\begin{equation} \label{eq:BCJR}
P(s_t | y^t) = \frac{\sum_{x_t,s_{t-1}} P(s_{t-1}|y^{t-1}) P(x_t|s_{t-1},q_{t-1}) P(s_t,y_t \mid x_t,s_{t-1})} {\sum_{s'_t,x_t,s_{t-1}} P(s_{t-1}|y^{t-1}) P(x_t|s_{t-1},q_{t-1}) P(s'_t,y_t \mid x_t,s_{t-1})}
\end{equation}
where, again, $q_{t-1} = \Phi_{t-1}(y^{t-1})$ is obtained from the given $Q$-graph. 
We view \eqref{eq:BCJR} as a forward-recursive equation: given the set of beliefs at time $t-1$, ${\bigl(P(s_{t-1}|y^{t-1})\bigr)}_{s_{t-1}\in\mathcal{S}}$, and a new channel output $y_t$, we can compute the set of beliefs at time $t$ ${\bigl(P(s_{t}|y^{t})\bigr)}_{s_t\in\mathcal{S}}$. In other words, for any fixed choice of input distribution $P(x | s,q)$, the BCJR equation \eqref{eq:BCJR} defines a mapping $\overline{F}:\mathcal P(\mathcal S)\mathcal \times\mathcal Y\to \mathcal{P}(\mathcal{S})$.

Recall from Section~\ref{subsec:Qgraph} that the input distribution $P(x | s,q)$ is always chosen so that the induced Markov chain on $\mathcal{S} \times \mathcal{Q}$ (see \eqref{SQ_Markov}) has a unique stationary distribution $\pi(s,q)$. Let $\pi_{S|Q=q}$ denote the conditional distribution on $\mathcal{S}$ given by $\pi_{S|Q=q}(s) \triangleq  \frac{\pi(s,q)}{\sum_{\tilde{s}} \pi(\tilde{s},q)}$. 
We say that an input distribution $P(x|s,q)$ is \emph{BCJR-invariant} if the corresponding set of conditional distributions $\{\pi_{S|Q=q}\}_{q \in \mathcal{Q}}$ satisfies
\begin{align}\label{eq:BCJR-inv}
    \overline{F}(\pi_{S|Q=q},y) = \pi_{S|Q=\Phi(q,y)}
\end{align}
for all $q \in \mathcal{Q}$ and $y \in \mathcal{Y}$. Here, as usual, $\Phi(q,y)$ is the $Q$-graph node obtained by following the edge labelled $y$ that leaves the node $q$.

An alternative definition for a BCJR-invariant input distribution is that it induces the Markov chain $$S^+-Q^+-(Q,Y).$$ Intuitively, a BCJR-invariant input implies that the belief of the state given the channel outputs depends only on the binning of the channel outputs, as specified by the given $Q$-graph.

We can now state the $Q$-graph lower bound on the feedback capacity of a unifilar channel.
\begin{theorem}[\cite{sabag2017single}, Theorem~3] \label{th:Q_LB}
For any $Q$-graph and a BCJR-invariant input distribution $P(x|s,q)$, the feedback capacity of a unifilar FSC is bounded below by
\begin{equation} \label{eq:Theorem_Lower}
\mathrm{C}^\mathrm{fb} \geq I(X,S;Y|Q),
\end{equation}
where $(S,Q,X,Y)$ follows the joint distribution specified in Theorem~\ref{th:Q_UB}. 
\end{theorem}
The idea of the proof is to show that a BCJR-invariant input distribution induces the Markov chain $Y_t - Q_{t-1} - Y^{t-1}$ for all $t$. This is then used to show that the expression 
$$\liminf_{n \to \infty} \frac{1}{n} \sum_{t=1}^n I(X_t,S_{t-1} ; Y_t \mid Y^{t-1})$$
on the right-hand side of the feedback capacity characterization in \eqref{eq:Cfb} reduces to 
$I(X, S; Y |Q)$. We refer the reader to \cite[Appendix~B]{sabag2017single} for the details.

Combining Theorems~\ref{th:Q_UB} and \ref{th:Q_LB}, we obtain the following useful corollary, which gives a sufficient condition for the $Q$-graph upper bound to be tight.
\begin{corollary} \label{cor:Q_bnds}
If, for some $Q$-graph, the supremum in $\sup\limits_{P(x|s,q)} I(X,S;Y|Q)$
is attained by a BCJR-invariant input distribution $P^*(x|s,q)$, then 
$$
\mathrm{C}^\mathrm{fb} = I(X,S;Y|Q),
$$
where the joint distribution on $(S,Q,X,Y)$ is given by $P(s,q,x,y) = \pi(s,q) P^*(x|s,q) P(y|x,s)$.
\end{corollary}

The typical way in which we use Corollary~\ref{cor:Q_bnds} is by first evaluating, for a given $Q$-graph, the upper bound $\sup I(X,S;Y|Q)$, and then checking if the input distribution $P(x|s,q)$ attaining the supremum is BCJR-invariant. Note that the upper bound can be evaluated by formulating it as a convex optimization problem, as summarized in Theorem~\ref{theorem:UB_convex}.
This strategy works, for example, for the binary Ising channel, for which it can be shown that the maximizing input distribution on the $Q$-graph of Figure~\ref{fig:Q_Ising} is BCJR-invariant.
We emphasize, however, that the BCJR-invariance condition is not necessary for the optimality of the $Q$-graph upper bound; there are instances where the upper bound is tight, but the BCJR-invariant property is not satisfied by the input distribution attaining the supremum. We refer the reader, for example, to the solution of the Trapdoor channel with delayed feedback in \cite{huleihel2023capacity}.

The lower bound of Theorem~\ref{th:Q_LB} is useful in its own right, as it is relatively straightforward to find a BCJR-invariant input distribution (if such a distribution exists) for a given $Q$-graph, and to subsequently evaluate $I(X,S;Y|Q)$. One can thus evaluate, at least numerically, the lower bound on many different $Q$-graphs, and pick the best one. In practice, one often chooses $Q$-graphs that correspond to state transition diagrams of $m$-th order Markov chains on the channel output alphabet $\mathcal{Y}$, for $m = 1,2,3,\ldots$. These are also known as de Bruijn graphs of order $m$.\footnote{A de Bruijn graph of order $m$ on the symbol alphabet $\mathcal{Y}$ is an edge-labeled directed graph on the vertex set $\mathcal{Y}^m$. There is a directed edge from vertex $y_1y_2\ldots y_m$ to vertex $y'_1y'_2\ldots y'_m$ iff $y_2 \ldots y_m = y'_1 \ldots y'_{m-1}$; this edge is labeled by $y'_m$.} As we increase the value of $m$, the resulting $Q$-graphs will typically yield upper and lower bounds (via Theorems~\ref{th:Q_UB} and \ref{th:Q_LB}, respectively) that get progressively closer to one another. In Section~\ref{subsec:Q-exploration}, we make additional useful observations on how to choose good $Q$-graphs. 
A BCJR-invariant input distribution also yields a simple and explicit coding scheme, which we describe in the following section. 

\subsection{A Coding Scheme}
Any BCJR-invariant input distribution gives rise to a construction of an explicit posterior matching coding scheme. We present the coding scheme construction with an informal statement of its achievable rate and discuss the missing technical details.

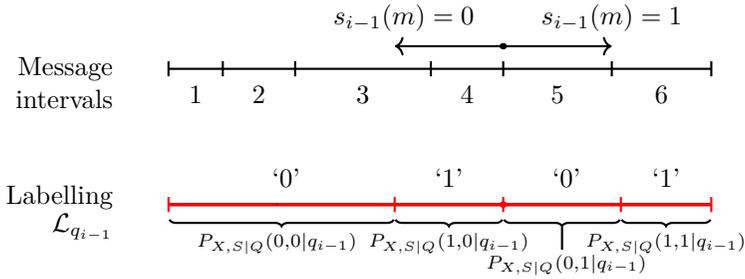
\begin{figure}[ht]
  \centering

\begin{tikzpicture}[every node/.style={font=\small}, xscale=1.2, yscale=1]
\usetikzlibrary{snakes}
  \def\yA{0}
  \def\yC{-1.8}

  \node[left] at (-0.5,\yA) {Message};
  \node[left] at (-0.5,\yA-0.4) {intervals};
  \node[left] at (-0.5,\yC+0.1) {Labelling};
  \node[left] at (-0.5,\yC-0.3) {$\mathcal{L}_{q_{i-1}}$};

  \draw[thick] (0,\yA) -- (6,\yA);
  \foreach \xstart/\xend/\lab in {
    0/0.6/1,
    0.6/1.4/2,
    1.4/2.9/3,
    2.9/3.7/4,
    3.7/4.9/5,
    4.9/6/6
  } {
    \draw[thick] (\xstart,\yA-0.1) -- (\xstart,\yA+0.1);
    \node at ({(\xstart+\xend)/2},\yA-0.35) {\lab};
  }
  \draw[thick] (6,\yA-0.1) -- (6,\yA+0.1);

\draw[<->, thick] (2.5,\yA+0.3) -- (4.9,\yA+0.3);
\filldraw[black] (3.7,\yA+0.3) circle (1pt);
\node[above=2pt] at (2.6,\yA+0.3) {$s_{i-1}(m) = 0$};
\node[above=2pt] at (4.9,\yA+0.3) {$s_{i-1}(m) = 1$};

\draw[very thick] (0,\yC) -- (6,\yC);

\draw[very thick, red] (0,\yC) -- (1.2,\yC);
\draw[very thick, red] (1.2,\yC) -- (3.7,\yC);
\draw[very thick, red] (3.7,\yC) -- (5,\yC);
\draw[very thick, red] (5,\yC) -- (6,\yC);

\filldraw[red] (3.7,\yC) circle (1pt);

\node at (1.3,\yC+0.35) {`0'};
\node at (3.1,\yC+0.35) {`1'};
\node at (4.4,\yC+0.35) {`0'};
\node at (5.5,\yC+0.35) {`1'};

\foreach \x in {0, 2.5, 3.7, 5, 6} {
  \draw[thick, red] (\x,\yC-0.1) -- (\x,\yC+0.1);
}

\draw[snake=brace, mirror snake, thick] (0.01,\yC-0.2) -- (2.49,\yC-0.2) ;
\draw[snake=brace, mirror snake, thick] (2.51,\yC-0.2) -- (3.69,\yC-0.2) ;
\draw[snake=brace, mirror snake, thick] (3.71,\yC-0.2) -- (4.99,\yC-0.2) ;
\draw[snake=brace, mirror snake, thick] (5.01,\yC-0.2) -- (5.99,\yC-0.2) ;
\node[below=7pt] at (1.2,\yC) {\scriptsize $P_{X,S|Q}(0,\! 0 | q_{i-1})$};
\node[below=7pt] at (3.1,\yC) {\scriptsize $P_{X,S|Q}(1,\! 0 | q_{i-1})$};
\node[below=15pt] at (4.4,\yC) {\scriptsize $P_{X,S|Q}(0,\! 1 | q_{i-1})$};
\node[below=7pt] at (5.5,\yC) {\scriptsize $P_{X,S|Q}(1,\! 1 | q_{i-1})$};
\draw[thick] (4.35,\yC-0.3) -- ++(0,-0.3) ;
\end{tikzpicture}
\caption{Illustration of the posterior matching scheme for a unifilar finite-state channel with binary alphabets. The length of each message interval reflects its posterior probability. The messages are separated based on their hypothetical channel state. Transmission is performed using a labeling function $\mathcal{L}_{q_{i-1}}$ that partitions the unit interval based on the conditional distribution $P_{X,S|Q}$. That is, the channel input is determined by the label of the true message interval.}
  \label{fig:example}
\end{figure}

Throughout the scheme, a $Q$-graph and an input distribution $P_{X|S,Q}$ are fixed ahead of communication. The scheme is given by a simple procedure that is iterated $n$ times. In each iteration, both the encoder and the decoder will keep track of the posterior probability (PP):
\begin{align}\label{eq:PP}
  \lambda(m)&\triangleq  P(m|y^{i-1}),
\end{align}
initialized with the prior distribution $P(m) = 2^{-nR}$. 
The PP corresponds to the decoder's belief regarding the message at time $i$, which is also available to the encoder due to the feedback. To encode, we use a posterior matching (described below) of the PP to an input distribution $P_{X|S=s,Q=q}$, where $(s,q)$ are determined as follows:
\begin{itemize}
    \item The $Q$-graph node $Q=q_{i-1}$ is determined from the channel outputs $y^{i-1}$.
    \item The state $S=s_{i-1}$ is determined for each message separately. Recall that, in a unifilar FSC, when the initial state $s_0$ is known, the channel state $s_{i-1}$ can be determined from $(x^{i-1},y^{i-1})$. Moreover, knowledge of the encoding procedure allows one to determine $x^{i-1}$ from $(m,y^{i-1})$. Therefore, for each $m$, both the encoder and the decoder can compute $s_{i-1}(m)$, which corresponds to the hypothetical channel state assuming $M=m$. 
\end{itemize}
An illustration of the posterior matching scheme is provided in Figure \ref{fig:example}.

We now present the transmission procedure, with $m^\ast$ denoting the correct message. \\[6pt]

\noindent \underline{Procedure that is iterated $n$ times}: 
\begin{enumerate}
 \item The encoder transmits
 \begin{align*}
 x(m^\ast) &= F_{X|S=s(m^*),Q=q}^{-1}[\Lambda(m^\ast)],
 \end{align*}
 where
  \begin{align*}
    \Lambda(m^\ast) &= \frac1{\pi_{S|Q}(s(m^*)|q)}\sum_{\{m< m^\ast: s(m)=s(m^*)\}} \lambda(m)
  \end{align*}
 \item The channel output $y$ is revealed. 
 \item The PP of each message is updated recursively as
\begin{align}\label{eq:PP_update}
\lambda^+(m) &= \frac{P_{Y|X,S}(y|F_{X|S=s(m),Q=q}^{-1}[\Lambda(m)],s(m))}{P_{Y|Q}(y|q)}\lambda(m)
\end{align}
 \item The graph node is updated:
\begin{align*}
q^+ &= g(q,y)
\end{align*}

 \item The state of each message is updated:
\begin{align*}
s^+(m) &= f(s(m),x(m),y)
\end{align*}
\end{enumerate}

\noindent \underline{Decoding (after $n$ iterations)}:
\begin{align*}
    \hat{m} = \arg\max \lambda(m).
\end{align*}

The following theorem records the achievable rate of the scheme.
\begin{theorem}[Informal]\label{theorem:scheme_informal}
For any BCJR-invariant input $P_{X|S,Q}$, the coding scheme above achieves $I(X,S;Y|Q)$.
\end{theorem}

The main takeaway from the above theorem is that the lower bound in Theorem~\ref{th:Q_LB} is always accompanied by a low-complexity coding scheme. Clearly, if the lower bound is tight, the corresponding coding scheme is capacity-achieving. 

The analysis of the coding scheme is omitted for the sake of brevity and due to the many technical details that are required to show Theorem~\ref{theorem:scheme_informal} precisely\footnote{Specifically, there is a need to dither the messages before transmitting channel inputs. Also, we need to use message splitting to maintain accurate behavior of the stationary distribution.}. In \cite{Sabag_BIBO_IT}, a rigorous proof is presented for the binary-input binary-output (BIBO) channel with input constraints where the state is $x^-$. By replacing $x^-$ with a general state, $s$, the analysis is identical and Theorem \ref{theorem:scheme_informal} is proved.

\section{Dual Capacity Upper Bound} \label{sec:duality_UB}

An alternative technique that yields computable upper bounds on feedback capacity is based on an extension of the so-called dual capacity upper bounding method for discrete memoryless channels \cite{topsoe67}, \cite[Theorem~8.4]{CsiszarKorner}. The resulting bounds are simpler to compute and optimize, both numerically and analytically, compared to the $Q$-graph upper bound of Theorem~\ref{th:Q_UB}. However, as we will see, the dual capacity bound requires additional parameterization of a $Q$-graph: besides the mapping $\Phi(q,y)$, it needs transition probabilities $P(y|q)$ to be defined on the edges.

The dual capacity upper bound for a discrete memoryless channel $P_{Y|X}$ states that, for an arbitrary \emph{test distribution}, $T_Y$, on the channel output alphabet, we have 
\begin{align}\label{eq:duality_DMC}
    \mathrm{C} &= \max_{P_X} I(X;Y)\nonumber\\
    & \le \max_{P_X} \DKL(P_{Y|X}\|T_Y \mid P_X)\nonumber\\
    &= \max_{x\in\mathcal{X}} \DKL(P_{Y|X=x}\|T_Y).
\end{align}
The inequality above arises from the identity $I(X;Y) = \DKL(P_{Y|X}\|T_Y \mid P_X) - \DKL(P_Y \| T_Y)$. Here, $\DKL(P_{Y|X}\|T_Y \mid P_X)$ is the conditional KL divergence, defined as the expected value with respect to $P_X$, of the KL divergences $\DKL(P_{Y|X=x} \| T_Y)$, $x \in \mathcal{X}$:
$$
\DKL(P_{Y|X}\|T_Y \mid P_X) = \sum_{x \in \mathcal{X}} P_X(x) \DKL(P_{Y|X=x} \| T_Y)
$$
The bound is known to be tight when $T_Y$ is chosen to be the output distribution $P_Y^*$ induced by any optimal (i.e., capacity-achieving) input distribution. It is known that the output distribution $P_Y^*$ is unique (while the optimal input distribution need not be). The challenge, of course, is to identify $P_Y^*$, but the bound itself is simple to compute as it is an optimization over the (typically finite) input alphabet $\mathcal X$.

Following a derivation similar in spirit to \eqref{eq:duality_DMC}, we can also obtain a dual capacity bound for DI \cite{Huleihel_Duality_FB_TIT25}. For this, we need an analogue of causal conditioning for deterministic assignments. For a sequence of deterministic functions $f_i:\mathcal{X}^{i-1} \times \mathcal{Y}^{i-1} \to \mathcal{X}$, $i = 1,2,\ldots,n$, we define the ``deterministic causal conditioning'' via
\begin{equation} \label{def:det_cc}
P^f(x^n \| y^{n-1}) \triangleq  \prod_{i=1}^n \mathbbm{1}\{x_i=f_i(x^{i-1},y^{i-1})\}.
\end{equation}
Note that $P^f(\cdot \| \cdot)$ is a $0/1$-valued function, and its support (i.e., all $(x^n,y^{n-1})$ for which $P^f(x^n \| y^{n-1}) = 1$) completely determines the component functions $f_i$. Note also that the dependence of $x_i = f_i(x^{i-1},y^{i-1})$ on $x^{i-1}$ is redundant since $y^{i-1}$ uniquely determines $x^{i-1}$. 

Given a deterministic causal conditioning $P^f(x^n\|y^{n-1})$, we define a distribution over the channel outputs
\begin{align} \label{def:Y||f}
P^f_{Y^n}(y^n) &\triangleq  \mathbb{E}_{P^f(x^n\|y^{n-1})}[P_{Y^n\|X^n}]\nonumber\\
&=\prod_{i=1}^n P\bigl(y_i \mid y^{i-1},x^i = f(x^i \| y^{i-1})\bigr),
\end{align}
where $f(x^i \| y^{i-1})$ is shorthand notation for the (unique) $x^i$ for which $f(x^i \| y^{i-1}) = 1$. We can now state and prove the dual capacity upper bound for DI.
\begin{theorem}
\label{th:DB_DI}
Given a joint distribution $P_{X^n,Y^n}$ and a test distribution $T_{Y^n}\in \mathcal P(\mathcal Y^n)$, we have
\begin{align}
    I(X^n\rightarrow Y^n)\le\max_{f(x^n\|y^{n-1})} \DKL\left(P^f_{Y^n} \, \| \, T_{Y^n}\right),
\end{align}
where the maximization is over all deterministic mappings $f(x^n \| y^{n-1})$, which affects in turn $P^f_{Y^n}$ via \eqref{def:Y||f}.
\end{theorem}

The proof of \eqref{th:DB_DI} is elegant and intuitive, e.g., [\cite{Huleihel_Duality_FB_TIT25}]:
\begin{align}
    I(X^n \to Y^n) \ & \le \ \sum_{x^n,y^n} P(x^n,y^n) \log \frac{P(y^n \| x^n)}{T(y^n)} \nonumber \\
    & \le \ \sum_{y^n} \max_{x^n: P(x^n \| y^{n-1}) > 0} P(y^n \| x^n) \log \frac{P(y^n \| x^n)}{T(y^n)} \nonumber \\
    & = \max_{f(x^n\|y^{n-1})} \underbrace{\sum_{y^n} P^f_{Y^n}(y^n) \log \frac{P^f_{Y^n}(y^n)}{T(y^n)}}_{=\ \DKL\left(P^f_{Y^n} \, \| \, T_{Y^n}\right)}. \nonumber
\end{align}    


To choose a test distribution, we take our cue from the $Q$-graph upper bound and choose \emph{graph-based test distributions} \cite{Huleihel_Sabag_DB}. 
For a fixed $Q$-graph, a graph-based test distribution is defined by a conditional probability kernel $T_{Y|Q}$ that allocates weights to the $Q$-graph edges. The test distribution defined by the kernel is given by
\begin{align}\label{eq:test_dist}
	T_{Y^n}(y^n) = \prod_{t=1}^n T_{Y|Q}(y_t|q_{t-1}),
\end{align}
where $q_{t-1}=\Phi_{t-1}(y^{t-1})$, initialized by an arbitrary Q-graph node $q_0$ (see \eqref{eq:Phi_i}).

\begin{theorem} \label{Th:DUB_Q}  
For any $Q$-graph and test distribution $T_{Y|Q}$, the feedback capacity of a strongly connected unifilar FSC is bounded by
\begin{align}
    \mathrm{C}^\mathrm{fb} \le \lim_{n\to\infty}\max_{f(x^n\|y^{n-1},s_0)}\frac{1}{n}\sum_{i=1}^n\mathbb{E}\left[\DKL\left(P_{Y|X,S}(\cdot|X_i,S_{i-1})\Bigg{\|}T\left(\cdot|Q_{i-1}\right)\right)\right],\nn
\end{align}
where the joint distribution is defined by
\begin{align}
     &P(s_i,y_i,x_i,q_i|s^{i-1},y^{i-1},x^{i-1},s_0,q_0)\nn\\
     &= P_{S^+,Y|X,S}(s_i,y_i|x_i,s_{i-1})\mathbbm{1}\{x_i=f_i(x^{i-1},y^{i-1},s_0)\}\mathbbm{1}\{q_i=\Phi_i(y^i)\}.\nn
\end{align}
Moreover, the upper bound corresponds to the average reward of a discrete MDP. 
\end{theorem}
The first part of the theorem presents a multi-letter upper bound that can be obtained by particularizing the general duality bound in Theorem \ref{th:DB_DI} for graph-based test distributions and unifilar FSCs. The main contribution is the formulation of this bound as a discrete MDP. In particular, we will present an MDP with discrete states and actions whose infinite-horizon average reward is equal to the upper bound. This implies that numerical evaluation and analytical solutions are feasible, as we illustrate next. 

Compared to the $Q$-graph upper bound of Theorem~\ref{th:Q_UB}, the computation of the upper bound of Theorem~\ref{Th:DUB_Q} is much easier since the optimization is over a discrete set. However, there is a cost to be paid: the input to the upper bound of Theorem~\ref{Th:DUB_Q} needs to be a $Q$-graph along with a test distribution, while in the $Q$-graph upper bound, one needed to only choose a $Q$-graph. In fact, these bounds complement each other; the convex optimization for the $Q$-graph upper bound can be evaluated to gain insights into the distribution $P_{Y|Q}$, e.g., if it has zeros or repetitive entries. We can then use this simpler parametrization of $P_{Y|Q}$ in the duality upper bound. 
We proceed to show the MDP formulation and its computation for our running example of the Ising channel. 

\subsection{The Duality Upper Bound as an MDP} \label{subsec:duality_DP} 
We present the MDP formulation of the dual capacity upper bound for unifilar FSCs. 
\begin{table}[t]
\centering
\caption{MDP Formulation for the Duality Bound}\label{table:duality_MDP}
 \begin{tabular}{|c|c|}
 \hline
MDP & Duality bound\\ [0.5ex] 
 \hline\hline
MDP state & $(s_{t-1},q_{t-1})$ \\ [0.3ex]
 \hline
Action & $x_t$ \\[0.3ex]
 \hline
Disturbance & $y_t$ \\[0.3ex]
 \hline
The reward & $\DKL\left(P_{Y|X,S}(\cdot|x_t,s_{t-1})\middle\|T_{Y|Q}(\cdot|q_{t-1})\right)$\\[0.3ex]
 \hline
\end{tabular}
\end{table}
Fix a $Q$-graph and a corresponding graph-based test distribution $T_{Y|Q}$. The MDP state at time $t$ is defined as $z_{t-1}\triangleq  (s_{t-1},q_{t-1})$. The action is defined as the channel input $u_t \triangleq  x_t$, and the disturbance is the channel output $w_t\triangleq  y_t$. The reward function is defined as
\begin{align} \label{reward_1}
	g(s_{t-1},q_{t-1},x_t) \triangleq  \DKL\left(P_{Y|X,S}(\cdot|x_t,s_{t-1})\middle\|T_{Y|Q}(\cdot|q_{t-1})\right).
\end{align}
The MDP formulation is summarized in Table \ref{table:duality_MDP}, and its infinite-horizon average-reward is
\begin{align} \label{eq: optimal_reward_1}
    &\rho^*=\sup\liminf_{n\to\infty}\min_{s_0,q_0}\frac{1}{n}\sum_{t=1}^{n}\mathbb{E}\left[g(S_{t-1},Q_{t-1},X_t)\right], 
\end{align}
where the supremum is over sequences of deterministic actions $\{f_i:\mathcal{X}^{i-1}\times\mathcal{Y}^{i-1}\rightarrow \mathcal{X}\}_{i\ge 1}$. The following theorem summarizes the relationship between the upper bound in Theorem \ref{Th:DUB_Q} and $\rho^\ast$.
\begin{theorem}\label{theorem: formulation}
The upper bound in Theorem \ref{Th:DUB_Q} is equal to the optimal average reward in \eqref{eq: optimal_reward_1}. That is, the capacity is upper bounded by the average reward of the MDP $\rho^*$.
\end{theorem}
The proof follows by showing that the formulation above constitutes a valid MDP and that its optimal average reward is equal to the duality upper bound on the capacity.

\subsubsection{Evaluation}
We note that the state, action, and disturbance are discrete. Consequently, solving the MDP numerically and finding the optimal policy is quite straightforward. For example, the policy iteration algorithm converges to the optimal policy in a finite time. For the resulting \emph{suspected} optimal policy, $x^*$ (as a function of the MDP state $(s,q)$), one can use the set of linear equations
\begin{align}\label{eq:duality_fixed_policy}
    \rho + V(s,q) &= \DKL\left(P_{Y|X=x^*,S=s}\lVert T_{Y|Q=q}\right) +\sum_{y}P(y|x^*,s)V\left(x^*,\Phi(q,y)\right).
\end{align}
to find $\rho$ and $V(s,q)$. All that is left then is to verify the Bellman equation
\begin{align}\label{eq:duality_bellman}
    \rho + V(s,q) &= \max_x \DKL\left(P_{Y|X=x,S=s}\lVert T_{Y|Q=q}\right) +\sum_{y}P(y|x,s)V\left(x,\Phi(q,y)\right).
\end{align}
for all $(s,q)$, which allows us to conclude that $\mathrm{C}^\mathrm{fb} \leq \rho^*$. We proceed to illustrate this method for our running example of the binary Ising channel.

\textbf{Example~\ref{ex:ising} (cont.)} 
Consider the $Q$-graph in Fig. \ref{fig:Q_Ising} and the test distribution
\begin{align} \label{eq:duality_test_ising}
    T(y=0|q) &=\
    \begin{cases}
    \frac{1-a}{2}, &q=1\\
    \frac{1-a}{1+a}, &q=2\\
    \frac{2a}{1+a}, &q=3\\
    \frac{1+a}{2}, &q=4
    \end{cases},
\end{align}
where $a\in[0,1]$ is a parameter to be optimized later. The structure of the test distribution and the $Q$-graph are based on numerical simulation of the $Q$-graph upper bound in Theorem \ref{theorem:UB_convex}. Applying the value iteration algorithm reveals the optimal policy $x^*(s,q)$. That is, for each $(s,q)$, we have $x\in\{0,1\}$ that solves the right hand side of \eqref{eq:duality_bellman}. Solving the set of linear equations in \eqref{eq:duality_fixed_policy} yields
\begin{align}\label{eq:rho_ISIng}
  \rho(a) 
  &= -0.5\log a,
\end{align}
and the value function $V(s,q)$:
\begin{align}
  V(0,1) &= V(1,4) = 1 - \log (1-a) \\
  V(0,2) &= V(1,3) = \log \frac{1+a}{1-a} \\
  V(0,3) &= V(0,3) = \log (1+a) + 2\rho(a) - 1\\
  V(0,4) &= V(1,1) = \rho(a),\label{eq:value_ISIng}
\end{align}
where $a\in[0,1]$.

Let $a^*$ be the unique solution in $[0,1]$ of $(1-a)^4=a^3$. Then, it is possible to show via a direct verification (the details of which we omit) that the Bellman equation \eqref{eq:duality_bellman} is satisfied by $\rho(a^*)$ and $V(s,q)$ given by \eqref{eq:rho_ISIng}-\eqref{eq:value_ISIng}. Thus, the capacity of the binary Ising channel is upper bounded by $\rho(a^*)$.

\subsection{$Q$-Graphs Exploration} \label{subsec:Q-exploration}
A natural question for the bounds provided so far is how to find $Q$-graphs that will yield the tightest bounds. The $Q$-graph upper bound was developed in \cite{sabag2017single} as an alternative proof technique to solving the Bellman equation. Thus, the most effective method for finding an optimal $Q$-graph for a feedback capacity problem is to study the associated MDP. For several examples of unifilar channels \cite{Permuter2008Trapdoor,ElishcoPermuter2014Ising,SabagPermuterKashyap2016BEC,Sabag_BIBO_IT,Ising_artyom_IT}, Monte Carlo simulations of the corresponding MDP under an estimated optimal policy (obtained, for example, through value or policy iteration) happened to produce a discrete histogram of the states visited by the MDP. This gave a strong hint that for each of these channels, under the optimal policy, the number of visited MDP states is indeed discrete and therefore can be described by a $Q$-graph. Each visited state is taken as a node in the $Q$-graph, and the labelled edges are obtained from the evolution of the MDP states as a function of the channel outputs. The $Q$-graphs obtained in this manner often yielded the exact feedback capacity via Corollary~\ref{cor:Q_bnds}. Unfortunately, this technique does not scale well with the dimension of the MDP state space, so we turn to coarser approximations obtained through reinforcement learning, which we describe in the following section.

Other methods for obtaining candidate $Q$-graphs include a systematic enumeration of all small $Q$-graphs (see  \cite{sabag2019graph}), the complexity of which explodes as the graph size increases, and constructions of de Bruijn graphs (Markov state transition diagrams) of small orders $m$, which we previously described at the end of Section~\ref{subsec:Qgraph-lobnd}.

\section{Capacity Evaluation via Reinforcement Learning}
In this section, we provide a brief background on reinforcement learning (RL), based on \cite{sutton2018reinforcement}, in order to formulate the feedback capacity of a unifilar FSC as an RL problem and provide algorithms to compute the capacity. In our case, the environment is known, so the main objective of RL methods is to avoid computational complexity issues that are encountered in the value and policy iteration algorithms for large channel state alphabets. In particular, we will avoid the curse of dimensionality encountered in tabular methods by approximating different components of the MDP setup using neural networks. Our first algorithm is based on the deep deterministic policy gradient (DDPG) algorithm along with some improvements. The second algorithm is policy optimization via unfolding (POU). 

The performance of these algorithms will be demonstrated via the running example of the Ising channel. The DDPG algorithm estimates both the value function and the policy, and  its results are easier to interpret. However, we found that it yielded better lower bounds (compared with POU) only for $|\cX|\leq 15$. The POU algorithm, on the other hand, directly estimates the policy and improved the performance of the DDPG algorithm for $|\cX|>15$.

\subsection{RL Setting}
\begin{figure}[t]
    \centering
    \subfigure[General RL setting]
    {
    \includegraphics[width=0.29\linewidth]{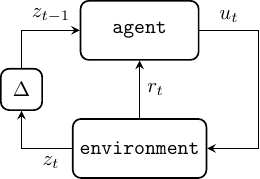}
    }
    \qquad\qquad
    \subfigure[Feedback capacity formulated in the RL setting]
    {
     \includegraphics[width=0.45\linewidth]{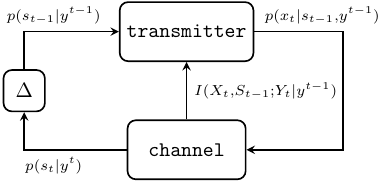}
    }
  \caption{A description of (a) the general RL setting and (b) the feedback capacity problem formulated in the RL setting.}
  \label{fig:rl-formulation}
\end{figure}

The RL setting comprises an agent that interacts with a state-dependent environment whose input is an action and the output is a state and a reward. 
Formally, at time $t$, the environment state is $z_{t-1}$, and an action $u_t\in\mathcal{U}$ is chosen by the agent. Then, a reward $r_t\in\mathcal{R}$ and a new state $z_t\in\mathcal{Z}$ are generated by the environment, and are made available to the agent (Figure \ref{fig:rl-formulation}). 
The environment is assumed to satisfy the Markov property 
\begin{equation} 
    P\left(r_t, z_t \lvert z^{t-1},u^t,r^{t-1}\right) = P\left(r_t,z_t \lvert z_{t-1},u_t\right), \label{eqn:markov}
\end{equation}
and hence, it can be characterized by the time-invariant distribution $P\left(r_t,z_t \lvert z_{t-1},u_t\right)$. The agent's \emph{policy} is defined as the sequence of actions $\pi = (u_1, u_2, \dots )$.
\begin{figure}[b]
    \centering
    \includegraphics[scale=1.2, trim=0 0 0 0, clip]{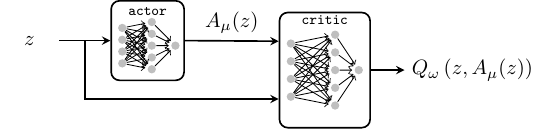}
    \caption{Depiction of actor and critic networks. The actor network comprises a NN that maps the state $z$ to an action $A_\mu(z)$. The critic NN maps the tuple $(z, u)$ to an estimate of expected future cumulative rewards.}
    \label{fig:actor-critic}
\end{figure}

The objective of the agent is to choose a policy that yields maximal accumulated rewards across a predetermined horizon $h\in\NN$. Here, we consider the infinite-horizon average reward regime, where the agent aims to maximize the average reward. The average reward of the agent is defined by 
\begin{equation}\label{eqn:average_reward}
    \rho(\pi) = \lim_{h\rightarrow\infty} \frac{1}{h}\sum_{t=1}^h \EE_\pi[R_t|Z_0],
\end{equation}
where the rewards depend on the initial state $Z_0$ and on the actions taken according the policy $\pi$.

The \emph{differential return} is defined by 
\begin{equation}
    G_t = R_t -\rho(\pi) + R_{t+1} - \rho(\pi) + R_{t+2} -\rho(\pi) + \cdots.
\end{equation}
Accordingly, the differential value function of state-action pairs is defined as 
\begin{equation}
    Q_\pi(z,u) = \EE_\pi \left[ {G_t \lvert Z_{t-1}=z, U_t=u} \right]. \label{eqn:Q}
\end{equation}
This function corresponds to the average reward achieved by a policy $\pi$ from a state $z$ but when taking the action $u$ at the first time $t$. Using the Markov property \eqref{eqn:markov}, we can write \eqref{eqn:Q} as the sum of the immediate reward and the subsequent differential value function as
\begin{align}
    Q_\pi(z,u) &= \EE \left[ R_t \lvert Z_{t-1}=z, U_t=u \right]- \rho(\pi) \nonumber\\
    & + \EE_\pi \left[ Q_\pi(Z_t, U_{t+1})\lvert Z_{t-1}=z, U_t=u\right], \label{eqn:Q_decompose}
\end{align}
which is the Bellman equation \cite{Bertsekas05}, and is essential for estimating the function $Q_{\pi}$. Given a policy and an estimation of its value function, we can improve the action taken at state $z \in \mathcal{Z}$ by computing the greedy action
\begin{equation}
    \pi'(z) = \arg\max_{u} Q_\pi(z, u).
\end{equation}

In RL, \emph{function approximators} are used to model the value function $Q_\pi(z,u)$ and the policy $\pi(z)$. The \emph{actor} is defined by $A_\mu(z)$, a parametric model of $\pi(z)$, whose parameters are $\mu$.
The \emph{critic} is defined by $Q_\omega(z,u)$, a parametric model with parameters $\omega$ of the state-action value function that corresponds to the policy $A_\mu(z)$.
Generally, in deep RL, the actor and critic are modeled by NNs, as shown in Fig.~\ref{fig:actor-critic}. 

\subsection{The Deep Deterministic Policy Gradient (DDPG) Algorithm}
We elaborate on the DDPG \cite{ddpg} algorithm and its improvement.

\subsubsection{Algorithm}
The DDPG algorithm \cite{ddpg} is a deep RL algorithm suitable for optimizing deterministic policies in environments with continuous states and continuous actions. The training procedure comprises $M$ episodes, where each episode includes $T$ steps. Each episode consists of two operations: (1) collecting experience from the environment, and (2) improving the actor and critic deep networks using the accumulated data.

\begin{figure}[!ht]
    \centering
    \includegraphics[scale=0.8, trim=0 0 0 0, clip]{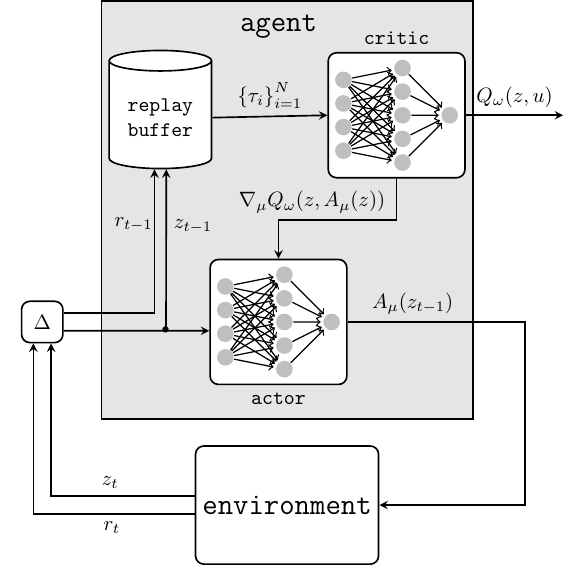}
    \caption{Depiction of the work flow of the DDPG algorithm. At each time step $t$, the agent samples a transition from the environment using an $\epsilon$-greedy policy and stores the transition in the replay buffer. Simultaneously, $N$ past transitions $\left\{\tau_i\right\}_{i=1}^N$ are drawn from the replay buffer and used to update the critic and actor NN according to \eqref{eqn:critic-update} and \eqref{eqn:actor-update}, respectively. }
    \label{fig:ddpg-workflow}
\end{figure}

In the first operation, the agent collects experience from the environment. 
When at state $z_{t-1}$, the agent chooses an action $u_t$ based on the actor with an element of exploration. The action is a probability distribution and therefore exploration is applied by adding noise to the actor network's last hidden layer, and not by adding noise to the network output (as done in \cite{ddpg}). We denote this action by $A_\mu(z_{t-1}; N_t)$, where $\{N_t\}$ is an i.i.d. Gaussian process with $N_t \sim \cN(0,\sigma^2)$. After taking the action $A_\mu(z_{t-1};N_t)$, the agent observes the incurred reward $r_t$ and the next state $z_t$. This describes a single transition and the tuple 
\begin{align}
  \tau = \left(z_{t-1},u_t,r_t,z_t\right)  
\end{align}
is stored in a \emph{replay buffer} that will be later used to improve the actor and critic neural networks.

\begin{algorithm}[ht]
  \caption{DDPG algorithm for feedback capacity of unifilar FSC \label{alg:ddpg}}
  \begin{algorithmic}
    \STATE Randomly generate the critic $Q_\omega$ and actor
    $A_\mu$
    \STATE Initialize target networks $Q^\prime_\omega$ and $A^\prime_\mu$ with $\omega^\prime 
    \leftarrow \omega$, $\mu^\prime \leftarrow \mu$
    \STATE Moving-average parameter $\alpha$
    \FOR{episode = 1:M}
      \STATE Generate random $\{N_t\}$ for exploration
      \STATE Set $\rho_{MC} = \frac{1}{T_{MC}}\sum_{t=1}^{T_{MC}} r_{t}$ by a Monte-Carlo evaluation of the actor $A_\mu$
      \STATE Randomize initial state $z_0$ from the $|\cX|$-simplex
      \FOR{step = 1:T}
        \STATE Execute action $u_t = A_\mu(z_{t-1} ; N_t) $ and observe $(r_t,z_t)$
        \STATE Store transition $(z_{t-1}, u_t,
                r_t, z_t)$ in $R$
        \STATE Sample a random batch of $N$ transitions from $R$
        \STATE Set $b_i = r_i - \rho_{MC} + Q^\prime_\omega\left(z_i, A^\prime_\mu(z_i)\right)$ \label{alg-step:td-target}

        \STATE Update critic by minimizing the loss:
               $L\left( \omega \right) = \frac{1}{N} \sum_{i=1}^{N
               } \left[Q_\omega\left(z_{i-1}, A_\mu(z_{i-1})\right) - b_i \right]^2$
        \STATE Update the actor policy using the sampled policy gradient:
        \begin{equation*}
            \frac{1}{N} \sum_{i=1}^{N} \nabla_a Q_\omega\left(z_{i-1}, a\right)\lvert_{a=A_\mu(z_{i-1})} \nabla_\mu A_\mu\left( z_{i-1}\right)
         \end{equation*}
        \STATE Update the target networks:
          \begin{align*}
            \omega^\prime &\leftarrow \alpha \omega + (1 - \alpha) \omega^\prime \\
            \mu^\prime &\leftarrow \alpha \mu +
                (1 - \alpha) \mu^\prime
          \end{align*}

        \ENDFOR
    \ENDFOR
    \RETURN $\rho_{MC}= \frac{1}{T_{MC}}\sum_{t=0}^{T_{MC}-1} r_{t+1}$
  \end{algorithmic}
\end{algorithm}

\par The second operation entails training the actor and critic networks.
First, $N$ tuples $\left\{ \tau_i \right\}_{i=1}^N$ are drawn uniformly from the replay buffer. For each tuple, a target $b_i$ is computed based on the right-hand-side of \eqref{eqn:Q_decompose}:
\begin{equation}\label{eqn:rl_sample_target}
    b_i = r_i - \rho_{MC} +  Q^\prime_\omega\left(z_i, A^\prime_\mu(z_i)\right), \quad i=1,\dots,N.
\end{equation}
The target corresponds to the estimate of future rewards; for numerical reasons, it is computed as a moving average of the networks $Q_\omega,A_\mu$, which are the target networks $Q^\prime_\omega,A^\prime_\mu$.
The term $\rho_{MC}$ is a Monte-Carlo estimate of the average reward, computed at the beginning of every episode and includes $T_{MC}$ samples. Then, we minimize the following objective with respect to the parameters of the critic network $\omega$ as given by
\begin{equation}\label{eqn:critic-update}
    L\left( \omega \right) = \frac{1}{N} \sum_{i=1}^{N} \left(Q_\omega\left(z_{i-1}, A_\mu(z_{i-1})\right) - b_i \right)^2.
\end{equation}
The aim of this update is to train the critic to comply with the Bellman update \eqref{eqn:Q_decompose}.
Afterwards, we train the actor to maximize the critic's estimation of future cumulative rewards. 
That is, we train the actor to choose actions that result in high cumulative rewards according to the critic's estimation.
The formula for the actor update is 
\begin{equation} \label{eqn:actor-update}
    \frac{1}{N} \sum_{i=1}^{N} \nabla_a Q_\omega\left(z_{i-1}, a\right)\lvert_{a=A_\mu(z_{i-1})} \nabla_\mu A_\mu\left( z_{i-1}\right).
\end{equation}
Finally, the agent updates its current state to be $z_t$ and moves to the next time step. 
\par To conclude, the algorithm alternates between improving the critic's estimation of future cumulative rewards and training the actor to choose actions that maximize the critic's estimation.
The algorithm is given in Algorithm \ref{alg:ddpg} and its workflow is depicted in Figure \ref{fig:ddpg-workflow}.

\subsubsection{Improvements}
Two improvements are proposed for the vanilla DDPG algorithm. The first utilizes the fact that the environment is known, so that in the estimation of $Q_\pi$ we can replace samples with expectations. That is, to estimate the right-hand-side of \eqref{eqn:Q_decompose}, we modify \eqref{eqn:rl_sample_target} with
\begin{align}\label{eqn:rl_exp_target}
    b_i &= r_i - \rho_{MC}  + \sum_{w \in \cY} P(w|z_{i-1},u_i)Q^\prime_\omega\left(z_i, A^\prime_\mu(z_i)\right),
\end{align}
where $z_i=f(z_{i-1},u_i,w)$. By computing the expectation over the next state, we reduce the error variance. 

The second improvement is a variant of importance sampling \cite{importance_sampling}, which is crucial due to the continuous state space and the presence of rarely visited states. The DDPG algorithm often struggles to improve the policy for these states because of their infrequent appearance. The proposed solution is to cluster transitions in the replay buffer, which ensures a more uniform update of the policy across all states, including those less frequently encountered. Before a new transition is stored in the replay buffer, its distance (e.g., max-norm) from all cluster centers is computed. If the distance to the closest cluster is smaller than a chosen threshold, the transition is stored in the corresponding cluster; otherwise, a new cluster is introduced with the transition as its center.

The sampling process from the replay buffer is modified as follows: a cluster is first sampled uniformly, and then a transition is sampled uniformly from the selected cluster. This modification increases the probability that rare states—those located in smaller clusters—will be sampled from the replay buffer. Consequently, this enhances the value function estimation for rare states, leading to better policies for those rarely visited states.

The implementation details appear in \cite{AharoniSabagRL} and its implementation is available in a public repository \cite{aharoni2024capacityrl}.

\subsection{Policy Optimization by Unfolding (POU) Algorithm}
The POU algorithm optimizes $n$ consecutive rewards with respect to time-invariant policies.
The main idea is to utilize the knowledge of the RL environment so as to create a differentiable mapping of the cumulative reward as a function of the policy and an arbitrary initial state. We emphasize that this is possible since the reward and the state evolution functions are known.

\begin{figure}[!b]
    \centering
    \includegraphics[scale=0.8, trim=0 0 0 0, clip]{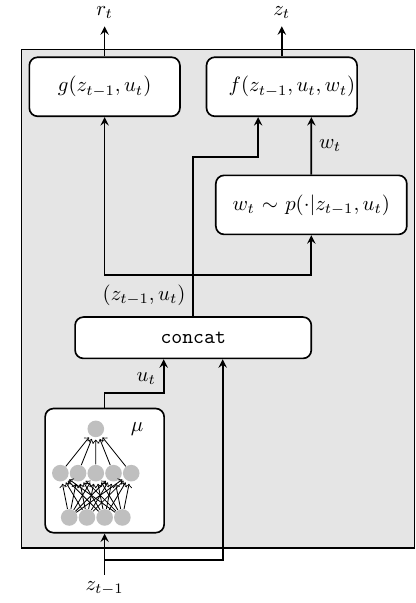}
    \caption{A single step of the environment. The input of the block is the current RL state $z_{t-1}$ and the outputs are the immediate reward $r_t$ and the next sampled state $z_t$. Initially, the block uses the actor to construct the tuple $(z_{t-1},u_t)$. Afterwards, it samples the disturbance from $w_t \sim P(\cdot|z_{t-1}, u_t)$, and finally, uses $g$ and $f$ to compute the reward and the next state, respectively.}
    \label{fig:policy-opt-cell}
\end{figure}

\subsubsection{Algorithm}
Denote the policy-dependent reward function by
\begin{align}
    R_\mu(z)&= g\big(z,\mu(z)\big),
\end{align}
which depends only on the state $z$ for a deterministic policy $\mu$. For some initial state $z_0$, the average reward over $n$ time instances is
\begin{align}\label{eq:POU_accum_reward}
    R^n_\mu(z_0)&= \frac{1}{n}\sum_{t=1}^n \mathbb{E} \big[ R_\mu(Z_{t-1}) \big] \nonumber\\
    &= \frac{1}{n}\left[R_\mu(z_0) + \sum_{t=2}^n\mathbb{E} \big[ R_\mu(Z_{t-1}) \big]\right],
\end{align}
where $Z_t$ is subject to the state evolution function $Z_t = f(Z_{t-1}, \mu(Z_{t-1}), W_t)$ so that the only randomness in the expectation is the conditional distribution $P_{W_{t}|Z_{t-1},\mu(Z_{t-1})}$ for $t=1,\dots,n$.

\begin{figure}[!t]
    \centering
    \includegraphics[scale=0.73, trim=10pt 0 0 0, clip]{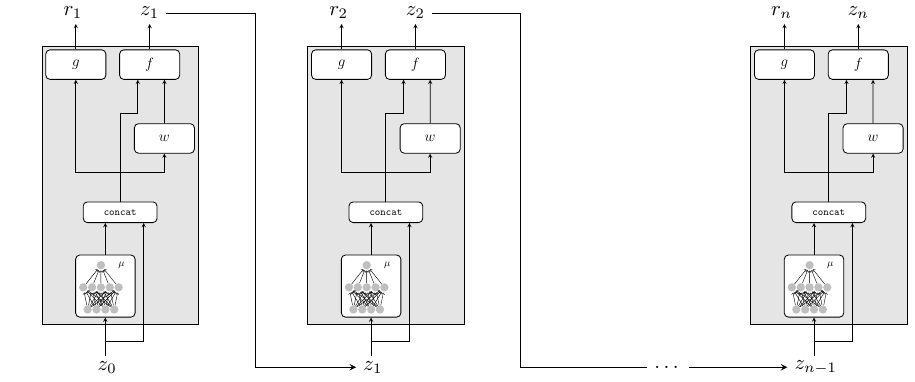}
    \caption{The interaction with the channel unrolled across subsequent time steps. The weights of the actor network are shared across time steps.}
    \label{fig:policy-opt}
\end{figure}

The choice of the horizon $n$ significantly affects the performance of the optimized policy; as $n$ increases, the policy takes into account more future rewards. For instance, $n=1$ translates to optimizing the immediate reward, which yields the greedy policy. In practice, an interaction over relatively small $n$, e.g., $n=20$, suffices to achieve policies with long-term high performance. However, the number of  MDP states grows exponentially as $|\cY|^n$ (recall that the disturbance is the channel output $Y$).

To resolve the exponential complexity, the POU algorithm proposes a simple, yet efficient, method to \emph{unfold} the interaction with the environment. Given a policy $\mu$ and an initial state $z_0$, we sample consecutive disturbances and their corresponding states as
\begin{align} \label{eqn:pou_single_time_step}
    W_{t}&\sim P_{W|Z,U}(\cdot |Z=z_{t-1},U = \mu(z_{t-1})),\nonumber\\
    z_{t}&= f(z_{t-1},\mu(z_{t-1}),w_{t}),
\end{align}
where $P_{W|Z,U}$ describes the channel output distribution conditioned on the previous MDP state and the action $\mu(Z_{t-1})$. This law is dictated by the RL environment and the chosen policy, but is not a function of the horizon $n$. For a single $t$, the law in \eqref{eqn:pou_single_time_step} describes a single step where the agent interacts with the environment, as shown in Figure \ref{fig:policy-opt-cell}. The interaction with the environment for $n$ consecutive steps is shown in Figure \ref{fig:policy-opt}.

Having sampled the disturbances sequence, a differentiable mapping between $z_0$ and the average reward is established. Specifically, we can compute the derivative of the average reward in \eqref{eq:POU_accum_reward} (without the expectation),
\begin{align}
    \nabla_\mu \left[ \frac{1}{n}\sum_{t=1}^n  R_\mu(z_{t-1}) \right],
\end{align}
where $z_1,\dots,z_{n-1}$ are computed by $z_{t}= f(z_{t-1},\mu(z_{t-1}),w_{t})$ and $w^{n-1}$ is the sampled disturbances sequence. 
Then, we update the policy $\mu$ with the standard gradient ascent update as:
\begin{align}
    \mu = \mu + \eta \nabla_\mu \left[ \frac{1}{n}\sum_{t=1}^n  R_\mu(z_{t-1}) \right],
\end{align}
where $\eta$ is the step size.
\begin{algorithm}[!ht]
  \caption{POU algorithm for feedback capacity of unifilar FSC \label{alg:pou}}
  \begin{algorithmic}
    \STATE Randomly generate a policy $A_\mu$
    \FOR{episode = 1:M}
      \STATE Sample $z_0$ uniformly from $(|\cS|-1)$-simplex
      \FOR{$t = 1:T$}
        \STATE Conditioned on $z_{0}$ and $A_\mu$, sample $(w_1,\dots,w_{n})$ according to \eqref{eqn:pou_single_time_step}
        \STATE Update the actor parameters using gradient ascent
        \begin{equation*}
            \mu = \mu + \eta \nabla_\mu \left[ \frac{1}{n}\sum_{i=1}^n  R_\mu(z_{i-1}) \right]
         \end{equation*}
        \STATE Update the initial state $z_0 = z_n$
        \ENDFOR
    \ENDFOR
    \RETURN $\rho_{MC}= \frac{1}{T_{MC}}\sum_{t=1}^{T_{MC}} r_{t}$
  \end{algorithmic}
\end{algorithm}
This procedure is repeated using the last state $z_n$ as the initial state of the next $n$ steps. This is shown in Algorithm \ref{alg:pou}. 

\subsection{The Ising Channel with Large Alphabet}
This section applies the RL methodology to the non-binary Ising channel. The state of the Ising channel takes values in the $|\cX|$-simplex. Implementing the value iteration algorithm requires iterations and memory handling of MDP states. The standard approach creates a uniform grid for each axis in the simplex to quantize the true state. However, even with $|\cX| = 3$, implementation is challenging. The RL algorithms discussed earlier can be evaluated for the Ising channel with a larger alphabet.


Both RL methods yield a policy parameterized by a neural network. While this does not provide deep insight into the analytical form of the optimal policy, we can use Monte Carlo simulations to analyze the state. For example, when $|\cX| = 3$, the histogram of visited states under the optimized policy appears discrete, allowing to extract a $Q$-graph for lower and upper bounds. Further discussion can be found in \cite{aharoni2024capacityrl}. However, these RL methods can be used to derive analytical solutions for the channel capacity. The following theorem relies on a direct extension of the graphs for $|\cX| = 2,3$ to arbitrary $|\cX|$. Surprisingly, this extension leads to a closed-form expression for the feedback capacity of the Ising channel for $ \left| \cX \right|\leq 8$.
\begin{theorem}[Feedback Capacity] \label{thm:ising_cfb}
    The feedback capacity of the Ising channel with $\left| \cX \right|\leq 8$ is given by 
    \begin{equation}
        C_\mathsf{FB}(\cX) = \max_{p \in [0,1]}  2\frac{H_2(p) + (1-p) \log\left(\lvert\cX\lvert-1\right)}{p+3}.
    \end{equation}
    Equivalently, the feedback capacity can be expressed as
    \begin{equation}
        C_\mathsf{FB}(\cX) = \frac{1}{2}\log\frac{1}{p},
    \end{equation}
    where $p$ is the unique solution of $x^4 - ((\lvert\cX\lvert-1)^4 + 4)x^3 +6x^2 -4x+1=0$ in $[0,1]$.
\end{theorem}

\begin{acknowledgements}
We thank the reviewers for their insightful observations and comments, which helped shape this manuscript.
The work of Haim H. Permuter and Dor Tsur was supported in part by
Israel Science Foundation (ISF) under Grant 899/21 and Grant 3211/23,
in part by the NSF-(Israel-USA) Binational Science Foundation (BSF)
Grant, and in part by the Israeli Innovation Authority.
The work of Gerhard Kramer was supported by the German Research Foundation (DFG) via the German-Israeli Project Cooperation (DIP) under project KR 3517/13-1.
The work of Oron Sabag was supported by the Israel Science Foundation (ISF) under grant No. 1096/23.
The work of Navin Kashyap was supported in part by the Science and Engineering Research Board (SERB), Government of India, under the MATRICS program, and in part by Qualcomm Innovation Fellowships.
\end{acknowledgements}

\backmatter  

\printbibliography

\end{document}